\documentclass[reqno,10pt,a4paper,dvips]{amsart}

\usepackage{amssymb,mathptmx,cite,psfrag,eucal,array,setspace,geometry,enumitem}
\usepackage[dvips]{graphicx}
\usepackage{tikz}

\numberwithin{equation}{section}

\newcolumntype{C}{>{$}c<{$}} 

\allowdisplaybreaks

\newcommand{\R}{\mathbb{R}}

\newcommand{\alg}[1]{\mathfrak{#1}}
\newcommand{\group}[1]{\mathsf{#1}}
\newcommand{\uealg}[1]{\mathcal{U} \bigl( #1 \bigr)}

\newcommand{\func}[2]{#1 \left( #2 \right)}
\newcommand{\tfunc}[2]{#1 \bigl( #2 \bigr)}

\newcommand{\brac}[1]{\left( #1 \right)}

\newcommand{\set}[1]{\left\{ #1 \right\}}
\newcommand{\tset}[1]{\bigl\{ #1 \bigr\}}
\newcommand{\st}{\mspace{5mu} : \mspace{5mu}}

\newcommand{\abs}[1]{\left| #1 \right|}

\newcommand{\ZZ}{\mathbb{Z}}

\newcommand{\RR}{\mathbb{R}}
\newcommand{\CC}{\mathbb{C}}

\newcommand{\dd}{\mathrm{d}}
\newcommand{\ii}{\mathfrak{i}}
\newcommand{\ee}{\mathsf{e}}

\newcommand{\eps}{\varepsilon}

\newcommand{\killing}[2]{\kappa \bigl( #1 , #2 \bigr)}

\newcommand{\affine}[1]{\widehat{#1}}

\newcommand{\comm}[2]{\bigl[ #1 , #2 \bigr]}
\newcommand{\acomm}[2]{\bigl\{ #1 , #2 \bigr\}}

\newcommand{\ket}[1]{\bigl\lvert #1 \bigr\rangle}

\newcommand{\normord}[1]{\mbox{${} : #1 : {}$}} 

\newcommand{\VerMod}[1]{\mathcal{V}_{#1}}

\newcommand{\ProjMod}[1]{\mathcal{P}_{#1}}
\newcommand{\TypMod}[1]{\mathcal{T}_{#1}}
\newcommand{\AtypMod}[1]{\mathcal{A}_{#1}}

\newcommand{\AffIrrMod}[1]{\affine{\mathcal{L}}_{#1}}
\newcommand{\AffVerMod}[1]{\affine{\mathcal{V}}_{#1}}

\newcommand{\AffProjMod}[1]{\affine{\mathcal{P}}_{#1}}
\newcommand{\AffTypMod}[1]{\affine{\mathcal{T}}_{#1}}
\newcommand{\AffAtypMod}[1]{\affine{\mathcal{A}}_{#1}}
\newcommand{\AffKerMod}[1]{\affine{\mathcal{K}}_{#1}}

\newcommand{\TwAffTypMod}[1]{\affine{\mathcal{T}}_{#1}^{\; \text{tw}}}

\newcommand{\CosIrrMod}[1]{\mathbb{L}_{#1}}
\newcommand{\CosOthMod}[1]{\mathbb{E}_{#1}}
\newcommand{\CosProjMod}{\mathbb{S}}

\newcommand{\SLA}[2]{\alg{#1} \left( #2 \right)}
\newcommand{\SLSA}[3]{\alg{#1} \left( #2 \middle\vert #3 \right)}
\newcommand{\AKMA}[2]{\affine{\alg{#1}} \left( #2 \right)}
\newcommand{\AKMSA}[3]{\affine{\alg{#1}} \left( #2 \middle\vert #3 \right)}
\newcommand{\SLG}[2]{\group{#1} \left( #2 \right)}
\newcommand{\SLSG}[3]{\group{#1} \left( #2 \middle\vert #3 \right)}

\newcommand{\ch}[2]{\tfunc{\chi_{\raisebox{-3pt}{$\scriptstyle #1$}}}{#2}}
\newcommand{\nch}[2]{\tfunc{\mathrm{ch} \bigl[ #1 \bigr]}{#2}}
\newcommand{\sch}[2]{\tfunc{\mathrm{sch} \bigl[ #1 \bigr]}{#2}}

\newcommand{\NCH}[1]{\mathrm{ch} \bigl[ #1 \bigr]}
\newcommand{\SCH}[1]{\mathrm{sch} \bigl[ #1 \bigr]}

\newcommand{\fuse}{\mathbin{\times}}
\newcommand{\fuscoeff}[3]{N_{#1 \; #2}^{\hphantom{#1 ; #2} #3}}
\newcommand{\sfuscoeff}[3]{\mathcal{N}_{#1 \; #2}^{\hphantom{#1 ; #2} #3}}

\newcommand{\Vertop}[1]{\ee^{#1}}
\newcommand{\vertop}[1]{\normord{\Vertop{#1}}}

\newcommand{\jth}[1]{\vartheta_{#1}}
\newcommand{\Jth}[2]{\tfunc{\jth{#1}}{#2}}

\newcommand{\qnum}[2]{\left( #1 \right)_{#2}}

\newcommand{\hypergeom}[4]{{}_2 {\phi}_1 \Bigl( \genfrac{}{}{0pt}{}{#1 \hphantom{#3} #2}{#3} ; \: #4 \Bigr)}

\newcommand{\dses}[3]{0 \longrightarrow #1 \longrightarrow #2 \longrightarrow #3 \longrightarrow 0}
\newcommand{\dres}[4]{\cdots \longrightarrow #4 \longrightarrow #3 \longrightarrow #2 \longrightarrow #1 \longrightarrow 0}

\newcommand{\eqnref}[1]{Equation~\eqref{#1}}
\newcommand{\eqnDref}[2]{Equations~\eqref{#1} and \eqref{#2}}
\newcommand{\eqnTref}[3]{Equations~\eqref{#1}, \eqref{#2} and \eqref{#3}}
\newcommand{\eqnQref}[4]{Equations~\eqref{#1}, \eqref{#2}, \eqref{#3} and \eqref{#4}}
\newcommand{\eqnsref}[2]{Equations~\eqref{#1} -- \eqref{#2}}
\newcommand{\secref}[1]{Section~\ref{#1}}

\newcommand{\appref}[1]{Appendix~\ref{#1}}
\newcommand{\figref}[1]{Figure~\ref{#1}}

\newcommand{\tabref}[1]{Table~\ref{#1}}

\newcommand{\cft}{conformal field theory}
\newcommand{\cfts}{conformal field theories}
\newcommand{\uea}{universal enveloping algebra}

\newcommand{\lcfts}{logarithmic conformal field theories}
\newcommand{\WZW}{Wess-Zumino-Witten}
\newcommand{\ope}{operator product expansion}
\newcommand{\opes}{operator product expansions}
\newcommand{\hws}{highest weight state}
\newcommand{\hwss}{highest weight states}

\newcommand{\hwm}{highest weight module}
\newcommand{\hwms}{highest weight modules}

\newcommand{\bb}{\beta}
\newcommand{\bg}{\gamma}
\newcommand{\bc}{\varkappa}

\newcommand{\BH}{\mathbf{H}}
\newcommand{\BZ}{\mathbf{Z}}
\newcommand{\BE}{\mathbf{E}}
\newcommand{\BF}{\mathbf{F}}
\newcommand{\Be}{\mathbf{e}}
\newcommand{\Bf}{\mathbf{f}}

\DeclareMathOperator{\id}{id}
\DeclareMathOperator{\vectspan}{span}
\DeclareMathOperator{\tr}{tr}

\newcommand{\traceover}[1]{\tr_{\raisebox{-3pt}{$\scriptstyle #1$}}}

\theoremstyle{plain}

\begin{document}

\title{Relating the Archetypes of Logarithmic Conformal Field Theory}

\author[T Creutzig]{Thomas Creutzig}

\address[T Creutzig]{
Department of Physics and Astronomy \\
University of North Carolina \\
Phillips Hall, CB 3255 \\
Chapel Hill, NC 27599-3255 \\
USA;
Fachbereich Mathematik \\
Technische Universit\"{a}t Darmstadt\\
Schlo\ss{}gartenstra\ss{}e 7\\
64289 Darmstadt\\ Germany
}

\email{tcreutzig@mathematik.tu-darmstadt.de}

\author[D Ridout]{David Ridout}

\address[David Ridout]{
Department of Theoretical Physics \\
Research School of Physics and Engineering;
and
Mathematical Sciences Institute;
Australian National University \\
Canberra, ACT 0200 \\
Australia
}

\email{david.ridout@anu.edu.au}

\thanks{\today}

\begin{abstract}
Logarithmic conformal field theory is a rich and vibrant area of modern mathematical physics with well-known applications to both condensed matter theory and string theory.  Our limited understanding of these theories is based upon detailed studies of various examples that one may regard as archetypal.  These include the $c=-2$ triplet model, the Wess-Zumino-Witten model on $\SLG{SL}{2 ; \RR}$ at level $k=-\tfrac{1}{2}$, and its supergroup analogue on $\SLSG{GL}{1}{1}$.  Here, the latter model is studied algebraically through representation theory, fusion and modular invariance, facilitating a subsequent investigation of its cosets and extended algebras.  The results show that the archetypes of logarithmic conformal field theory are in fact all very closely related, as are many other examples including, in particular, the $\SLSG{SL}{2}{1}$ models at levels $1$ and $-\tfrac{1}{2}$.  The conclusion is then that the archetypal examples of logarithmic conformal field theory are practically all the same, so we should not expect that their features are in any way generic.  Further archetypal examples must be sought.
\end{abstract}

\maketitle

\onehalfspacing

\section{Introduction} \label{secIntro}

Conformal field theories defined in terms of non-linear sigma models on supergroups and their cosets represent an interesting and very active area of current research.  They appear in many important problems in string theory and condensed matter physics, for example the AdS/CFT correspondence \cite{Maldacena:1997re} and the theory of disordered systems \cite{Zirnbauer:1999ua,Guruswamy:1999hi}.  Questions about these problems may often be related to WZW models and their cosets through appropriate perturbations \cite{Quella:2007sg,Ashok:2009xx,Creutzig:2010hr,Candu:2011hu}.

Over the last few years, there has been significant progress in the understanding of WZW models on supergroups.  Correlation functions of WZW models on type I supergroups can be computed in principle \cite{Quella:2007hr}.  There are relations to theories with an extended superconformal symmetry algebra \cite{Hikida:2007sz,Creutzig:2011qm} and to twisted super-CFTs \cite{Creutzig:2010ne}. Furthermore, there are results on branes \cite{Creutzig:2008ag}, partial results on some specific examples \cite{Gotz:2006qp,Saleur:2006tf,Creutzig:2010zp}, and extensions to supergroup quotients \cite{Giribet:2009eb,Creutzig:2009fh}.  The best understood example is however the $\SLSG{GL}{1}{1}$ (or $\SLSG{U}{1}{1}$) WZW model.  This theory was first studied by Rozansky and Saleur twenty years ago \cite{Rozansky:1992td,RozQua92}.  Later, this model was revisited with the aim of computing correlation functions in the bulk \cite{SalGL106} and on the boundary \cite{CS09}.  Moreover, boundary states have been constructed \cite{Creutzig:2007jy} and the WZW model was shown to be an orbifold of the theory generated by a pair of symplectic fermions and two free bosons \cite{CR09}.  Nevertheless, there are still many unanswered questions, even for this simplest of supergroup CFTs.

Aside from theories based on the family $\SLSG{OSP}{1}{2m}$, supergroup WZW models represent a fundamental family of logarithmic CFTs \cite{RozQua92,GurLog93}.  The importance of such logarithmic theories to modern physics has recently gained widespread appreciation, due in part to the realisation that they describe non-local observables in statistical lattice models \cite{CarLog99,PirPre04,PeaLog06,ReaAss07,RidPer07,RidPer08,VasInd11}.  They also arise in the description of the algebraic structure of Schramm-Loewner evolution martingales \cite{KytFro08} and certain chiral gravity models \cite{GruIns08} (we refer to \cite{FloBit03} for further applications).  Aside from supergroup WZW models, fundamental classes of logarithmic CFTs include the so-called logarithmic minimal models \cite{GabInd96,GabRat96,FeiLog06,EbeVir06,GabLog06,AdaTri08,RidLog07,PeaInt08}, the fractional level WZW models \cite{GabFus01,LesLog04,AdaCon05,RidFus10} and ghost theories \cite{LesSU202,RidSL208}.  There are also indications that sigma models with non-compact target spaces may also display logarithmic behaviour \cite{FjeDua11}.  The remarkably rich mathematical structures exhibited by logarithmic CFTs are studied in \cite{RohRed96,FjeLog02,FeiMod06,FeiKaz06,HuaLog07,RidSta09,AdaStr10}.

The first aim of this article is to comprehensively review the representation theory of the affine Kac-Moody superalgebra $\AKMSA{gl}{1}{1}$ as it pertains to CFT.  We discuss general Verma modules (which coincide with Kac modules \cite{Kac:1977em,Kac:1977qb}) and determine their submodule structure completely.  The methods required are surprisingly elementary, relying on basic algebra and the induced action of the so-called spectral flow automorphisms.  This should be contrasted with \cite{KacSup87} in which this structure was deduced from a Kac-Kazhdan formula for the determinant of the Shapovalov form.  The structure of the Verma modules is then used to derive the fusion rules of the irreducible \hwms{}, thereby confirming the rules stated in \cite{Creutzig:2007jy}.  We also show that the characters of the irreducibles admit a (projective) action of the modular group $\SLG{SL}{2 ; \ZZ}$ and that the Verlinde formula correctly reproduces the projection of the fusion rules onto the Grothendieck ring of characters.

Our second aim is to investigate the $\AKMA{u}{1}$ cosets of $\AKMSA{gl}{1}{1}$ and to identify them with interesting known logarithmic conformal field theories. After carefully decomposing the representations and characters of $\AKMSA{gl}{1}{1}$, we find that one such coset can be identified as the $\beta\gamma$ ghost system of central charge $c=-1$ \cite{LesSU202,RidSL208}. We then confirm this result through an explicit free field computation using the realization of $\AKMSA{gl}{1}{1}$ used in \cite{CR09}.  This is not the only interesting logarithmic CFT one can obtain as a $\AKMSA{gl}{1}{1}/\AKMA{u}{1}$-coset.  We also find the $\mathcal{N} = 2$ super-Virasoro algebra of central charge $c=-1$ and the Bershadsky-Polyakov W-algebra $W^{(2)}_3$ at level $k=0$.

Our third aim is to investigate extended algebras for $\AKMSA{gl}{1}{1}$, the inspiration coming from the expectations of real forms and modular invariance.  In a conformal field theory, the choice of real form determines the adjoint and is reflected in the spectrum.  Experience suggests that sigma models defined on compact target spaces possess discrete spectra while non-compact spaces give rise to continuous spectra.  Moreover, one expects that the characters of the irreducible modules of the spectrum carry a representation of the modular group.  In the case of $\AKMSA{gl}{1}{1}$, this turns out to be possible only when the spectrum is continuous.  We therefore have to question how this may be reconciled with our expectation that there are choices of real forms which should lead to discrete, or at least semi-discrete, spectra.

An answer may be found by extending the symmetry algebra.  We propose that the true symmetry algebra for a certain real form, that we call the $\SLSG{U}{1}{1}$ WZW model, should in fact be taken to be significantly larger than $\AKMSA{gl}{1}{1}$.  Infinitely many characters of $\AKMSA{gl}{1}{1}$ are combined into each character of the extended algebra, leading to the possibility of a discrete set of representations whose characters transform nicely under the modular group action.

We therefore turn to a systematic investigation of a class of extended algebras for $\AKMSA{gl}{1}{1}$, finding for each positive integer $n$, a set of $n$ distinct algebras generated by four fields of dimension $\tfrac{n}{2}$.  We compute these algebras explicitly up to dimension $\tfrac{3}{2}$.  They are as follows:  The dimension $\tfrac{1}{2}$ extension is the direct sum of the $\beta \gamma$ ghost algebra and that of the complex free fermion.  The two extensions of dimension $1$ may be identified with the affine Lie superalgebra $\AKMSA{sl}{2}{1}$ at levels $k=1$ and $k=-\tfrac{1}{2}$.  The dimension $\tfrac{3}{2}$ extensions are new $W$-superalgebras, containing both the Bershadsky-Polyakov and the $\mathcal{N} = 2$ super-Virasoro algebras.  For the dimension $\tfrac{1}{2}$ extension, we identify the corresponding real form as $\AKMSA{u}{1}{1}$ and use the fusion rules of $\AKMSA{gl}{1}{1}$ to determine the characters of the extended algebra.  This allows us to construct a discrete chiral spectrum (both at the level of $\AKMSA{gl}{1}{1}$ and the extended algebra) which is closed under fusion and for which the extended algebra characters carry an action of the modular group.

Combining our extended algebra and coset studies, we obtain relations between a number of well-known logarithmic conformal field theories.
$\AKMSA{gl}{1}{1}$ is a simple current extension of its $\ZZ_2$-orbifold (or maximal bosonic subalgebra) $\mathcal{W}$, an unfamiliar $W$-algebra generated by fields of dimensions $1$, $1$, $2$, $3$, $3$ and $3$.  $\AKMA{u}{1}$-cosets of $\AKMSA{gl}{1}{1}$ and $\mathcal{W}$ lead to the $\beta\gamma$-ghosts and $\AKMA{sl}{2}_{-1/2}$, respectively.  The former is a simple current extension of the latter \cite{RidSL208}.  Further, $\AKMA{u}{1}$-cosets of these give rise \cite{RidSL210} to the $c=-2$ triplet model $\func{\mathcal{W}}{1,2}$ of \cite{GabRat96} and its simple current extension, the symplectic fermions $\AKMSA{psl}{1}{1}$ \cite{KauSym00}. We summarize this in the following diagram:
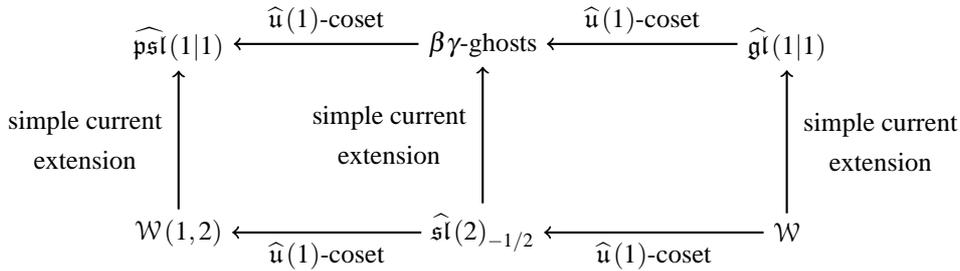
\begin{equation*}
\parbox[c]{0.9\textwidth}{
\begin{center}
\begin{tikzpicture}[auto,thick,
	nom/.style={rectangle,draw=black!20,fill=black!20,minimum height=20pt}
	]
\node (topleft) at (0,1.25) [] {$\AKMSA{psl}{1}{1}$};
\node (botleft) at (0,-1.25) [] {$\func{\mathcal{W}}{1,2}$};
\node (topmiddle) at (4,1.25) [] {$\beta\gamma$-ghosts};
\node (botmiddle) at (4,-1.25) [] {$\AKMA{sl}{2}_{-1/2}$};
\node (topright) at (8,1.25) [] {$\AKMSA{gl}{1}{1}$};
\node (botright) at (8,-1.25) [] {$\mathcal{W}$};
\draw [->] (botleft) to node {\parbox[c]{22mm}{\centering simple current extension}}(topleft);
\draw [->] (botmiddle) to node {\parbox[c]{22mm}{\centering simple current extension}}(topmiddle);
\draw [->] (botright) to node [right] {\parbox[c]{22mm}{\centering simple current extension}}(topright);
\draw [->] (topright) to node [above] {$\AKMA{u}{1}$-coset} (topmiddle);
\draw [->] (botright) to node [below] {$\AKMA{u}{1}$-coset} (botmiddle);
\draw [->] (topmiddle) to node [above] {$\AKMA{u}{1}$-coset} (topleft);
\draw [->] (botmiddle) to node [below] {$\AKMA{u}{1}$-coset} (botleft);
\end{tikzpicture}
\end{center}
} .
\end{equation*}
We wish to strongly emphasise that this diagram, coupled with the realisation of $\AKMSA{sl}{2}{1}_1$ and $\AKMSA{sl}{2}{1}_{-1/2}$ as extended algebras of $\AKMSA{gl}{1}{1}$, indicates that practically all of the best understood logarithmic CFTs are in fact all extremely closely related.  This suggests that any confidence we may have in our knowledge of the behaviour of general logarithmic theories may be unjustified.  We regard this as the most important conclusion of the work reported here.

In a sense, one has three families of relatively accessible logarithmic CFTs:  Those based on affine Lie superalgebras, the admissible level \WZW{} models and the logarithmic minimal models $\func{\mathcal{W}}{1,p}$.  The three theories present in the above diagram are among the ``smallest'' representatives of each of these families while still capturing most of the essential features.  We therefore think of them as \emph{archetypes} for a large class of logarithmic theories sharing these features.  Of course, one is used to coincidences between ``small'' examples in algebra and representation theory.  Our results therefore illustrate such a coincidence in detail, though we expect that there are further relations to be uncovered involving more sophisticated families from the same class of logarithmic CFTs.

The article is organized as follows.  In \secref{secFinite}, we begin with the finite-dimensional Lie superalgebra $\SLSA{gl}{1}{1}$, discussing its structure, real forms and representations.  \secref{secAffine} then repeats this for the affine algebra $\AKMSA{gl}{1}{1}$.  This is complemented by a thorough derivation of the fusion ring and the modular properties of the (super)characters.  In particular, we observe here that while the fusion ring has infinitely many discrete subrings, we need a continuous set of characters to obtain a representation of the modular group.

\secref{secCoset} begins with a review of the $\beta\gamma$ ghost system at $c=-1$.  We then consider a general coset of the form $\AKMSA{gl}{1}{1}/\AKMA{u}{1}$ and show how to choose the boson so as to obtain the ghost algebra.  The $\AKMSA{gl}{1}{1}$-modules are decomposed and the coset characters are identified as $\beta\gamma$ ghost modules.  From the fusion rules of $\AKMSA{gl}{1}{1}$ and this identification, we recover the $\beta \gamma$ ghost fusion rules derived in \cite{RidFus10}, thereby providing a strong consistency check of the former.  We conclude by listing some other coset algebras that are of interest and discuss briefly the implications of coset identifications for the question of discrete spectra.

In \secref{secExtAlg}, we search for extended algebras. For this we employ the well known free field realization used, for example, in \cite{CR09}.  We construct the extensions explicitly when the extension fields have dimension less than or equal to $\tfrac{3}{2}$.  In the case of the dimension $\tfrac{1}{2}$ extension, we identify the real form as compact and find a discrete set of extended algebra modules which close under fusion and whose characters close under modular transformations.  We conclude with a short outlook in \secref{secoutlook}.  Various conventions and technical identities are collected in a series of appendices.

\section{$\SLSA{gl}{1}{1}$ and its Representations} \label{secFinite}

In this section, we review the structure and representation theory of the finite-dimensional Lie superalgebra $\SLSA{gl}{1}{1}$.  Although it is not a simple Lie superalgebra, $\SLSA{gl}{1}{1}$ possesses many features in common with more general simple superalgebras.  Its advantage is that its analysis requires only elementary methods.  Further details may be found in
\cite{SalGL106,Creutzig:2009zz}.

\subsection{Algebraic Structure} \label{secFinAlg}

The endomorphisms of the super vector space $\CC^{1 \mid 1}$ have a basis consisting of the matrices
\begin{equation} \label{eqngl11DefRep}
N = \frac{1}{2} 
\begin{pmatrix}
1 & 0 \\
0 & -1
\end{pmatrix}
, \qquad E = 
\begin{pmatrix}
1 & 0 \\
0 & 1
\end{pmatrix}
, \qquad \psi^+ = 
\begin{pmatrix}
0 & 1 \\
0 & 0
\end{pmatrix}
\qquad \text{and} \qquad \psi^- = 
\begin{pmatrix}
0 & 0 \\
1 & 0
\end{pmatrix}
.
\end{equation}
Note that $N$ and $E$ are parity-preserving (bosonic) whereas $\psi^+$ and $\psi^-$ are parity-reversing (fermionic).  This endomorphism algebra admits the structure of a Lie superalgebra because
\begin{equation} \label{eqngl11Rels}
\comm{N}{\psi^{\pm}} = \pm \psi^{\pm} \qquad \text{and} \qquad \acomm{\psi^+}{\psi^-} = E,
\end{equation}
with all other brackets vanishing.  The abstract Lie superalgebra defined by these relations is called the \emph{general linear superalgebra} $\SLSA{gl}{1}{1}$.  This superalgebra is not simple because $E$ is central, and hence generates a one-dimensional ideal.  It is easy to check that this ideal has no complement and so $\SLSA{gl}{1}{1}$ does not decompose as a direct sum of ideals.  Equivalently, the adjoint representation of $\SLSA{gl}{1}{1}$ is reducible, but indecomposable.

We define a bilinear form $\killing{\cdot}{\cdot}$ on $\SLSA{gl}{1}{1}$ as the supertrace of the product in the defining representation \eqref{eqngl11DefRep}.  This bilinear form is invariant, consistent and supersymmetric.  Moreover, it is non-degenerate, unlike the form obtained by taking the supertrace in the adjoint representation.\footnote{Some authors reserve the term \emph{Killing form} for the supertrace in the adjoint representation.  To avoid confusion, we will therefore refer to $\killing{\cdot}{\cdot}$ as the \emph{invariant form}.  It is not unique, however, even up to constant multiples, because $\SLSA{gl}{1}{1}$ is not simple.}  Specifically, we have
\begin{equation}
\killing{N}{E} = \killing{E}{N} = 1 \qquad \text{and} \qquad \killing{\psi^+}{\psi^-} = -\killing{\psi^-}{\psi^+} = 1,
\end{equation}
with all other combinations vanishing.  One standard application of constructing such an invariant form is to compute the quadratic Casimir $Q \in \uealg{\SLSA{gl}{1}{1}}$.  In this case, $E$ qualifies as a Casimir element of degree $1$, so the quadratic Casimir is only defined modulo polynomials in $E$.  We choose it to be
\begin{equation} \label{eqnDefCasimir}
Q = NE + \psi^- \psi^+.
\end{equation}
We remark that a redefinition of $N$ by shifts of $E$ does not affect the defining relations \eqref{eqngl11Rels} and only changes the entry $\killing{N}{N}$.  It follows that we can choose this latter quantity arbitrarily through a simple change of basis.  We prefer, however, to keep the basis \eqref{eqngl11DefRep} so as to leave $\killing{N}{N}$ as $0$.

It remains to choose an adjoint and there are several to choose from.  While this has little bearing on the representation theory that we shall shortly turn to, the choice of adjoint reflects the real form that is relevant to geometric descriptions of the theory.  We discuss the real forms of $\SLSA{gl}{1}{1}$ in \appref{sec:real}.  Moreover, one must choose an adjoint for the triangular decomposition \eqref{eqngl11TriDecomp} below --- this plays, for example, a subtle role in the construction of extended algebras (see \secref{secExtAlg}).  The most natural choice is that suggested by the defining relations \eqref{eqngl11DefRep}:
\begin{equation}\label{eqnu11Adjoint}
N^{\dag} = N, \qquad E^{\dag} = E, \qquad \brac{\psi^{\pm}}^{\dag} = \psi^{\mp}.
\end{equation}
The corresponding real form is the unitary superalgebra $\SLSA{u}{1}{1}$.

\subsection{Representation Theory} \label{secFinRep}

There is an obvious triangular decomposition respecting the adjoint \eqref{eqnu11Adjoint}:
\begin{equation} \label{eqngl11TriDecomp}
\SLSA{gl}{1}{1} = \vectspan \set{\psi^-} \oplus \vectspan \set{N,E} \oplus \vectspan \set{\psi^+} \qquad \text{(as subalgebras)}.
\end{equation}
Thus, we regard $\psi^+$ as a raising (annihilation) operator, $\psi^-$ as a lowering (creation) operator, and $N$ and $E$ as generating the Cartan subalgebra.  
A \hws{} of a $\SLSA{gl}{1}{1}$-representation is then defined to be an eigenstate of $N$ and $E$ which is annihilated by $\psi^+$.  Such states generate Verma modules in the usual way.\footnote{Note that $\SLSA{gl}{1}{1}$ is a rather special Lie superalgebra in that its Verma modules coincide with its Kac modules \cite{Kac:1977em,Kac:1977qb}.  We will use the former nomenclature here as it is better suited to fusion studies (\secref{secAffFus}).}  If $n$ and $e$ are the eigenvalues of $N$ and $E$ (respectively) on the \hws{}, we denote the corresponding Verma module by $\VerMod{n-1/2,e}$.  Note that as $\psi^{\pm}$ squares to zero in any representation, every Verma module has dimension $2$.  The shift of $n$ by $-\tfrac{1}{2}$ in the Verma module index then amounts to characterising these modules by the \emph{average} $N$-eigenvalue.  This convention leads to a useful simplification of many of the formulae to follow.

Suppose that $\ket{v}$ is a (generating) \hws{} of $\VerMod{n,e}$.  Its $N$- and $E$-eigenvalues are then $n + \tfrac{1}{2}$ and $e$, respectively.  Moreover, it satisfies
\begin{equation}
\psi^+ \psi^- \ket{v} = \acomm{\psi^+}{\psi^-} \ket{v} = E \ket{v} = e \ket{v},
\end{equation}
so that the descendant $\psi^- \ket{v} \neq 0$ is a singular vector if and only if $e=0$.  We see then that Verma modules are irreducible when $e \neq 0$, and when $e = 0$, the irreducible quotient module has dimension $1$.  The irreducibles with $e \neq 0$ are said to be \emph{typical} and those with $e = 0$ are said to be \emph{atypical}.  We will denote a typical irreducible by $\TypMod{n,e} \cong \VerMod{n,e}$ and an atypical irreducible by $\AtypMod{n}$.  Our convention of labelling modules by their average $N$-eigenvalue leads us to define the latter to be the irreducible quotient of $\VerMod{n-1/2,0}$.  This is neatly summarised by the short exact sequence
\begin{equation} \label{ESFinV}
\dses{\AtypMod{n-1/2}}{\VerMod{n,0}}{\AtypMod{n+1/2}}
\end{equation}
and structure diagram
\begin{equation}
\parbox[c]{0.3\textwidth}{
\begin{tikzpicture}[auto,thick,
	nom/.style={circle,draw=black!20,fill=black!20,inner sep=2pt}
	]
\node (q1) at (0,0) {$\AtypMod{n+1/2}$};
\node (s1) at (3,0) {$\AtypMod{n-1/2}$};
\node at (1.5,-1) [nom] {$\VerMod{n,0}$};
\draw [->] (q1) to node {$\psi^-$} (s1);
\end{tikzpicture}
} \ .
\end{equation}
Such diagrams illustrate the decomposition of a module into its irreducible composition factors, with arrows indicating (schematically) the action of the algebra.

The representation ring of $\SLSA{gl}{1}{1}$ is defined by introducing the \emph{graded} tensor product of two representations.  This is given by the action
\begin{equation} \label{eqnDefGTP}
J \Bigl[ \ket{v} \otimes \ket{w} \Bigr] = J \ket{v} \otimes \ket{w} + \brac{-1}^{\abs{J} \abs{v}} \ket{v} \otimes J \ket{w} \qquad \text{($J = N, E, \psi^{\pm}$),}
\end{equation}
where $\abs{J}$ and $\abs{v}$ are the parities of $J$ and $\ket{v}$, respectively, understanding that linearity allows us to restrict to elements of definite parity.  The notion of parity for states $\ket{v}$ deserves some comment.  There is an obvious notion of relative parity for states, so parity makes sense once we choose that of a generator (here we restrict ourselves to indecomposable modules without loss of generality).  \emph{A priori}, this requires us to distinguish between otherwise identical modules based on whether the generator is chosen to be bosonic or fermionic.  We shall not bother to make such distinctions explicit at this point.\footnote{This becomes important when considering modular transformations (\secref{sec:modular}).  Surprisingly, this is also relevant when considering extended algebras (\secref{secDim1/2Ext}).}  This is generally permissible because it is easy to check that \eqref{eqnDefGTP} allows us to swap the parity of a module by tensoring with the atypical irreducible $\AtypMod{0}$ spanned by a single fermionic state.

We will be interested in the representations generated by the irreducible modules.  Tensor products involving the atypical irreducibles $\AtypMod{n}$ are rather trivial to analyse:
\begin{equation} \label{eqnTPL0xLe}
\AtypMod{n} \otimes \AtypMod{n'} = \AtypMod{n+n'}, \qquad \AtypMod{n} \otimes \TypMod{n',e'} = \TypMod{n+n',e'}.
\end{equation}
The corresponding result for the typical irreducibles follows from a straight-forward computation \cite{RozQua92}:
\begin{equation} \label{eqnTPLexLe}
\TypMod{n,e} \otimes \TypMod{n',e'} = \TypMod{n+n'+1/2,e+e'} \oplus \TypMod{n+n'-1/2,e+e'} \qquad \text{($e+e' \neq 0$).}
\end{equation}
The restriction that $e+e' \neq 0$, meaning that the result is again in the typical r\'{e}gime, is crucial.  When $e+e' = 0$, one observes that the result is not a direct sum, but is rather a reducible yet indecomposable module:
\begin{equation} \label{eqnDefPbar}
\TypMod{n,e} \otimes \TypMod{n',-e} = \ProjMod{n+n'}.
\end{equation}
The structure of these new indecomposable modules is given by the diagram
\begin{equation} \label{picStaggered}
\parbox[c]{0.3\textwidth}{
\begin{center}
\begin{tikzpicture}[auto,thick,
	nom/.style={circle,draw=black!20,fill=black!20,inner sep=2pt}
	]
\node (top) at (0,1.5) [] {$\AtypMod{n}$};
\node (left) at (-1.5,0) [] {$\AtypMod{n+1}$};
\node (right) at (1.5,0) [] {$\AtypMod{n-1}$};
\node (bot) at (0,-1.5) [] {$\AtypMod{n}$};
\node at (0,0) [nom] {$\ProjMod{n}$};
\draw [->] (top) to node [swap] {$\psi^+$} (left);
\draw [->] (top) to node {$\psi^-$} (right);
\draw [->] (left) to node [swap] {$\psi^-$} (bot);
\draw [->] (right) to node {$-\psi^+$} (bot);
\end{tikzpicture}
\end{center}
}
.
\end{equation}
The symbol $\ProjMod{}$ is chosen to reflect the projective nature of these modules and this is what they are generally referred to in the physical literature.\footnote{The fact that the $\ProjMod{n}$ (and the typical irreducibles $\TypMod{n,e}$) are projective in the category of finite-dimensional weight modules is often remarked upon.  We will not have any need for this property here, so we content ourselves with remarking that proofs follow easily from a mild generalisation of the standard arguments for the non-super category $\mathcal{O}$ case (see \cite{HumRep08} for example).}  We mention, however, that the structure diagram shows that the $\ProjMod{n}$ may be viewed as particularly simple examples of \emph{staggered modules} \cite{RidSta09}.  Indeed, they may be regarded as extensions of \hwms{} via the exact sequence
\begin{equation} \label{ESFinP}
\dses{\VerMod{n+1/2,0}}{\ProjMod{n}}{\VerMod{n-1/2,0}},
\end{equation}
and one can verify that the Casimir $Q$ acts non-diagonalisably on $\ProjMod{n}$, taking the generator associated with the top $\AtypMod{n}$ factor to the generator of the bottom $\AtypMod{n}$ factor and annihilating the other states.  The $\ProjMod{n}$ are not \hwms{}, but their appearance in the representation ring is not unnatural.  Indeed, we note that the adjoint representation of $\SLSA{gl}{1}{1}$ is isomorphic to $\ProjMod{0}$.

To summarise, we have seen that the set consisting of direct sums of irreducible modules does not close under tensor products in the representation ring.  Rather, one is forced to admit the reducible but indecomposable modules of the form $\ProjMod{n}$.  Combining associativity and \eqref{eqnDefPbar} shows that tensoring these projectives with the irreducibles or themselves yields no new types of indecomposables:
\begin{subequations} \label{eqnTPAss}
\begin{align}
\AtypMod{n,0} \otimes \ProjMod{n'} &= \ProjMod{n+n'}, \\
\TypMod{n,e} \otimes \ProjMod{n'} &= \TypMod{n+n'+1,e} \oplus 2 \TypMod{n+n',e} \oplus \TypMod{n+n'-1,e}, \\
\ProjMod{n} \otimes \ProjMod{n'} &= \ProjMod{n+n'+1} \oplus 2 \ProjMod{n+n'} \oplus \ProjMod{n+n'-1}.
\end{align}
\end{subequations}
There are further indecomposables which may be constructed as submodules and quotients of the $\ProjMod{n}$ (and by taking tensor products of these submodules and quotients).  We will have no use for them in the following; see \cite{GotRep07} for further discussion.

\section{$\AKMSA{gl}{1}{1}$ and its Representations} \label{secAffine}

The affine Kac-Moody superalgebra $\AKMSA{gl}{1}{1}$ is defined in the usual manner as a central extension of the loop algebra of $\SLSA{gl}{1}{1}$.  In this section, we discuss its structure and representation theory, culminating in a description of the fusion ring generated by its irreducible highest weight modules and the modular properties of the associated Grothendieck ring of characters.  This puts the results stated in \cite{RozQua92,SalGL106,Creutzig:2007jy,Creutzig:2009zz} on a solid footing.

\subsection{Algebraic Structure} \label{secAffAlg}

Our conventions for $\SLSA{gl}{1}{1}$ carry over to its affinisation $\AKMSA{gl}{1}{1}$ in the usual way.  Explicitly, the non-vanishing brackets are
\begin{equation}
\comm{N_r}{E_s} = r k \delta_{r+s,0}, \qquad \comm{N_r}{\psi^{\pm}_s} = \pm \psi^{\pm}_{r+s} \qquad \text{and} \qquad \acomm{\psi^+_r}{\psi^-_s} = E_{r+s} + r k \delta_{r+s,0},
\end{equation}
where $k \in \RR$ is called the level.  
The adjoint of the unitary superalgebra extends from \eqref{eqnu11Adjoint} to
\begin{equation} \label{eqnDefAdjoint}
N_r^{\dag} = N_{-r}, \qquad E_r^{\dag} = E_{-r}, \qquad \text{and} \qquad \brac{\psi^{\pm}_r}^{\dag} = \psi^{\mp}_{-r}.
\end{equation}
Note however, that for $k \neq 0$, we can rescale the generators so as to normalise $k$ to $1$ as follows:
\begin{equation} \label{eqnLevelScaling}
N_r \longrightarrow N_r, \qquad E_r \longrightarrow \frac{E_r}{k}, \qquad \psi^{\pm}_r \longrightarrow \frac{\psi^{\pm}_r}{\sqrt{k}}.
\end{equation}
As with $\AKMA{u}{1}$, it follows that the actual value of a non-critical $\AKMSA{gl}{1}{1}$ level $k \neq 0$ is not physical.  Nevertheless, we shall continue to give formulae for general $k$ in order to emphasise the presence throughout of the scaled combinations in \eqref{eqnLevelScaling}.  We mention that \eqref{eqnLevelScaling} with $k$ negative will introduce an imaginary scaling factor for the fermions which will change their adjoints by a sign.\footnote{Equivalently, the adjoint will be twisted by the automorphism $\omega_{-1}$; see \eqnref{eqnautomorphisms}.}

The Virasoro generators are constructed using (a modification of) the Sugawara construction.  Because the quadratic Casimir of $\SLSA{gl}{1}{1}$ is only defined modulo polynomials in $E$, one tries the ansatz \cite{RozQua92}
\begin{equation} \label{eqnDefT}
\func{T}{z} = \alpha \func{\normord{NE + EN - \psi^+ \psi^- + \psi^- \psi^+}}{z} + \beta \func{\normord{EE}}{z},
\end{equation}
where $\alpha$ and $\beta$ are constants to be determined.  This turns out to satisfy the correct \ope{} if and only if $\alpha = 1/2k$ and $\beta = 1/2k^2$.  Then, the $\AKMSA{gl}{1}{1}$ currents $\func{N}{z}$, $\func{E}{z}$ and $\func{\psi^{\pm}}{z}$ are Virasoro primaries of conformal dimension $1$, as required, and the central charge is found to vanish.  We record a convenient form for the Virasoro zero-mode for future use:
\begin{equation} \label{eqnDefL0}
L_0 = \frac{1}{k} \sum_{r \in \ZZ} \normord{N_r E_{-r} - \psi^+_r \psi^-_{-r}} - \frac{1}{2k} E_0 + \frac{1}{2k^2} \sum_{r \in \ZZ} \normord{E_r E_{-r}}.
\end{equation}

There are some interesting automorphisms of $\AKMSA{gl}{1}{1}$ which will be useful in the following.  
First, there is the automorphism $\mathsf{w}$ which defines the notion of conjugation.  Explicitly,
\begin{equation} \label{eqnDefConj}
\func{\mathsf{w}}{N_r} = -N_{r}, \quad \func{\mathsf{w}}{E_r} = -E_{r}, \quad \func{\mathsf{w}}{\psi^+_r} = \psi^-_{r}, \quad \func{\mathsf{w}}{\psi^-_r} = -\psi^+_{r} \quad \text{and} \quad \func{\mathsf{w}}{L_0} = L_0.
\end{equation}
Note that $\mathsf{w}^2$ acts as the identity on the bosons and minus the identity on the fermions.  Second, we have the \emph{spectral flow} automorphisms $\sigma^{\ell}$, $\ell \in \ZZ$, defined by \cite{SalGL106}
\begin{equation} \label{eqnSpectralFlow}
\func{\sigma^{\ell}}{N_r} = N_r, \quad \func{\sigma^{\ell}}{E_r} = E_r - \ell k \delta_{r,0}, \quad \func{\sigma^{\ell}}{\psi^{\pm}_r} = \psi^{\pm}_{r \mp \ell} \quad \text{and} \quad \func{\sigma^{\ell}}{L_0} = L_0 - \ell N_0.
\end{equation}
Both $\mathsf{w}$ and $\sigma$ may be used to construct new modules $\func{\mathsf{w}^*}{\mathcal{M}}$ and $\func{\sigma^*}{\mathcal{M}}$ from an arbitrary $\AKMSA{gl}{1}{1}$-module $\mathcal{M}$ by taking the underlying vector space to remain the same and twisting the action of the algebra:
\begin{equation} \label{eqnInducedAction}
J \cdot \tfunc{\mathsf{w}^*}{\ket{v}} = \func{\mathsf{w}^*}{\tfunc{\mathsf{w}^{-1}}{J} \ket{v}} \qquad \text{and} \qquad J \cdot \tfunc{\sigma^*}{\ket{v}} = \func{\sigma^*}{\tfunc{\sigma^{-1}}{J} \ket{v}} \qquad \text{($J \in \AKMSA{gl}{1}{1}$).}
\end{equation}
Note that $\func{\mathsf{w}^*}{\mathcal{M}}$ is precisely the module conjugate to $\mathcal{M}$.  We will generally not bother to write the distinguishing superscript ``$*$'' in what follows --- whether the automorphism acts on the algebra or one of its modules will be clear from the context.

\subsection{Representation Theory} \label{secAffRep}

We can now define \hwss{}, Verma modules $\AffVerMod{n,e}$, and their irreducible quotients as before.  By \eqnref{eqnDefL0}, the affine \hws{} $\ket{v_{n,e}}$ of $\AffVerMod{n,e}$, whose $N_0$- and $E_0$-eigenvalues are taken to be $n + \tfrac{1}{2}$ and $e$ respectively, has conformal dimension
\begin{equation} \label{eqnConfDim}
\Delta_{n,e} = n \frac{e}{k} + \frac{1}{2} \frac{e^2}{k^2}.
\end{equation}
Of course, this formula also applies to singular vectors.  We emphasise that the label $n$ refers to the average $N_0$-eigenvalue of the zero-grade subspace of $\AffVerMod{n,e}$, generalising the labelling convention of \secref{secFinRep}.  Verma modules for $\AKMSA{gl}{1}{1}$ are infinite-dimensional, but a Poincar\'{e}-Birkhoff-Witt basis is easy to write down:
\begin{equation} \label{eqnPBW}
\brac{\psi^-_{-m}}^{a_m} N_{-m}^{b_m} E_{-m}^{c_m} \brac{\psi^+_{-m}}^{d_m} \cdots \brac{\psi^-_{-2}}^{a_2} N_{-2}^{b_2} E_{-2}^{c_2} \brac{\psi^+_{-2}}^{d_2} \brac{\psi^-_{-1}}^{a_1} N_{-1}^{b_1} E_{-1}^{c_1} \brac{\psi^+_{-1}}^{d_1} \brac{\psi^-_{0}}^{a_0} \ket{v_{n,e}}.
\end{equation}
Here, the $a_i$ and $d_i$ are restricted to $0$ or $1$ (because the $\psi^{\pm}_{-m}$ square to zero), whereas the $b_i$ and $c_i$ (and $m$) may take any non-negative integer values.  It follows that the character of a Verma module is given by
\begin{equation} \label{eqnCharVerma}
\ch{\AffVerMod{n,e}}{z;q} = \traceover{\AffVerMod{n,e}} z^{N_0} q^{L_0} = z^{n+1/2} q^{\Delta_{n,e}} \prod_{i=1}^{\infty} \frac{\brac{1 + z q^i} \brac{1 + z^{-1} q^{i-1}}}{\brac{1 - q^i}^2}.
\end{equation}

Let us restrict now to the Verma modules $\AffVerMod{n,0}$.  Since $E_0$ is central, it annihilates every state in this module.  In particular, it annihilates every singular vector, hence the singular vectors of $\AffVerMod{n,0}$ have conformal dimension $0$ by \eqnref{eqnConfDim}.  This proves that the only non-trivial singular vector of $\AffVerMod{n,0}$ is $\psi^-_0 \ket{v_{n,0}}$.  Taking the quotient by the module it generates gives a module with a one-dimensional zero-grade subspace.  The only singular vector is then the \hws{}, so this quotient module is irreducible.\footnote{Equivalently, the Verma modules $\AffVerMod{n,0}$ have no (non-singular) subsingular vectors.}  We denote it by $\AffAtypMod{n+1/2,0}$ as its zero-grade subspace has $N_0$-eigenvalue $n+\tfrac{1}{2}$.  The character of this atypical irreducible is now obtained by imposing $a_0 = 0$ in \eqnref{eqnPBW}.  This leads to
\begin{equation} \label{eqnCharVac}
\ch{\AffAtypMod{n,0}}{z;q} = z^n \prod_{i=1}^{\infty} \frac{\brac{1 + z q^i} \brac{1 + z^{-1} q^i}}{\brac{1 - q^i}^2}.
\end{equation}
Identifying the vacuum as the unique \hws{} which is annihilated by all the zero-modes of $\AKMSA{gl}{1}{1}$, we obtain the vacuum character by setting $n = 0$.

We mention that the submodule of $\AffVerMod{n,0}$ generated by the singular vector $\psi^-_0 \ket{v_{n,0}}$ is not isomorphic to a Verma module because $\bigl( \psi^-_0 \bigr)^2 \ket{v_{n,0}} = 0$.  It must therefore be isomorphic to a quotient of $\AffVerMod{n-1,0}$ and, by the above argument, the only non-trivial such quotient is the irreducible $\AffAtypMod{n-1/2,0}$.  It follows that we have an exact sequence for $\AffVerMod{n,0}$:
\begin{equation} \label{SESVerma}
\dses{\AffAtypMod{n-1/2,0}}{\AffVerMod{n,0}}{\AffAtypMod{n+1/2,0}}.
\end{equation}
Splicing this with the analogous exact sequence for $\AffVerMod{n-1,0}$ and iterating, we obtain a \emph{resolution} of the atypical irreducible $\AffAtypMod{n,0}$ in terms of Verma modules:
\begin{equation}
\dres{\AffAtypMod{n,0}}{\AffVerMod{n-1/2,0}}{\AffVerMod{n-3/2,0}}{\AffVerMod{n-7/2,0} \longrightarrow \AffVerMod{n-5/2,0}}.
\end{equation}
The character of the irreducible $\AffAtypMod{n,0}$ must therefore satisfy the formal relation
\begin{equation}\label{eq:characteratypical}
\ch{\AffAtypMod{n,0}}{z;q} = \sum_{j=0}^{\infty} \brac{-1}^j \ch{\AffVerMod{n-1/2-j,0}}{z;q}.
\end{equation}
It is of course trivial to verify this from the formulae \eqref{eqnCharVerma} and \eqref{eqnCharVac}.

Let us now turn to the singular vectors of the Verma module $\AffVerMod{n,e}$ with $e \neq 0$.  Suppose that there is a singular vector of conformal dimension $\Delta_{n,e} + m$ and $N_0$-eigenvalue $n+j+\tfrac{1}{2}$.  Then, one has $-m-1 \leqslant j \leqslant m$.  In fact, it follows from $\bigl( \psi^-_n \bigr)^2 = 0$ that $j = -m-1$ can only be achieved when $m=0$ or $1$, in which case the candidate singular vector can only be $\psi^-_0 \ket{v_{n,e}}$ or $\psi^-_{-1} \psi^-_0 \ket{v_{n,e}}$.  But, it is easy to verify that these vectors are never singular.  We may therefore assume that $\abs{j} \leqslant m$.  Equating $\Delta_{n,e} + m$ with $\Delta_{n+j,e}$ now gives
\begin{equation}
m = j \frac{e}{k}.
\end{equation}
Thus, a non-trivial singular vector can only exist if $\abs{e/k} \neq 0$ is rational and at least $1$.  In particular, we can conclude that $\AffVerMod{n,e}$ is irreducible for $0 < \abs{e/k} < 1$.  Regarding these as typical irreducibles, and denoting them by $\AffTypMod{n,e}$, their characters therefore coincide with those given in \eqnref{eqnCharVerma} for Verma modules.

To study the reducibility of the Verma modules for $\abs{e/k} \geqslant 1$, we make use of the spectral flow automorphism $\sigma$, defined in \eqnref{eqnSpectralFlow}, and its induced action \eqref{eqnInducedAction} on modules.  As $\sigma$ obviously maps non-trivial submodules to non-trivial submodules, the submodule structure will be preserved.  In particular, irreducibles will be mapped to irreducibles.  Moreover, the image under spectral flow of a state $\ket{v}$ with $E_0$-eigenvalue $e$ will satisfy
\begin{equation}
E_0 \tfunc{\sigma^{\ell}}{\ket{v}} = \tfunc{\sigma^{\ell}}{\brac{E_0 + \ell k} \ket{v}} = \brac{e + \ell k} \tfunc{\sigma^{\ell}}{\ket{v}}.
\end{equation}
Along with
\begin{equation} \label{eqnSpecFlowChar}
\ch{\tfunc{\sigma^{\ell}}{\mathcal{M}}}{z;q} = \traceover{\tfunc{\sigma^{\ell}}{\mathcal{M}}} z^{N_0} q^{L_0} = \traceover{\mathcal{M}} z^{N_0} q^{L_0 + \ell N_0} = \ch{\mathcal{M}}{zq^{\ell} ; q},
\end{equation}
which holds for any $\AKMSA{gl}{1}{1}$-module $\mathcal{M}$, this will allow us to deduce the complete set of irreducible characters as well as the singular vector structure of general Verma modules.

Let us first determine what happens to the label $n$ under the (induced) action of the spectral flow.  Denoting an arbitrary irreducible \hwm{} with average $N_0$-eigenvalue $n$ and $E_0$-eigenvalue $e$ by $\AffIrrMod{n,e}$ and its \hws{} by $\ket{n,e}$, the key observation is that $\tfunc{\sigma}{\ket{n,e}}$ is only a \hws{} for $e=0$.  Indeed, the $\psi^+_{r-1}$, $N_r$, $E_r$ and $\psi^-_{r+1}$ with $r \geqslant 1$ annihilate this state, but
\begin{equation}
\psi^-_1 \tfunc{\sigma}{\ket{n,e}} = \tfunc{\sigma}{\psi^-_0 \ket{n,e}},
\end{equation}
which vanishes if and only if $e=0$.  However, it is not hard to check that for $e \neq 0$, $\psi^-_1 \tfunc{\sigma}{\ket{n,e}}$ is itself a \hws{}, hence by irreducibility, it is \emph{the} \hws{} of $\tfunc{\sigma}{\AffIrrMod{n,e}}$.  This analysis therefore identifies the result of applying $\sigma$ to all irreducibles:
\begin{equation}
\tfunc{\sigma}{\AffIrrMod{n,e}} = 
\begin{cases}
\AffIrrMod{n-1/2,e+k} & \text{if $e/k = 0, -1$,} \\
\AffIrrMod{n-1,e+k} & \text{otherwise.}
\end{cases}
\end{equation}
One can deduce the full spectral flow action from this, but we will only make the following explicit:
\begin{equation} \label{eqnIrrSpecFlow}
\tfunc{\sigma^{\ell}}{\AffIrrMod{n,e}} = \AffIrrMod{n - \ell ,e + \ell k} \quad \text{($e/k \notin \ZZ$)} \qquad \text{and} \qquad \tfunc{\sigma^{\ell}}{\AffIrrMod{n,0}} = \AffIrrMod{n - \ell + \func{\eps}{\ell} , \ell k}.
\end{equation}
Here, we introduce a convenient variant $\eps$ of the sign function on $\ZZ$:
\begin{equation} \label{eqnDefEps}
\func{\eps}{\ell} = 
\begin{cases}
+\frac{1}{2} & \text{if $\ell = +1,+2,+3,\ldots$ ,} \\
0 & \text{if $\ell = 0$,} \\
-\frac{1}{2} & \text{if $\ell = -1,-2,-3,\ldots$}
\end{cases}
\end{equation}
This function will prove very useful in \secref{secAffFus}.

Consider now the effect of spectral flow on the Verma modules $\AffVerMod{n,e}$ with $0 < \abs{e/k} < 1$.  As these modules are irreducible, it follows from \eqref{eqnIrrSpecFlow} that we already know that their images under $\sigma^{\ell}$ are the irreducible \hwms{} $\AffIrrMod{n - \ell ,e + \ell k}$.  Our question is whether these images are in fact the Verma modules $\AffVerMod{n - \ell ,e + \ell k}$.  In other words, we ask if every Verma module $\AffVerMod{n,e}$ with $e/k \notin \ZZ$ is irreducible.  This may be settled using \eqref{eqnConfDim}, \eqref{eqnCharVerma} and \eqref{eqnSpecFlowChar}:
\begin{align} \label{eqnSpecFlowVermaChar}
\ch{\tfunc{\sigma^{\ell}}{\AffVerMod{n,e}}}{z;q} &= \ch{\AffVerMod{n,e}}{zq^{\ell} ; q} = z^{n+1/2} q^{\Delta_{n,e} + \brac{n+1/2} \ell} \prod_{i=1}^{\infty} \frac{\brac{1 + z q^{i + \ell}} \brac{1 + z^{-1} q^{i - \ell - 1}}}{\brac{1 - q^i}^2} \notag \\
&= z^{n - \ell +1/2} q^{\Delta_{n,e} + \brac{n+1/2} \ell - \ell \brac{\ell + 1} / 2} \prod_{i=1}^{\infty} \frac{\brac{1 + z q^i} \brac{1 + z^{-1} q^{i-1}}}{\brac{1 - q^i}^2} = \ch{\AffVerMod{n - \ell,e + \ell k}}{z;q}.
\end{align}
It follows that $\tfunc{\sigma^{\ell}}{\AffVerMod{n,e}}$ is an irreducible \hwm{} whose character coincides with that of $\AffVerMod{n - \ell,e + \ell k}$. The image \emph{is} therefore this Verma module, proving that all Verma modules with $e/k \notin \ZZ$ are irreducible.  We will therefore regard such irreducibles as typical, denoting them by $\AffTypMod{n,e}$.

To analyse the remaining case $e/k \in \ZZ$, it is convenient to resort to somewhat more abstract methods.  Since spectral flow preserves the submodule structure, we can apply $\sigma^{\ell}$ to the exact sequence \eqref{SESVerma} to obtain another exact sequence of the form
\begin{equation}
\dses{\AffIrrMod{n-1/2 - \ell + \func{\eps}{\ell},\ell k}}
{\tfunc{\sigma^{\ell}}{\AffVerMod{n,0}}}{\AffIrrMod{n+1/2 - \ell + \func{\eps}{\ell},\ell k}}.
\end{equation}
As $\AffVerMod{n,0}$ is indecomposable, this exhibits $\tfunc{\sigma^{\ell}}{\AffVerMod{n,0}}$ (with $\ell \neq 0$) as an indecomposable with two irreducible composition factors.  Since \eqref{eqnSpecFlowVermaChar} applies, the result of the spectral flow will be the Verma module $\AffVerMod{n+1/2 - \ell + \func{\eps}{\ell} , \ell k}$ if and only if $\Delta_{n-1/2 - \ell + \func{\eps}{\ell},\ell k} \geqslant \Delta_{n+1/2 - \ell + \func{\eps}{\ell},\ell k}$, that is, if and only if $\ell \leqslant 0$.  Otherwise, we obtain an indecomposable in which the conformal dimension of the singular vector is strictly less that that of the generating state (which is no longer highest weight).  With the conformal dimension increasing as one descends, the difference in structure is as follows:
\begin{equation} \label{picVermaAndDual}
\parbox[c]{0.65\textwidth}{
\begin{center}
\begin{tikzpicture}[auto,thick,
	nom/.style={rectangle,draw=black!20,fill=black!20,minimum height=20pt}
	]
\node (topleft) at (0,1) [] {$\AffIrrMod{n - \ell,\ell k}$};
\node (botleft) at (0,-1) [] {$\AffIrrMod{n-1 - \ell,\ell k}$};
\node (topright) at (4,1) [] {$\AffIrrMod{n - \ell,\ell k}$};
\node (botright) at (4,-1) [] {$\AffIrrMod{n+1 - \ell,\ell k}$};
\node at (-2,0) [nom] {$\ell < 0 \st$};
\node at (6,0) [nom] {$\st \ell > 0$};
\draw [->] (topleft) to (botleft);
\draw [->] (botright) to (topright);
\end{tikzpicture}
\end{center}
}
.
\end{equation}
In fact, the indecomposable on the right is the (twisted) contragredient dual of the Verma module on the left. In summary,
\begin{subequations} \label{eqnSpecFlowVerma}
\begin{equation}
\tfunc{\sigma^{\ell}}{\AffVerMod{n,0}} = \AffVerMod{n - \ell,\ell k} \qquad \text{($\ell = -1,-2,-3,\ldots$).}
\end{equation}

This proves that every Verma module $\AffVerMod{n,e}$ with $e/k \in \ZZ_-$ is reducible with two irreducible composition factors.  We will therefore regard the irreducible quotients as atypical and denote them by $\AffAtypMod{n,e}$.  To obtain the same result for Verma modules with $e/k \in \ZZ_+$ (and their atypical irreducible quotients), we can repeat the above manipulations for the spectral flow of the conjugate Verma module $\tfunc{\mathsf{w}}{\AffVerMod{n,0}}$, obtaining
\begin{equation} \label{eqnSpecFlowConjVerma}
\tfunc{\sigma^{\ell}}{\tfunc{\mathsf{w}}{\AffVerMod{n,0}}} = \AffVerMod{-n - \ell,\ell k} \qquad \text{($\ell = 1,2,3,\ldots$).}
\end{equation}
\end{subequations}

To summarise, we have shown that the irreducible \hwms{} split naturally into two classes:  Atypicals $\AffAtypMod{n,e}$ with $e/k \in \ZZ$ and typicals $\AffTypMod{n,e}$ with $e/k \notin \ZZ$.  The label $n$ always refers to the average $N_0$-eigenvalue of the zero-grade states.  The Verma modules with $e/k \notin \ZZ$ are irreducible, $\AffVerMod{n,e} = \AffTypMod{n,e}$, and those with $e/k \in \ZZ$ are reducible but indecomposable with exact sequence
\begin{subequations}
\begin{align}
&\dses{\AffAtypMod{n+1,e}}{\AffVerMod{n,e}}{\AffAtypMod{n,e}} & &\text{($e/k = +1,+2,+3,\ldots$),} \\
&\dses{\AffAtypMod{n-1,e}}{\AffVerMod{n,e}}{\AffAtypMod{n,e}} & &\text{($e/k = -1,-2,-3,\ldots$).}
\end{align}
\end{subequations}
(The exact sequence for $e=0$ was given in \eqref{SESVerma}.)  It follows that the unique non-trivial singular vector of an atypical Verma module $\AffVerMod{n,e}$ appears at grade $\abs{e/k}$ and differs in $N_0$-eigenvalue from that of the (generating) \hws{} by $\pm 1$.\footnote{One can even use the spectral flow to derive, \emph{ab initio}, explicit formulae for these singular vectors.  We will not require such formulae here.  They may be found in \cite{SalGL106}.}  It also follows that there are no \emph{subsingular} vectors in the Verma modules of $\AKMSA{gl}{1}{1}$.  The characters of the typical irreducibles are given by 
\begin{subequations} \label{eqnChars}
\begin{equation} \label{eqnCharTyp}
\ch{\AffTypMod{n,e}}{z;q} = z^{n+1/2} q^{ne/k + e^2/2k^2} \prod_{i=1}^{\infty} \frac{\brac{1 + z q^i} \brac{1 + z^{-1} q^{i-1}}}{\brac{1 - q^i}^2}
\end{equation}
and the atypical irreducibles have characters given by
\begin{equation} \label{eqnCharAtyp}
\ch{\AffAtypMod{n,\ell k}}{z;q} = 
\begin{cases}
\displaystyle \frac{z^{n+1/2} q^{n \ell + \ell^2 / 2}}{1 + zq^{\ell}} \prod_{i=1}^{\infty} \frac{\brac{1 + z q^i} \brac{1 + z^{-1} q^{i-1}}}{\brac{1 - q^i}^2} & \text{if $\ell = +1,+2,+3,\ldots$ ,} \\
\displaystyle \frac{z^{n+1/2} q^{n \ell + \ell^2 / 2}}{1 + z^{-1} q^{-\ell}} \prod_{i=1}^{\infty} \frac{\brac{1 + z q^i} \brac{1 + z^{-1} q^{i-1}}}{\brac{1 - q^i}^2} & \text{if $\ell = -1,-2,-3,\ldots$}
\end{cases}
\end{equation}
\end{subequations}
For $\ell = 0$, the characters were given in \eqnref{eqnCharVac}.  The latter with $\ell \neq 0$ follow readily from this and \eqnref{eqnSpecFlowChar}.  These results were first obtained in \cite{KacSup87} using far less elementary methods.

\subsection{Fusion} \label{secAffFus}

Now that the Verma module structure has been determined, we can turn to an investigation of the fusion rules of their irreducible quotients.  As exploring the representation ring of $\SLSA{gl}{1}{1}$ leads to the introduction of non-highest weight indecomposables $\ProjMod{n}$, it is natural to expect that analogous indecomposables will appear in the fusion ring of $\AKMSA{gl}{1}{1}$.  It is often claimed that an affine superalgebra fusion ring may be computed by merely truncating the representation ring of the horizontal subalgebra by the constraints afforded by spectral flow.  While this is unlikely to be true in general, we can prove this for $\AKMSA{gl}{1}{1}$ using concepts underlying the fusion algorithm of Nahm and Gaberdiel-Kausch \cite{NahQua94,GabInd96}.

Roughly speaking, this algorithm allows one to construct a representation $\delta$ of the chiral algebra on the tensor product (over $\CC$) of the modules to be fused.  The vanishing of singular vectors then amounts to imposing relations upon this product (the \emph{spurious states} \cite{NahQua94}).  Decomposing the resulting quotient representation finally gives the fusion rule.  For $\AKMSA{gl}{1}{1}$, we will see that we only need the action of $\delta$ on the affine zero-modes:
\begin{equation} \label{eqnDeltaZeroModes}
\func{\delta}{J_0} \Bigl[ \ket{v} \otimes \ket{w} \Bigr] = J_0 \ket{v} \otimes \ket{w} + \brac{-1}^{\abs{J} \abs{v}} \ket{v} \otimes  J_0 \ket{w} \qquad \text{($J = N, E, \psi^{\pm}$).}
\end{equation}
Note that this precisely coincides with the (graded) tensor product action of $\SLSA{gl}{1}{1}$ given in \eqnref{eqnDefGTP} (explaining why the affine fusion rules are truncations of the horizontal subalgebra's tensor product rules).

We begin by discussing the fusion rules involving the atypical irreducibles $\AffAtypMod{n,0}$.  It follows from the general theory that the zero-grade subspace of $\AffAtypMod{n,0} \fuse \AffAtypMod{n',0}$ is a quotient of the tensor product of the zero-grade subspaces of $\AffAtypMod{n,0}$ and $\AffAtypMod{n',0}$.  While the corresponding Verma modules have singular vectors (at grade $0$), these do not result in spurious states because they are used to define the zero-grade subspaces.  Thus, there is no truncation and the zero-grade fusion product is one-dimensional.  Applying \eqnref{eqnDeltaZeroModes} gives the eigenvalues of $N_0$ and $E_0$ as $n+n'$ and $0$, respectively, on the fusion product.  We conclude that
\begin{equation} \label{eqnFusL0xL0}
\AffAtypMod{n,0} \fuse \AffAtypMod{n',0} = \AffAtypMod{n+n',0},
\end{equation}
because the result is manifestly a \hwm{}\footnote{We mention that for $\AKMSA{gl}{1}{1}$, the identification of a fusion product is particularly simple because we do not need to consider modules whose zero-grade subspace is trivial, meaning $\set{0}$.  Examples of this arise when a module has states whose conformal dimensions are unbounded above and below.  Modules with trivial zero-grade subspaces significantly complicate the identification of fusion products for other affine algebras and superalgebras (such issues are detailed for $\AKMA{sl}{2}$ in \cite{GabFus01,RidFus10}).} with a one-dimensional zero-grade subspace.

The fusion rules involving the typical irreducibles $\AffTypMod{n,e}$ with $e/k \notin \ZZ$ are even easier to elucidate.  Because their parent Verma modules are irreducible, there are no vanishing singular vectors, hence no spurious states.  Repeating the previous arguments, we see that the zero-mode action on the zero-grade subspace of the fusion products $\AffAtypMod{n,0} \fuse \AffTypMod{n',e}$ and $\AffTypMod{n,e} \fuse \AffTypMod{n',e'}$ is identical to the $\SLSA{gl}{1}{1}$ tensor product action.  \eqnsref{eqnTPL0xLe}{eqnDefPbar} therefore give
\begin{subequations} \label{eqnFusEasy}
\begin{align}
\AffAtypMod{n,0} \fuse \AffTypMod{n',e'} &= \AffTypMod{n+n',e'}, \label{eqnFusL0xLe} \\
\AffTypMod{n,e} \fuse \AffTypMod{n',e'} &= \AffTypMod{n+n'+1/2,e+e'} \oplus \AffTypMod{n+n'-1/2,e+e'} & &\text{($\brac{e+e'}/k \notin \ZZ$)} \label{eqnFusLexLe} \\
\text{and} \qquad \AffTypMod{n,e} \fuse \AffTypMod{n',-e} &= \AffProjMod{n+n',0}. \label{eqnDefP}
\end{align}
\end{subequations}
We have excluded $\brac{e+e'}/k = \pm 1, \pm 2, \pm 3, \ldots$ from \eqref{eqnFusLexLe} and \eqref{eqnDefP} because in these cases, the zero-grade calculations suggest atypical modules whose Verma parents have singular vectors of \emph{positive} grade.  Our calculations cannot yet detect whether these singular vectors vanish or not.  Moreover, this exclusion lets us deduce that the sum in \eqref{eqnFusLexLe} is direct because $\Delta_{n+n'+1/2,e+e'} - \Delta_{n+n'-1/2,e+e'} = \brac{e+e'}/k \notin \ZZ$.

\eqnref{eqnDefP} effectively defines the $\AKMSA{gl}{1}{1}$-module $\AffProjMod{n,0}$ whose zero-grade subspace is the indecomposable $\SLSA{gl}{1}{1}$-module $\ProjMod{n}$ described in \eqref{picStaggered} (with average $N_0$-eigenvalue $n$ and $E_0$-eigenvalue $0$).  Because $\VerMod{n+1/2,0}$ appears as a submodule of $\ProjMod{n}$, we may conclude that there is a submodule of $\AffProjMod{n,0}$ isomorphic to a quotient of $\AffVerMod{n+1/2,0}$.  As the singular vectors of the latter are all at zero-grade, inspection allows us to conclude that this submodule is in fact isomorphic to $\AffVerMod{n+1/2,0}$.  Similarly, we deduce that the quotient of $\AffProjMod{n,0}$ by this submodule is isomorphic to $\AffVerMod{n-1/2,0}$.  This gives the exact sequence
\begin{equation}
\dses{\AffVerMod{n+1/2,0}}{\AffProjMod{n,0}}{\AffVerMod{n-1/2,0}}.
\end{equation}
We remark that $L_0$ acts non-diagonalisably on $\AffProjMod{n,0}$.  In fact, $L_0$ annihilates every state except those associated with the top $\AffAtypMod{n,0}$ factor in the structure diagram
\begin{equation} \label{picAffineStaggered}
\parbox[c]{0.28\textwidth}{
\begin{center}
\begin{tikzpicture}[auto,thick,
	nom/.style={circle,draw=black!20,fill=black!20,inner sep=2pt}
	]
\node (top) at (0,1.5) [] {$\AffAtypMod{n,0}$};
\node (left) at (-1.5,0) [] {$\AffAtypMod{n+1,0}$};
\node (right) at (1.5,0) [] {$\AffAtypMod{n-1,0}$};
\node (bot) at (0,-1.5) [] {$\AffAtypMod{n,0}$};
\node at (0,0) [nom] {$\AffProjMod{n,0}$};
\draw [->] (top) to (left);
\draw [->] (top) to (right);
\draw [->] (left) to (bot);
\draw [->] (right) to (bot);
\end{tikzpicture}
\end{center}
}
.
\end{equation}
These states are instead mapped into the bottom $\AffAtypMod{n,0}$ factor.  $\AffProjMod{n,0}$ therefore qualifies as a staggered $\AKMSA{gl}{1}{1}$-module and any \cft{} in which this module appears is \emph{logarithmic}.

To investigate the fusion rules of the $\AffAtypMod{n,e}$ with $e \neq 0$, we will exploit the symmetry of the fusion rules under the induced action of the spectral flow automorphisms:\footnote{In principle, one can use the Nahm-Gaberdiel-Kausch algorithm.  However, the presence of positive-grade singular vectors in the corresponding Verma modules would result in a tedious search for spurious states.  Moreover, the zero-grade subspace cannot distinguish $\AffAtypMod{n,e}$ from $\AffVerMod{n,e}$ when $e \neq 0$, hence one would have to resort to further tedious computations in order to identify the fusion result.}
\begin{equation} \label{eqnSpecFlowFus}
\tfunc{\sigma^{\ell}}{\mathcal{M}_1} \fuse \tfunc{\sigma^{\ell'}}{\mathcal{M}_2} = \tfunc{\sigma^{\ell + \ell'}}{\mathcal{M}_1 \fuse \mathcal{M}_2}.
\end{equation}
Strictly speaking, this symmetry is only a conjectured relation for affine Kac-Moody (super)algebras, although it has been tested for $\AKMA{sl}{2}$ \cite{GabFus01,RidFus10} and is well-known for integrable modules \cite{DiFCon97}.  We have checked several fusion rules for $\AKMSA{gl}{1}{1}$ using the full Nahm-Gaberdiel-Kausch algorithm and found that \eqref{eqnSpecFlowFus} holds in every case.

Combining \eqnTref{eqnIrrSpecFlow}{eqnFusL0xL0}{eqnSpecFlowFus} then gives the fusion rules of the general atypicals $\AffAtypMod{n,e}$.  The result is best summarised by introducing the combination
\begin{equation}
\func{\eps}{\ell , \ell'} = \func{\eps}{\ell} + \func{\eps}{\ell'} - \func{\eps}{\ell + \ell'},
\end{equation}
where we recall the definition of the sign function $\func{\eps}{\ell}$ from \eqnref{eqnDefEps}.  We find that
\begin{equation} \label{eqnFusL0xL0SF}
\AffAtypMod{n,\ell k} \fuse \AffAtypMod{n',\ell' k} = \AffAtypMod{n+n'-\func{\eps}{\ell,\ell'},\brac{\ell + \ell'} k}.
\end{equation}
Similarly, applying spectral flow to \eqref{eqnFusL0xLe} yields
\begin{equation} \label{eqnFusL0xLeSF}
\AffAtypMod{n,\ell k} \fuse \AffTypMod{n',e'} = \AffTypMod{n+n'-\func{\eps}{\ell},e' + \ell k}.
\end{equation}
However, repeating this for \eqref{eqnDefP} requires us to identify the images of the $\AffProjMod{n,0}$ under spectral flow.  From \eqnref{eqnIrrSpecFlow} and the diagram \eqref{picAffineStaggered}, we see that the structure diagram of $\tfunc{\sigma^{\ell}}{\AffProjMod{n,0}}$ may be obtained from that of $\AffProjMod{n,0}$ by setting the $E_0$-eigenvalue to $\ell k$ and replacing $n$ by $n - \ell + \func{\eps}{\ell}$.  We therefore write
\begin{equation}
\tfunc{\sigma^{\ell}}{\AffProjMod{n,0}} = \AffProjMod{n - \ell + \func{\eps}{\ell},\ell k},
\end{equation}
thereby extending the definition of the indecomposables $\AffProjMod{n,e}$ to all $e/k \in \ZZ$.  With this notation, we can apply spectral flow to the fusion rules \eqref{eqnFusLexLe} and \eqref{eqnDefP}, obtaining
\begin{equation} \label{eqnFusLexLeSF}
\AffTypMod{n,e} \fuse \AffTypMod{n',e'} = 
\begin{cases}
\AffProjMod{n+n'+\func{\eps}{\brac{e+e'} / k},e+e'} & \text{if $\brac{e+e'}/k \in \ZZ$,} \\
\AffTypMod{n+n'+1/2,e+e'} \oplus \AffTypMod{n+n'-1/2,e+e'} & \text{otherwise.}
\end{cases}
\end{equation}
The $\AffProjMod{n,e}$ may also be identified as staggered modules on which $L_0$ acts non-diagonalisably.  However, this non-diagonalisable action is not apparent until grade $\abs{e/k}$.

\eqnTref{eqnFusL0xL0SF}{eqnFusL0xLeSF}{eqnFusLexLeSF} therefore give the complete collection of fusion rules for the irreducible modules of $\AKMSA{gl}{1}{1}$.  It remains to compute the fusion rules involving the indecomposables $\AffProjMod{n,e}$.  As with the tensor products \eqref{eqnTPAss}, these follow readily from associativity:
\begin{subequations} \label{eqnFusxP}
\begin{align}
\AffAtypMod{n,\ell k} \fuse \AffProjMod{n',\ell' k} &= \AffProjMod{n+n'-\func{\eps}{\ell,\ell'},\brac{\ell + \ell'} k}, \\
\AffTypMod{n,e} \fuse \AffProjMod{n',\ell' k} &= \AffTypMod{n+n'+1-\func{\eps}{\ell'},e+\ell' k} \oplus 2 \AffTypMod{n+n'-\func{\eps}{\ell'},e+\ell' k} \oplus \AffTypMod{n+n'-1-\func{\eps}{\ell'},e+\ell' k}, \\
\AffProjMod{n,\ell k} \fuse \AffProjMod{n',\ell' k} &= \AffProjMod{n+n'+1-\func{\eps}{\ell,\ell'},\brac{\ell + \ell'} k} \oplus 2 \AffProjMod{n+n'-\func{\eps}{\ell,\ell'},\brac{\ell + \ell'} k} \oplus \AffProjMod{n+n'-1-\func{\eps}{\ell,\ell'},\brac{\ell + \ell'} k}.
\end{align}
\end{subequations}

\subsection{Characters and the Modular Group} \label{sec:modular}

For rational \cfts{}, in which the spectrum consists of irreducible modules, the fusion rules may be computed from the Verlinde formula.  This relates the structure constants of the fusion operation to the modular transformation properties of the characters of the irreducible modules.  In \lcfts{}, this link between fusion and modular properties does not hold in general, essentially because the (non-trivial) indecomposable modules cannot be distinguished from a direct sum of irreducibles by the characters alone.  What one therefore expects is that the Verlinde formula will reproduce the structure constants of the \emph{Grothendieck ring of characters} (see \cite{FucNon04} for an early example of this).  This is the projection of the fusion ring onto the abelian group generated by the irreducible characters.  The ring structure is determined by the projection of the fusion operation, assuming of course that this projection gives a well-defined multiplication.\footnote{Distinguishing the ring of characters from the fusion ring may seem to be of purely mathematical interest, but it turns out to be very useful.  For example, this shift of viewpoint completely explains the long-standing issue of interpreting ``negative fusion coefficients'' in fractional level \WZW{} models \cite{RidSL208}.}  In this section, we verify that for typical $\AKMSA{gl}{1}{1}$-modules, the Verlinde formula does indeed reproduce the structure constants of the Grothendieck ring.  As the verification for the atypical modules is very similar, it is omitted.

First, we check that the Grothendieck ring is well-defined, meaning that fusion descends to a well-defined operation on the characters.  This will be so if the kernel of the projection, the linear combinations of modules whose characters cancel, forms an ideal of the fusion ring.  It is clear that this kernel is generated by the combinations of the form
\begin{equation}
\AffKerMod{n,\ell k} = \AffAtypMod{n-1,\ell k} \oplus 2 \AffAtypMod{n,\ell k} \oplus \AffAtypMod{n+1,\ell k} \ominus \AffProjMod{n,\ell k}.
\end{equation}
We therefore check from \eqnQref{eqnFusL0xL0SF}{eqnFusL0xLeSF}{eqnFusLexLeSF}{eqnFusxP} that
\begin{equation}
\AffAtypMod{n,\ell k} \fuse \AffKerMod{n',\ell' k} = \AffKerMod{n + n' - \func{\eps}{\ell,\ell'}, \brac{\ell + \ell'} k} \qquad \text{and} \qquad \AffTypMod{n,e} \fuse \AffKerMod{n',\ell' k} = \AffProjMod{n,\ell k} \fuse \AffKerMod{n',\ell' k} = 0.
\end{equation}
The kernel is therefore an ideal of the fusion ring and the Grothendieck ring of characters is isomorphic to the quotient of the fusion ring by this ideal.

We now turn to the modular properties of the irreducible characters in order to verify the Verlinde formula.
To prepare for this, we define the \emph{normalised character} of a module $\mathcal{M}$ to be (recall that $c=0$)
\begin{equation} \label{eqnDefNormChar}
\nch{\mathcal{M}}{x;y;z;q} = \traceover{\mathcal{M}} x^k y^{E_0} z^{N_0} q^{L_0}.
\end{equation}
We remark that we have deliberately broken the invariance \eqref{eqnLevelScaling} under a rescaling of the level by writing $x^k y^{E_0}$ rather than $x y^{E_0 / k}$ above.  This follows the tradition of including a generator $x$ for the level, necessary for a well-defined action of the modular group.
The normalised character of a typical irreducible then follows from \eqnref{eqnCharTyp}:
\begin{subequations}
\begin{align}
\nch{\AffTypMod{n,e}}{x;y;z;q} &= x^k y^e z^{n+1/2} q^{ne/k+e^2/2k^2} \prod_{i=1}^{\infty} \frac{\brac{1 + z q^i} \brac{1 + z^{-1} q^{i-1}}}{\brac{1 - q^i}^2} \notag \\
&= x^k y^e z^n q^{\Delta_{n,e}} \frac{\Jth{2}{z;q}}{\func{\eta}{q}^3}.
\end{align}
Instead of taking the trace in \eqref{eqnDefNormChar}, we could have taken the supertrace, thereby obtaining the supercharacter. For typicals, this is 
\begin{equation}
\sch{\AffTypMod{n,e}}{x;y;z;q} = \pm \ii x^k y^e z^n q^{\Delta_{n,e}} \frac{\Jth{1}{z;q}}{\func{\eta}{q}^3},
\end{equation}
where the choice of sign $\pm$ is correlated with whether the \hws{} is chosen to be bosonic or fermionic.
To obtain the other Jacobi theta functions, we must twist the typical modules by half-integral powers of the spectral flow automorphism $\sigma$.  The fermionic currents $\psi^{\pm}$ will then be half-integer moded, so this generalises the Neveu-Schwarz sector familiar from free fermion theories.\footnote{Indeed, this procedure amounts to considering the modular transformation properties of the \emph{bosonic subalgebra} $\mathcal{W}$ of the \uea{} of $\AKMSA{gl}{1}{1}$, generated by $N$, $E$ and $\psi^{\pm} \partial \psi^{\pm}$.  For the free fermion, this bosonic subalgebra is just the Virasoro algebra at $c=\tfrac{1}{2}$.  In this case, adding twisted sectors is equivalent to reproducing the modular properties of the Ising model.}
Defining $\TwAffTypMod{n,e} = \tfunc{\sigma^{-1/2}}{\AffTypMod{n-1/2,e+k/2}}$, and noting that $\sigma^{-1/2}$ preserves the parity of the \hws{} (unlike $\sigma^{1/2}$), we obtain the characters of the twisted typicals from a slight generalisation of \eqref{eqnSpecFlowChar}:
\begin{equation}
\nch{\TwAffTypMod{n,e}}{x;y;z;q} = \nch{\AffTypMod{n-1/2,e+k/2}}{x y^{-1/2}; y; z q^{-1/2}; q} = x^k y^e z^n q^{\Delta_{n,e}} \frac{\Jth{3}{z;q}}{\func{\eta}{q}^3}.
\end{equation}
Similarly, the corresponding supercharacters are
\begin{equation}
\sch{\TwAffTypMod{n,e}}{x;y;z;q} = \pm x^k y^e z^n q^{\Delta_{n,e}} \frac{\Jth{4}{z;q}}{\func{\eta}{q}^3}.
\end{equation}
\end{subequations}
The fusion rules for twisted modules follow from those given in \secref{secAffFus} for standard (Ramond sector) modules by applying \eqnref{eqnSpecFlowFus}.  We note that $\TwAffTypMod{n,e}$ is a \hwm{} with a one-dimensional zero-grade subspace and that its \hws{} has weight $\brac{n,e}$ and conformal dimension $\Delta_{n,e} - \tfrac{1}{8}$.

The modular transformations of the typical characters can be formally derived as follows.  First, write $x = \ee^{2 \pi \ii t}$, $y = \ee^{2 \pi \ii \nu}$, $z = \ee^{2 \pi \ii \mu}$ and $q = \ee^{2 \pi \ii \tau}$.  The modular $T$-transformation then maps $(t,\nu,\mu,\tau)$ to $(t,\nu,\mu,\tau + 1)$ and induces the following transformation of characters:\footnote{The fact that $\mathsf{T}$ acts non-diagonally on the twisted (super)characters is standard for theories with fermions.}
\begin{subequations} \label{eqnTMod}
\begin{align}
\nch{\AffTypMod{n,e}}{\func{\mathsf{T}}{t \vert \nu \vert \mu \vert \tau}} &= \ee^{2 \pi \ii \Delta_{n,e}} \nch{\AffTypMod{n,e}}{t \vert \nu \vert \mu \vert \tau}, \\
\sch{\AffTypMod{n,e}}{\func{\mathsf{T}}{t \vert \nu \vert \mu \vert \tau}} &= \ee^{2 \pi \ii \Delta_{n,e}} \sch{\AffTypMod{n,e}}{t \vert \nu \vert \mu \vert \tau}, \\
\nch{\TwAffTypMod{n,e}}{\func{\mathsf{T}}{t \vert \nu \vert \mu \vert \tau}} &= \ee^{2 \pi \ii \brac{\Delta_{n,e} - 1/8}} \sch{\TwAffTypMod{n,e}}{t \vert \nu \vert \mu \vert \tau}, \\
\sch{\TwAffTypMod{n,e}}{\func{\mathsf{T}}{t \vert \nu \vert \mu \vert \tau}} &= \ee^{2 \pi \ii \brac{\Delta_{n,e} - 1/8}} \nch{\TwAffTypMod{n,e}}{t \vert \nu \vert \mu \vert \tau}.
\end{align}
\end{subequations}
Here and below, (super)characters will correspond to typical modules with bosonic \hwss{}.  If the opposite parity is required, one merely negates the supercharacters.  It is clear that changing the parity in \eqref{eqnTMod} only changes the twisted formulae by an overall factor of $-1$ on the right hand side.

For the $S$-transformation, mapping $\brac{t, \nu, \mu, \tau}$ to $\brac{t - \nu \mu / \tau, \nu / \tau, \mu / \tau, -1 / \tau}$, we recall that
\begin{equation} \label{eqnGaussInt}
\int_{\RR} \dd x \: \ee^{-ax^2+bx} = \sqrt{\frac{\pi}{a}} \ \ee^{b^2/4a}
\end{equation}
when the real part of $a$ is positive.  We will define this integral for all $a$ through analytic continuation.  Using \eqref{eqnGaussInt} twice, we compute that
\begin{align}
\iint_{\RR^2} &\dd n \: \dd\brac{e/k} \: \nch{\AffTypMod{n,e}}{\func{\mathsf{S}}{t \vert \nu \vert \mu \vert \tau}} \ee^{2 \pi \ii \brac{ne'/k + n'e/k + ee'/k^2}} \notag \\
&= x^k \ee^{-2 \pi \ii k \nu \mu / \tau} \frac{\Jth{2}{\mu / \tau \vert -1 / \tau}}{\func{\eta}{-1 / \tau}^3} \iint_{\RR^2} \dd n \: \dd \brac{e/k} \: \ee^{2 \pi \ii \brac{\nu e + \mu n - \Delta_{n,e}} / \tau} \ee^{2 \pi \ii \brac{ne'/k + n'e/k + ee'/k^2}} \notag \\
&= x^k \frac{\ee^{\ii \pi \mu^2 / \tau - 2 \pi \ii k \mu \nu / \tau}}{-\ii \tau} \frac{\Jth{4}{\mu \vert \tau}}{\func{\eta}{\tau}^3} \int_{\RR} \dd n \: \ee^{2 \pi \ii \brac{\mu / \tau + e'/k} n} \int_{\RR} \dd \brac{e/k} \: \ee^{-\ii \pi \brac{e/k}^2 / \tau - 2 \pi \ii \brac{n / \tau - \nu k / \tau - n' - e'/k} e/k} \notag \\
&= x^k \frac{\ee^{\ii \pi \mu^2 / \tau - 2 \pi \ii k \mu \nu / \tau}}{\sqrt{-\ii \tau}} \frac{\Jth{4}{\mu \vert \tau}}{\func{\eta}{\tau}^3} \ee^{\ii \pi \tau \brac{\nu k / \tau +n' +e'/k}^2} \int_{\RR} \dd n \: \ee^{\ii \pi n^2 / \tau + 2 \pi \ii \brac{\mu / \tau - \nu k / \tau - n'} n} \notag \\
&= \omega \: x^k \ee^{2 \pi \ii \nu e'} \ee^{2 \pi \ii \mu n'} \ee^{2 \pi \ii \tau \Delta_{n',e'}} \frac{\Jth{4}{\mu \vert \tau}}{\func{\eta}{\tau}^3} \notag \\
&= \omega \: \sch{\TwAffTypMod{n',e'}}{t \vert \nu \vert \mu \vert \tau}.
\end{align}
Here, we have used the analytic continuation of \eqref{eqnGaussInt} with $a = -\ii \pi / \tau$ to evaluate the integral over $n$.  The additional square root of $-1$ which we obtain beyond the desired factor of $\sqrt{-\ii \tau}$ is denoted by $\omega$.  Which square root is obtained depends upon how the analytic continuation is performed, but the actual value of $\omega$ will turn out to be irrelevant for what follows.
Inverting this Fourier transform of the characters then gives
\begin{subequations} \label{eqnSMod}
\begin{align}
\nch{\AffTypMod{n,e}}{\func{\mathsf{S}}{t \vert \nu \vert \mu \vert \tau}} &= \omega \iint_{\RR^2} \dd n' \: \dd \brac{e'/k} \: \sch{\TwAffTypMod{n',e'}}{t \vert \nu \vert \mu \vert \tau} \ee^{-2 \pi \ii \brac{ne'/k + n'e/k + ee'/k^2}}.
\intertext{Similarly, we obtain}
\sch{\AffTypMod{n,e}}{\func{\mathsf{S}}{t \vert \nu \vert \mu \vert \tau}} &= -\ii \omega \iint_{\RR^2} \dd n' \: \dd \brac{e'/k} \: \sch{\AffTypMod{n',e'}}{t \vert \nu \vert \mu \vert \tau} \ee^{-2 \pi \ii \brac{ne'/k + n'e/k + ee'/k^2}}, \\
\nch{\TwAffTypMod{n,e}}{\func{\mathsf{S}}{t \vert \nu \vert \mu \vert \tau}} &= \omega \iint_{\RR^2} \dd n' \: \dd \brac{e'/k} \: \nch{\TwAffTypMod{n',e'}}{t \vert \nu \vert \mu \vert \tau} \ee^{-2 \pi \ii \brac{ne'/k + n'e/k + ee'/k^2}}, \\
\sch{\TwAffTypMod{n,e}}{\func{\mathsf{S}}{t \vert \nu \vert \mu \vert \tau}} &= \omega \iint_{\RR^2} \dd n' \: \dd \brac{e'/k} \: \nch{\AffTypMod{n',e'}}{t \vert \nu \vert \mu \vert \tau} \ee^{-2 \pi \ii \brac{ne'/k + n'e/k + ee'/k^2}}.
\end{align}
\end{subequations}

Regardless of the value of $\omega$, it squares to $-1$ so that $\func{\mathsf{C}}{t \vert \nu \vert \mu \vert \tau} = \brac{t \vert -\nu \vert -\mu \vert \tau}$ induces
\begin{subequations}
\begin{align}
\nch{\AffTypMod{n,e}}{t \vert -\nu \vert -\mu \vert \tau} &= +\nch{\AffTypMod{-n,-e}}{t \vert \nu \vert \mu \vert \tau}, \\
\sch{\AffTypMod{n,e}}{t \vert -\nu \vert -\mu \vert \tau} &= -\sch{\AffTypMod{-n,-e}}{t \vert \nu \vert \mu \vert \tau}, \\
\nch{\TwAffTypMod{n,e}}{t \vert -\nu \vert -\mu \vert \tau} &= +\nch{\TwAffTypMod{-n,-e}}{t \vert \nu \vert \mu \vert \tau}, \\
\sch{\TwAffTypMod{n,e}}{t \vert -\nu \vert -\mu \vert \tau} &= +\sch{\TwAffTypMod{-n,-e}}{t \vert \nu \vert \mu \vert \tau},
\end{align}
\end{subequations}
as $\jth{1}$ is odd as a function of $\mu$ whereas the other theta functions are even.  This indeed agrees with the identification of the conjugate of $\AffTypMod{n,e}$ with $\AffTypMod{-n,-e}$ --- the sign for the untwisted supercharacter reflects the fact that typical modules change parity under conjugation so that conjugate fields have the same parity.  Similarly, one can check that the signs are correct for the twisted modules, essentially because the zero-grade subspace is then only one-dimensional.

We remark that the factors of $\omega$ appearing in the $S$-transformations are required for a well-defined modular action.  However, our results show that the (un)twisted (super)characters of $\AKMSA{gl}{1}{1}$ always carry a \emph{projective} representation of $\SLG{SL}{2 ; \ZZ}$, for which questions about $\omega$ are irrelevant.  Moreover, this is sufficient for the application of the Verlinde formula as the $\tau$-dependent phases will cancel.

It moreover appears that we have to include all typical modules, or at least a set of full (Lebesgue) measure, to get closure under the modular group action.\footnote{Clearly, we should exclude $e/k \in \ZZ$ from the integration ranges in this section.}  The modular transformations of the irreducible atypical modules $\AffAtypMod{n,\ell k}$ now follow from \eqref{eq:characteratypical} and its spectral flow versions.  In particular, we find that the vacuum character and supercharacter transform under $\mathsf{S}$ as
\begin{subequations}
\begin{align}
\nch{\AffAtypMod{0,0}}{\func{\mathsf{S}}{t \vert \nu \vert \mu \vert \tau}} &= \sum_{j=0}^{\infty} \brac{-1}^j \nch{\AffVerMod{-j-1/2,0}}{\func{\mathsf{S}}{t \vert \nu \vert \mu \vert \tau}} \notag \\
&= \omega \sum_{j=0}^{\infty} \brac{-1}^j \iint_{\RR^2} \dd n \: \dd \brac{e/k} \: \sch{\TwAffTypMod{n,e}}{t \vert \nu \vert \mu \vert \tau} \ee^{2 \pi \ii \brac{j+1/2} e/k} \notag \\
&= \omega \iint_{\RR^2} \dd n \: \dd \brac{e/k} \: \frac{\sch{\TwAffTypMod{n,e}}{t \vert \nu \vert \mu \vert \tau}}{\ee^{\ii \pi e/k} + \ee^{-\ii \pi e/k}}, \\
\sch{\AffAtypMod{0,0}}{\func{\mathsf{S}}{t \vert \nu \vert \mu \vert \tau}} &= \sum_{j=0}^{\infty} \sch{\AffVerMod{-j-1/2,0}}{\func{\mathsf{S}}{t \vert \nu \vert \mu \vert \tau}} \notag \\
&= \ii \omega \iint_{\RR^2} \dd n \: \dd \brac{e/k} \: \frac{\sch{\AffTypMod{n,e}}{t \vert \nu \vert \mu \vert \tau}}{\ee^{\ii \pi e/k} - \ee^{-\ii \pi e/k}}.
\end{align}
\end{subequations}
The generalisation to the twisted vacuum module and more general atypicals is straight-forward.  These results also suggest that a set of typical modules of full measure is required for modular invariance.

We now use these transformations to verify the Verlinde formula for typical characters. As we have a continuous spectrum, this formula will involve an integral over the typical ``$S$-matrix coefficients''
\begin{subequations}
\begin{align}
\mathsf{S}_{\brac{n,e} \brac{n',e'}} &= \brac{-\ii}^{\delta} \omega \: \ee^{-2 \pi \ii \brac{ne'/k + n'e/k + ee'/k^2}} \label{eqnSCoeff}
\intertext{and the vacuum coefficient}
\mathsf{S}_{\brac{0,0} \brac{n,e}} &= \frac{\ii^{\delta'} \omega}{\ee^{\ii \pi e/k} + \brac{-1}^{\delta'} \ee^{-\ii \pi e/k}}. \label{eqnVacCoeff}
\end{align}
\end{subequations}
In \eqref{eqnSCoeff}, $\delta$ vanishes unless $\brac{n,e}$ and $\brac{n',e'}$ both correspond to untwisted supercharacters, in which case it is $1$.\footnote{In fact, these coefficients should be understood to vanish unless $\brac{n,e}$ and $\brac{n',e'}$ correspond to an untwisted character and a twisted supercharacter or both correspond to either an untwisted supercharacter or a twisted character (see \eqnref{eqnSMod}).}  In \eqref{eqnVacCoeff}, $\delta' = 1$ for the vacuum supercharacter and $0$ for its character.
The Verlinde coefficients for typical characters are then
\begin{align}
\fuscoeff{\AffTypMod{n_1,e_1}}{\AffTypMod{n_2,e_2}}{\AffTypMod{n_3,e_3}} &= \iint_{\RR^2} \dd n \: \dd \brac{e/k} \: \frac{\mathsf{S}_{\brac{n_1,e_1} \brac{n,e}} \mathsf{S}_{\brac{n_2,e_2} \brac{n,e}} \mathsf{S}_{\brac{n_3,e_3} \brac{n,e}}^*}{\mathsf{S}_{\brac{0,0} \brac{n,e}}} \notag \\
&= \iint_{\RR^2} \dd n \: \dd \brac{e/k} \: \ee^{-2 \pi \ii \bigl( n \brac{e_1+e_2-e_3} / k + e \brac{n_1+n_2-n_3} / k + e \brac{e_1+e_2-e_3} / k^2 \bigr)} \brac{\ee^{\ii \pi e/k} + \ee^{-\ii \pi e/k}} \notag \\
&= \tfunc{\delta}{\brac{e_1+e_2-e_3}/k} \Bigl[ \tfunc{\delta}{n_1+n_2-n_3+1/2} + \tfunc{\delta}{n_1+n_2-n_3-1/2} \Bigr].
\end{align}
They therefore reproduce the fusion rule \eqref{eqnFusLexLeSF}, projected onto the Grothendieck ring of characters:
\begin{align}
\NCH{\AffTypMod{n_1,e_1}} \fuse \NCH{\AffTypMod{n_2,e_2}} &= \iint_{\RR^2} \dd n_3 \: \dd \brac{e_3/k} \: \fuscoeff{\AffTypMod{n_1,e_1}}{\AffTypMod{n_2,e_2}}{\AffTypMod{n_3,e_3}} \NCH{\AffTypMod{n_3,e_3}} \notag \\
&= \NCH{\AffTypMod{n_1+n_2+1/2,e_1+e_2}} + \NCH{\AffTypMod{n_1+n_2-1/2,e_1+e_2}}.
\end{align}
Note that the factors $\omega$ cancel in the final result.  Similarly, the Verlinde coefficients for typical supercharacters are
\begin{gather}
\sfuscoeff{\AffTypMod{n_1,e_1}}{\AffTypMod{n_2,e_2}}{\AffTypMod{n_3,e_3}} = \tfunc{\delta}{\brac{e_1+e_2-e_3}/k} \Bigl[ \tfunc{\delta}{n_1+n_2-n_3+1/2} - \tfunc{\delta}{n_1+n_2-n_3-1/2} \Bigr] \\
\Rightarrow \qquad \SCH{\AffTypMod{n_1,e_1}} \fuse \SCH{\AffTypMod{n_2,e_2}} = \SCH{\AffTypMod{n_1+n_2+1/2,e_1+e_2}} - \SCH{\AffTypMod{n_1+n_2-1/2,e_1+e_2}}.
\end{gather}
This correctly reflects the parity of the components of \eqref{eqnFusLexLeSF}.

\section{Coset Theories} \label{secCoset}

As we have seen, theories with $\AKMSA{gl}{1}{1}$-symmetry have vanishing central charge.  Any coset by a $\AKMA{u}{1}$-subalgebra will therefore have $c=-1$ and the coset by the $\AKMA{u}{1} \oplus \AKMA{u}{1}$-subalgebra generated by the $N_r$ and $E_r$ will have $c=-2$.  Our claim is that one may choose the $\AKMA{u}{1}$-coset so as to obtain the well-known $\beta \gamma$ ghost system.  It will then follow from \cite{RidSL210} that the $\AKMA{u}{1} \oplus \AKMA{u}{1}$-coset will realise the theory of symplectic fermions (or one of its orbifolds \cite{KauSym00}).  In this section, we demonstrate this claim after first reviewing the $\beta \gamma$ ghost system, following \cite{RidSL208,RidSL210,RidFus10}.  This is followed by a brief discussion of some other interesting cosets that can be obtained from $\AKMSA{gl}{1}{1}$ and what this may mean for $\AKMSA{gl}{1}{1}$ spectra.

\subsection{$\beta \gamma$ Ghosts} \label{secGhost}

The $c=-1$ $\beta \gamma$ ghost system is generated by two bosonic fields whose \opes{} are
\begin{equation} \label{eqnGhostOPEs}
\func{\beta}{z} \func{\beta}{w} = 2 \func{e}{w} + \ldots, \quad \func{\beta}{z} \func{\gamma}{w} = \frac{-1}{z-w} + \func{h}{w} + \ldots \quad \text{and} \quad \func{\gamma}{z} \func{\gamma}{w} = 2 \func{f}{w} + \ldots
\end{equation}
The fields $\func{e}{w}$, $\func{h}{w}$ and $\func{f}{w}$ generate an $\AKMA{sl}{2}$-subalgebra at level $k = -\tfrac{1}{2}$.  With respect to the zero-mode $h_0$, the ghost fields $\func{\beta}{z}$ and $\func{\gamma}{w}$ have weights $1$ and $-1$, respectively.  The energy-momentum tensor of the theory is of the Sugawara form and is given compactly by
\begin{equation}
\func{t}{z} = \frac{1}{2} \normord{\func{\partial \beta}{z} \func{\gamma}{z} - \func{\beta}{z} \func{\partial \gamma}{z}}.
\end{equation}
The conformal dimension of the ghost fields is $\tfrac{1}{2}$.

This ghost system is a free field theory in the sense that its Verma modules are irreducible.  However, the underlying $\AKMA{sl}{2}$-symmetry restricts the spectrum within the highest weight category to just two modules, the vacuum module $\CosIrrMod{0}$ with highest weight $0$ and the module $\CosIrrMod{-1/2}$ with highest weight $-\tfrac{1}{2}$ (where ``weight'' refers to the eigenvalue of $h_0$).  On the former, the ghost modes carry half-integer indices, whereas on the latter the indices are integers.  Closure under conjugation and fusion forces us to extend the spectrum by the modules obtained under the induced action of the spectral flow automorphism
\begin{subequations} \label{eqnGhostSpecFlow}
\begin{equation}
\tfunc{\varsigma}{\beta_r} = \beta_{r-1/2} \qquad \text{and} \qquad \tfunc{\varsigma}{\gamma_r} = \gamma_{r+1/2}
\end{equation}
which swaps the integer and half-integer mode sectors.  Indeed, $\tfunc{\varsigma}{\CosIrrMod{0}} = \CosIrrMod{-1/2}$ and the module conjugate to $\tfunc{\varsigma^{\ell}}{\CosIrrMod{0}}$ is $\tfunc{\varsigma^{-\ell}}{\CosIrrMod{0}}$.  Extending this automorphism to the $\AKMA{sl}{2}$ zero-modes gives, in particular,
\begin{equation}
\tfunc{\varsigma}{h_0} = h_0 + \frac{1}{2} \qquad \text{and} \qquad \tfunc{\varsigma}{\ell_0} = \ell_0 - \frac{1}{2} h_0 - \frac{1}{8},
\end{equation}
\end{subequations}
where $\ell_0$ is the zero-mode of the energy-momentum tensor $\func{t}{z}$.

We therefore obtain an infinite family of irreducible ghost modules $\tfunc{\varsigma^{\ell}}{\CosIrrMod{0}}$, all related by spectral flow.  For $\abs{\ell} > 1$, the conformal dimensions of the states of these modules are not bounded below.  Enlarging the collection of irreducibles by admitting relaxed\footnote{The definition of a \emph{relaxed} \hws{} is obtained from that of a standard \hws{} by omitting the requirement that the state be annihilated by the zero-mode raising operators.  In this case, we no longer require annihilation under $\beta_0$.  A relaxed \hwm{} is then a module which is generated by a relaxed \hws{}.} \hwms{} \cite{SemEmb97} (and their images under spectral flow) leads to a continuum of families of irreducible modules $\tfunc{\varsigma^{\ell}}{\CosOthMod{\lambda}}$ with $-\tfrac{1}{2} < \lambda < \tfrac{1}{2}$.  The modules $\CosOthMod{\lambda}$ are all relaxed \hwms{} whose zero-grade states have weights belonging to $\ZZ + \lambda$ and conformal dimension $-\tfrac{1}{8}$.

The free field nature of this theory makes its characters particularly easy to determine.  For the vacuum module $\CosIrrMod{0}$, the obvious Poincar\'{e}-Birkhoff-Witt basis implies that
\begin{subequations}
\begin{equation} \label{eqnGhostVacChar}
\ch{\CosIrrMod{0}}{z;q} = \traceover{\CosIrrMod{0}} z^{h_0} q^{\ell_0} = \prod_{i=1}^{\infty} \frac{1}{\brac{1 - z^{-1} q^{i-1/2}} \brac{1 - z q^{i-1/2}}} = \sum_{m_1,m_2 = 0}^{\infty} \frac{q^{\brac{m_1 + m_2} / 2}}{\qnum{q}{m_1} \qnum{q}{m_2}} z^{m_1 - m_2},
\end{equation}
where $\qnum{q}{m} = \prod_{i=1}^m \brac{1-q^i}$ and we use the standard identity \cite[Eq.~(1.3.15)]{GasBas04}.  Iterating \eqnref{eqnGhostSpecFlow} then gives the characters under spectral flow:
\begin{equation} \label{eqnGhostVacSFChar}
\ch{\tfunc{\varsigma^{\ell}}{\CosIrrMod{0}}}{z;q} = z^{-\ell / 2} q^{-\ell^2 / 8} \ch{\CosIrrMod{0}}{zq^{\ell / 2};q} = \sum_{m_1,m_2 = 0}^{\infty} \frac{q^{\brac{m_1 + m_2} / 2 + \brac{m_1 - m_2} \ell / 2 - \ell^2 / 8}}{\qnum{q}{m_1} \qnum{q}{m_2}} z^{m_1 - m_2 - \ell / 2}.
\end{equation}
Similarly, the characters of the $\CosOthMod{\lambda}$ and their spectral flow images are
\begin{gather}
\ch{\tfunc{\varsigma^{\ell}}{\CosOthMod{\lambda}}}{z;q} = \frac{q^{\brac{\ell^2 - 1}/8}}{\qnum{q}{\infty}^2} \sum_{m \in \ZZ + \lambda - \ell / 2} z^{m} q^{\ell m / 2}. \label{eqnGhostESFChar}
\end{gather}
\end{subequations}

The fusion ring generated by these irreducible ghost modules has been computed in \cite{RidSL210}, assuming only the validity of the analogue of \eqnref{eqnSpecFlowFus} (with $\sigma$ replaced by $\varsigma$).  With this assumption, we can restrict to the fusion rules involving $\CosIrrMod{0}$ and the $\CosOthMod{\lambda}$, $-\tfrac{1}{2} < \lambda < \tfrac{1}{2}$, without loss of generality.  The ghost vacuum module $\CosIrrMod{0}$ is found to be the fusion identity, whereas
\begin{subequations}
\begin{gather}
\CosOthMod{\lambda} \fuse \CosOthMod{\mu} = 
\begin{cases}
\CosProjMod & \text{if $\lambda + \mu = 0 \pmod{1}$,} \\
\tfunc{\varsigma}{\CosOthMod{\lambda + \mu + 1/2}} \oplus \tfunc{\varsigma^{-1}}{\CosOthMod{\lambda + \mu - 1/2}} & \text{if $\lambda + \mu \neq 0 \pmod{1}$.}
\end{cases}
\label{eqnFusExE} \\
\CosOthMod{\lambda} \fuse \CosProjMod = \tfunc{\varsigma^2}{\CosOthMod{\lambda}} \oplus 2 \CosOthMod{\lambda} \oplus \tfunc{\varsigma^{-2}}{\CosOthMod{\lambda}} \qquad \text{and} \qquad \CosProjMod \fuse \CosProjMod = \tfunc{\varsigma^2}{\CosProjMod} \oplus 2 \CosProjMod \oplus \tfunc{\varsigma^{-2}}{\CosProjMod}. \label{eqnFusExS,SxS}
\end{gather}
\end{subequations}
The addition of the indices of ghost modules will always be understood to be taken \emph{modulo} $1$.  The ghost module $\CosProjMod$ appearing in these fusion rules is an indecomposable staggered module of structure
\begin{equation} \label{picGhostStaggered}
\parbox[c]{0.28\textwidth}{
\begin{center}
\begin{tikzpicture}[auto,thick,
	nom/.style={circle,draw=black!20,fill=black!20,inner sep=2pt}
	]
\node (top) at (0,1.5) [] {$\CosIrrMod{0}$};
\node (left) at (-1.5,0) [] {$\tfunc{\varsigma^2}{\CosIrrMod{0}}$};
\node (right) at (1.5,0) [] {$\tfunc{\varsigma^{-2}}{\CosIrrMod{0}}$};
\node (bot) at (0,-1.5) [] {$\CosIrrMod{0}$};
\node at (0,0) [nom] {$\CosProjMod$};
\draw [->] (top) to (left);
\draw [->] (top) to (right);
\draw [->] (left) to (bot);
\draw [->] (right) to (bot);
\end{tikzpicture}
\end{center}
}
\ .
\end{equation}

\subsection{A $\AKMA{u}{1}$-Coset} \label{secgl11/u1}

We now turn to the analysis of a coset \cft{} of the form
\begin{equation} \label{Coset}
\frac{\AKMSA{gl}{1}{1}}{\AKMA{u}{1}}.
\end{equation}
It is, however, evident that $\AKMSA{gl}{1}{1}$ has many $\AKMA{u}{1}$-subalgebras including, in particular, those generated by the fields $\func{\partial \varphi_{\zeta,\eta}}{z} = \zeta \func{N}{z} + \eta \func{E}{z}$ with $\zeta,\eta \neq 0$.  These fields satisfy the \opes{}
\begin{equation} \label{eqnBosonOPE}
\func{\partial \varphi_{\zeta,\eta}}{z} \func{\partial \varphi_{\zeta,\eta}}{w} = \frac{2 \zeta \eta k}{\brac{z-w}^2} + \text{ regular terms},
\end{equation}
hence the $\AKMA{u}{1}$ energy-momentum tensors and the corresponding conformal dimensions of the $\AKMA{u}{1}$-\hwss{} are given by
\begin{equation}
\func{\widetilde{T}}{z} = \frac{1}{4 \zeta \eta k} \normord{\func{\partial \varphi_{\zeta,\eta}}{z} \func{\partial \varphi_{\zeta,\eta}}{z}} \qquad \text{and} \qquad \widetilde{\Delta}_{n,e} = \frac{1}{4 \zeta \eta k} \brac{\zeta n + \eta e}^2.
\end{equation}

In order to obtain the $\beta \gamma$ ghosts, we will identify the correct $\func{\partial \varphi_{\zeta,\eta}}{z}$ as follows:  
First, define the \emph{extremal states} of a $\AKMSA{gl}{1}{1}$-module to be any state whose conformal dimension is minimal among those states of the module with the same $N_0$-eigenvalue.  Similarly, the extremal states of a $\beta \gamma$ ghost module are the states whose conformal dimensions are minimal among those states of the module with the same $h_0$-eigenvalue.  For example, the extremal states of the $\AKMSA{gl}{1}{1}$ and ghost vacuum modules $\AffAtypMod{0,0}$ and $\CosIrrMod{0}$ are
\begin{equation}
\begin{aligned}
\psi^+_{-r} \cdots \psi^+_{-2} \psi^+_{-1} \ket{0,0}, \\
\psi^-_{-r} \cdots \psi^-_{-2} \psi^-_{-1} \ket{0,0} \phantom{,}
\end{aligned}
\qquad \text{and} \qquad
\begin{aligned}
\beta_{-1/2}^r \ket{0}, \\
\gamma_{-1/2}^r \ket{0} \phantom{,}
\end{aligned}
\qquad \text{($r = 0, 1, 2, \ldots$),}
\end{equation}
respectively.  Now, the extremal states of a $\AKMSA{gl}{1}{1}$-module are obviously $\AKMA{u}{1}$-\hwss{} regardless of the choice of $\zeta$ and $\eta$.  Moreover, given such an extremal state, any other $\AKMA{u}{1}$-\hws{} with the same $N_0$-eigenvalue will have the same $\AKMA{u}{1}$-conformal dimension.  Hence, the extremal states give rise to the states of minimal conformal dimension in the coset theory among states of constant $N_0$-eigenvalue.  It follows that extremal states in $\AffAtypMod{0,0}$ will represent extremal states in the coset vacuum module (which we intend to be $\CosIrrMod{0}$) if we can consistently identify $N_0$-eigenvalues with $h_0$-eigenvalues.

To show that this is possible, note that there is precisely one field (up to scalar multiples) which corresponds to a $\AKMA{u}{1}$-\hws{} in $\AffAtypMod{0,0}$ with $N_0$-eigenvalue $0$ and coset conformal dimension $1$:  $\func{\partial \varphi_{\zeta,-\eta}}{z} = \zeta \func{N}{z} - \eta \func{E}{z}$.  This must therefore represent $\func{h}{z}$ in the coset theory.  Normalising by $\zeta = 1$, it follows from $e=0$ for $\AffAtypMod{0,0}$ that the eigenvalues of $N_0$ and $h_0$ may be identified on their respective vacuum modules.  To determine the constant $\eta$, note that the identity field of the $\AKMSA{gl}{1}{1}$ theory must map to the identity field in the coset theory.  Comparing
\begin{equation} \label{eqnComparingOPEs}
\func{\partial \varphi_{1,-\eta}}{z} \func{\partial \varphi_{1,-\eta}}{w} = \frac{-2 \eta k}{\brac{z-w}^2} + \ldots \qquad \text{with} \qquad \func{h}{z} \func{h}{w} = \frac{-1}{\brac{z-w}^2} + \ldots
\end{equation}
then gives $\eta = \tfrac{1}{2k}$.  The appropriate $\AKMA{u}{1}$-subalgebra for constructing the coset theory is now fixed by requiring that the $\AKMSA{gl}{1}{1}$ field $\func{\psi^+}{z}$ is represented in the coset theory by a field of $h_0$-weight $1$ and conformal dimension $\tfrac{1}{2}$ (which we will identify with $\beta$).  This $\AKMA{u}{1}$-subalgebra is therefore generated by
\begin{subequations} \label{eqnDefBosons}
\begin{equation} \label{eqnDefVarPhi}
\func{\partial \varphi}{z} \equiv \func{N}{z} + \frac{1}{2k} \func{E}{z}.
\end{equation}
We remark that this defines a euclidean boson, whereas the bosonic field
\begin{equation} \label{eqnDefPhi}
\func{\partial \phi}{z} \equiv \func{N}{z} - \frac{1}{2k} \func{E}{z}
\end{equation}
\end{subequations}
which represents $\func{h}{z}$ in the coset theory is lorentzian.  Let $\alpha_0$ be the zero-mode of $\func{\partial \varphi}{z}$.

To prove that the coset theory \eqref{Coset} admits the $\beta \gamma$ ghost algebra as its chiral algebra, we will first show that the decomposition of the $\AKMSA{gl}{1}{1}$ vacuum character into $\AKMA{u}{1}$-characters results in the $\beta \gamma$ ghost vacuum character.  Writing the $\AKMSA{gl}{1}{1}$ vacuum character \eqref{eqnCharVac} in the form
\begin{align}
\ch{\AffAtypMod{0,0}}{z;q} &= \frac{1}{\qnum{q}{\infty}^2} \sum_{m_1,m_2 = 0}^{\infty} \frac{q^{m_1 \brac{m_1 + 1} / 2 + m_2 \brac{m_2 + 1} / 2}}{\qnum{q}{m_1} \qnum{q}{m_2}} z^{m_1 - m_2} \notag \\
&= \frac{1}{\qnum{q}{\infty}} \sum_{m_1,m_2 = 0}^{\infty} \frac{q^{\brac{m_1 - m_2}^2 / 2}}{\qnum{q}{\infty}} \frac{q^{m_1 m_2 + \brac{m_1 + m_2} / 2}}{\qnum{q}{m_1} \qnum{q}{m_2}} z^{m_1 - m_2},
\end{align}
using the well-known identity \cite[Eq.~(1.3.16)]{GasBas04} and noting that the first factor inside the sum is precisely the character of an irreducible $\AKMA{u}{1}$-module with $\alpha_0$-eigenvalue $m_1 - m_2$, we see that the coset character is
\begin{equation}
\frac{1}{\qnum{q}{\infty}} \sum_{m_1,m_2 = 0}^{\infty} \frac{q^{m_1 m_2 + \brac{m_1 + m_2} / 2}}{\qnum{q}{m_1} \qnum{q}{m_2}} z^{m_1 - m_2}.
\end{equation}
This will match the ghost vacuum character \eqref{eqnGhostVacChar} if the following remarkable identity holds:
\begin{equation} \label{eqnToBeProven}
\sum_{j = \abs{m}}^{\infty} \frac{q^{j^2 - m^2}}{\qnum{q}{\infty}} \frac{q^j}{\qnum{q}{j-m} \qnum{q}{j+m}} = \sum_{j = \abs{m}}^{\infty} \frac{q^j}{\qnum{q}{j-m} \qnum{q}{j+m}} \qquad \text{($m \in \tfrac{1}{2} \ZZ$).}
\end{equation}
Here, the sums either run over integer values or half-integer values so that $j \pm m$ is always integral.  We verify \eqref{eqnToBeProven} in \appref{appIdentity}.

To complete the proof that the coset algebra of \eqref{Coset} is that of the $\beta \gamma$ ghosts, we need only remark that the chiral algebra is in this case completely determined by the coset vacuum module.  Explicitly, the zero-mode $\alpha_0 = N_0 + \tfrac{1}{2k} E_0$ of the $\AKMA{u}{1}$-subalgebra has a well-defined action (grading) on the coset theory, hence the dimension $\tfrac{1}{2}$ fields of the latter theory are constrained to satisfy
\begin{equation}
\func{\widetilde{\beta}}{z} \func{\widetilde{\beta}}{w} \sim 0, \qquad \func{\widetilde{\beta}}{z} \func{\widetilde{\gamma}}{w} \sim \frac{a}{z-w} \quad \text{and} \qquad \func{\widetilde{\gamma}}{z} \func{\widetilde{\gamma}}{w} \sim 0,
\end{equation}
for some constant $a$.  Clearly, we can scale the generators to get $a=-1$.  The statistics of $\widetilde{\beta}$ and $\widetilde{\gamma}$ are then fixed by requiring that the coset energy-momentum tensor $\widetilde{t} = T - \widetilde{T}$ has central charge $-1$.  The proof is therefore complete.

It remains to determine the decomposition of the remaining $\AKMSA{gl}{1}{1}$-modules into $\AKMA{u}{1}$-components.  We have already seen that the vacuum module $\AffAtypMod{0,0}$ decomposes into the coset vacuum module $\CosIrrMod{0}$.  It is simple to generalise the above analysis to the modules $\AffAtypMod{n,0}$, finding that
\begin{equation}
\ch{\AffAtypMod{n,0}}{z;q} = \frac{1}{\qnum{q}{\infty}} \sum_{m_1,m_2 = 0}^{\infty} \frac{q^{\brac{n + m_1 - m_2}^2}}{\qnum{q}{\infty}} \frac{q^{m_1 m_2 + \brac{m_1 + m_2} / 2 - n \brac{m_1 - m_2} - n^2 / 2}}{\qnum{q}{m_1} \qnum{q}{m_2}} z^{n + m_1 - m_2},
\end{equation}
noting only that the additional factor of $z^n$ means that we must extract the $\AKMA{u}{1}$-character with $\alpha_0$-eigenvalue $n + m_1 - m_2$.  With \eqref{eqnToBeProven}, this allows us to identify the corresponding coset character as that of $\tfunc{\varsigma^{-2n}}{\CosIrrMod{0}}$ (see \eqnref{eqnGhostVacSFChar}).

To extend this to the remaining atypical irreducibles $\AffAtypMod{n,\ell k}$, we use the spectral flow \eqref{eqnIrrSpecFlow} to write
\begin{equation} \label{eqnAtypCharDecomp}
\ch{\AffAtypMod{n,\ell k}}{z;q} = \ch{\AffAtypMod{n + \ell - \func{\eps}{\ell} , 0}}{z q^{\ell} ; q} = \frac{1}{\qnum{q}{\infty}^2} \sum_{m_1,m_2 = 0}^{\infty} \frac{q^{\brac{m_1^2 + m_1 + m_2^2 + m_2} / 2 + \ell \brac{n + \ell - \func{\eps}{\ell} + m_1 - m_2}}}{\qnum{q}{m_1} \qnum{q}{m_2}} z^{n + \ell - \func{\eps}{\ell} + m_1 - m_2},
\end{equation}
where $\func{\eps}{\ell}$ was defined in \eqnref{eqnDefEps}.  The analysis is slightly more subtle than before because the $N_0$-eigenvalue of a state in $\AffAtypMod{n,\ell k}$ differs from its $\alpha_0$-eigenvalue by $\ell / 2$.  We must therefore extract the $\AKMA{u}{1}$-character with $\alpha_0$-eigenvalue $n + \tfrac{3}{2} \ell - \func{\eps}{\ell} + m_1 - m_2$.  Proceeding then gives the coset character as
\begin{equation}
\frac{1}{\qnum{q}{\infty}} \sum_{m_1,m_2 = 0}^{\infty} \frac{q^{m_1 m_2 + \brac{m_1 + m_2} / 2 - \brac{n + \ell / 2 - \func{\eps}{\ell}} \brac{m_1 - m_2} - \brac{n + \ell / 2 - \func{\eps}{\ell}}^2 / 2}}{\qnum{q}{m_1} \qnum{q}{m_2}} z^{n + \ell - \func{\eps}{\ell} + m_1 - m_2}.
\end{equation}
But here the exponent of $z$ is still measuring the $N_0$-eigenvalue rather than the $h_0$-eigenvalue we need for the coset.  Since $h_0$ has been identified with $N_0 - \tfrac{1}{2k} E_0$ (\secref{secGhost}), the coset character must be corrected by a factor of $z^{-\ell / 2}$.  With this observation, the final result is seen to be the character of $\tfunc{\varsigma^{-2n - \ell + 2 \func{\eps}{\ell}}}{\CosIrrMod{0}}$.

The analysis of the typical irreducibles $\AffTypMod{n,e}$ ($e/k \notin \ZZ$) is similar.  We use Jacobi's triple product identity \cite[Eq.~(1.6.1)]{GasBas04} to write the character \eqref{eqnCharVerma} in the form
\begin{equation}
\ch{\AffTypMod{n,e}}{z;q} = \frac{z^{n+1/2} q^{\Delta_{n,e}}}{\qnum{q}{\infty}^3} \sum_{m \in \ZZ} q^{m \brac{m+1} / 2} z^m.
\end{equation}
Extracting the $\AKMA{u}{1}$-character with $\alpha_0$-eigenvalue $m+n+1/2+e/2k$ and correcting by the factor $z^{-e/2k}$ gives the coset character as
\begin{equation}
\frac{q^{\brac{2n + e/k + 1} \brac{2n + e/k - 1} / 8}}{\qnum{q}{\infty}^2} \sum_{m \in \ZZ + n + 1/2 - e/2k} q^{-\brac{2n + e/k} m / 2} z^m,
\end{equation}
which we identify from \eqnref{eqnGhostESFChar} as the character of $\tfunc{\varsigma^{-2n - e/k}}{\CosOthMod{1/2 - e/k}}$.

To summarise, we have decomposed each irreducible $\AKMSA{gl}{1}{1}$-module into $\AKMA{u}{1}$-modules and identified the resulting coset modules under the action of the $\beta \gamma$ ghost algebra.  Specifically, we find that the correspondence between $\AKMSA{gl}{1}{1}$-modules and ghost modules is given by
\begin{subequations} \label{eqnCosetCorrespondence}
\begin{equation} \label{eqnCosetCorrIrr}
\AffAtypMod{n,\ell k} \longrightarrow \tfunc{\varsigma^{-2n - \ell + 2 \func{\eps}{\ell}}}{\CosIrrMod{0}} \qquad \text{and} \qquad \AffTypMod{n,e} \longrightarrow \tfunc{\varsigma^{-2n - e/k}}{\CosOthMod{1/2 - e/k}}.
\end{equation}
Because our $\AKMA{u}{1}$-subalgebra acts semisimply on every $\AKMSA{gl}{1}{1}$-module that we consider, we also obtain the correspondence for the indecomposables $\AffProjMod{n,e}$ (with $e/k \in \ZZ$).  Specifically, applying \eqref{eqnCosetCorrIrr} to the structural diagram \eqref{picAffineStaggered} (with the $E_0$-eigenvalues set to $e$ rather than $0$) and comparing with the diagram \eqref{picGhostStaggered}, we deduce that
\begin{equation}
\AffProjMod{n,\ell k} \longrightarrow \tfunc{\varsigma^{-2n - \ell + 2 \func{\eps}{\ell}}}{\CosProjMod}.
\end{equation}
\end{subequations}
This follows because the structure diagram of $\CosProjMod$ and the non-diagonalisable action of the coset Virasoro mode $\ell_0$ completely determine it as a ghost module \cite{RidSL210}.  It is tedious but elementary to check that applying this correspondence to the fusion rules of $\AKMSA{gl}{1}{1}$ (given in \secref{secAffFus}) reproduces the fusion rules of the ghost algebra (given in \secref{secGhost}).

\subsection{Other Cosets} \label{secOtherCosets}

Mathematicians define a general coset algebra $\affine{\alg{g}} / \affine{\alg{h}}$ to be the commutant of the vertex algebra associated with $\affine{\alg{h}}$ in the vertex algebra associated with $\affine{\alg{g}}$.  Roughly speaking, this means that the coset fields may be identified with the fields of $\affine{\alg{g}}$ which have regular \opes{} with every field of $\affine{\alg{h}}$.  This definition of coset algebra is, however, usually far smaller than what physicists would generally regard as the coset symmetry algebra.  One is then required to extend the commutant to a physically relevant coset algebra.  For example, applying this to the coset theory determined in \secref{secgl11/u1} leads to the \emph{zero $\AKMA{u}{1}$-charge subalgebra} of the ghost algebra, generated by the product $\normord{\beta \gamma}$ of the ghost fields.  Despite this limitation, the mathematician's definition has the advantage that the coset algebra is already equipped with the algebraic structure.\footnote{This may be understood by noting that the conformal dimensions of the commutant fields do not change when passing from the parent theory to its coset.  The form of their \opes{} therefore remain invariant.  For physically relevant cosets, one needs a way of extending this to fields whose conformal dimensions do change (because the coset energy-momentum tensor is not the same as that of its parent).  This quest for an (interesting) algebraic structure underlies coset studies in general.}  By contrast, the construction of \secref{secgl11/u1} proceeded by determining the character of the coset vacuum module and then deducing the algebraic structure from this.  This is easy when the conformal dimensions of the generators are small, but non-trivial in general.

An alternative means to construct the coset symmetry algebra becomes available when one has a (suitable) free field realisation of $\affine{\alg{g}}$.  For the commutant of the induced free field realisation of $\affine{\alg{h}}$ within that of $\affine{\alg{g}}$ is usually large, yielding (free field realisations of) coset algebras that are usually satisfactory to physicists.  The cohomological issues arising with general free field realisations can cause difficulties in this approach, but in favourable circumstances, the computations proceed relatively painlessly.  This is especially so when the free field realisation is closely related to the coset being examined.

The coset discussed in \secref{secgl11/u1} falls into the category in which circumstances are favourable.  In the familiar free field realisation of $\AKMSA{gl}{1}{1}$ in terms of a pair of symplectic fermions and a pair of free bosons of opposite signature (see \appref{appFreeFields}), the field $\partial \varphi = N + E/2k$ generating the subalgebra may be identified with the euclidean boson.  At the level of free fields then, the coset theory consists of the symplectic fermions and lorentzian boson algebras.  This is of course a free field realisation of the $\beta \gamma$ ghosts \cite{LesSU202}.  Note however that the free field commutant is considerably larger than what is required to construct the ghost fields.  The symplectic fermion fields, for example, do not correspond to anything in the ghost theory in this approach.  Instead, one has to identify the ghost fields as a subalgebra of this commutant by combining the fermions with certain vertex operators for the remaining boson.  In general, this is guided by the character decompositions of the previous section.

For a less familiar example, suppose we take the coset by the subalgebra generated by the lorentzian boson $\partial \phi = N - E/2k$, so that the free field commutant combines the symplectic fermion and euclidean boson algebras.  Character methods suggest that the physically relevant subalgebra will be generated by two fields $\mathsf{g}^{\pm}$ of dimension $\tfrac{3}{2}$ and charges $\pm 1$ (under $\partial \varphi$).  Constructing such fields explicitly as $\mathsf{g}^{\pm} = \vertop{\pm \varphi} \chi^{\pm}$ (see \appref{appFreeFields} for our free field notations) leads to the identification of this subalgebra as the \emph{Bershadsky-Polyakov algebra} $W_3^{\brac{2}}$ of level $0$ and central charge $-1$.  This algebra is reviewed in \appref{appSpin3/2Alg}.

It is clear that there is an infinite number of coset theories that one can construct based on $\AKMSA{gl}{1}{1}$, many of which we may recognise.  In addition to those already discussed, the computations of \secref{secExtAlg} suggest several more.  We summarise a selection of these in \tabref{tabCosets}, omitting proofs and decomposition formulae for brevity.  It is remarkable that $\AKMSA{gl}{1}{1}$ is connected to so many other interesting theories in this way.
{
\begin{table}[h]\label{table:cosets}
\begin{center}
\setlength{\extrarowheight}{4pt}
\begin{tabular}{|C|C|C|}
\hline
\affine{\alg{h}} & \AKMA{u}{1} \text{-current(s)} & \text{Symmetry Algebra of } \AKMSA{gl}{1}{1} / \affine{\alg{h}} \\
\hline\hline
\AKMA{u}{1} & N+E/2k & \beta \gamma \text{ Ghosts} \\
\hline
\AKMA{u}{1} & N-E/2k &  W_3^{\brac{2}} \text{ at level zero} \\
\hline
\AKMA{u}{1} \oplus \AKMA{u}{1} & N \pm E/2k & \AKMSA{psl}{1}{1} \\
\hline
\AKMA{u}{1} & 3N/2+E/k & \alg{sVir}^{\brac{2}} \text{ at central charge } -1 \\
\hline 
\end{tabular}
\vspace{3mm}
\caption{Symmetry algebras of some $\AKMSA{gl}{1}{1}$ cosets by a subalgebra $\affine{\alg{h}}$.  $\AKMSA{psl}{1}{1}$ is the symplectic fermion algebra, $W_3^{\brac{2}}$ is the Bershadsky-Polyakov algebra and $\alg{sVir}^{\brac{2}}$ is the $\mathcal{N} = 2$ superconformal algebra (see \appref{appSpin3/2Alg}).} \label{tabCosets}
\end{center}
\end{table} 
}		

\subsection{Cosets and $\AKMSA{gl}{1}{1}$ Spectra}

The identification of the coset $\AKMSA{gl}{1}{1} / \AKMA{u}{1}$ with the $\beta \gamma$ ghost system suggests some non-trivial possibilities for the nature of the possible spectra of $\AKMSA{gl}{1}{1}$.  The ghost system is believed to admit an infinite number of compactified (or orbifolded) theories, of which two are of particular interest.  The first is a non-logarithmic theory built from the irreducible vacuum ghost module $\CosIrrMod{0}$ and its spectral flow images $\tfunc{\varsigma^m}{\CosIrrMod{0}}$ with $m \in \ZZ$.  These modules decompose into the set of admissible modules of Kac and Wakimoto (and their spectral flow images) for the underlying $\AKMA{sl}{2}_{-1/2}$-subalgebra \cite{RidSL210}.  Indeed, this set is the smallest set of $\AKMA{sl}{2}_{-1/2}$-modules containing the admissibles which is closed under conjugation and fusion.  The second compactification is a logarithmic theory built from the $\tfunc{\varsigma^m}{\CosIrrMod{0}}$ and the images of the relaxed highest weight ghost module $\CosOthMod{0}$ under $\varsigma^m$, again with $m \in \ZZ$.  The interest in this larger set of modules arises because, upon decomposing again into $\AKMA{sl}{2}_{-1/2}$-modules, the relaxed \hwms{} are required in order to recover the full spectrum of the $c=-2$ triplet model under a further $\AKMA{u}{1}$ coset \cite{RidSL210}.  We recall that the fusion of $\CosOthMod{0}$ with itself gives rise to the indecomposable $\CosProjMod$, whence the logarithmic nature of the theory \cite{RidFus10}.

The coset decompositions derived in \eqref{eqnCosetCorrespondence} now suggest that it \emph{must} be possible to consistently truncate the spectrum of $\AKMSA{gl}{1}{1}$ so as to obtain the interesting ghost spectra discussed above.  Working backwards, it follows readily that the non-logarithmic ghost theory allows only the atypical irreducibles $\AffAtypMod{n,\ell k}$ with $n \in \tfrac{1}{2} \ZZ$ to be present in the parent $\AKMSA{gl}{1}{1}$-theory.  The logarithmic ghost theory is somewhat more accommodating, allowing in addition the typical irreducibles $\AffTypMod{n,e}$ with $e/k \in \ZZ + \tfrac{1}{2}$ and $n \in \tfrac{1}{2} \ZZ + \tfrac{1}{4}$, and the indecomposables $\AffProjMod{n,\ell k}$ with $n \in \tfrac{1}{2} \ZZ$.  We illustrate these two spectra diagrammatically in \figref{figSpec}.  Note that they are both discrete.  Our quest to justify such discrete spectra motivates the study of extended algebras for $\AKMSA{gl}{1}{1}$, to which we now turn.

\begin{figure}
\centering
\begin{tikzpicture}[auto]
\draw [<->,gray] (-2.5,0) -- (2.5,0) node [above,black] {$\scriptstyle e$};
\draw [<->,gray] (0,-1.5) -- (0,1.5) node [right,black] {$\scriptstyle n$};
\foreach \x in {-2,-1,0,1,2}
 \foreach \y in {-1,-0.5,0,0.5,1}
  \filldraw [black] (\x,\y) circle (1.5pt);
\foreach \y in {0,0.5,1}
 \draw [black,thick] (-2.5,\y+0.25) -- (-1,\y-0.5) -- (1,\y-0.5) -- (2.5,\y-1.25);
\draw (-2.5,1.25) node [left] {$\scriptstyle \tfunc{\varsigma^{-1}}{\CosIrrMod{0}}$};
\draw (-2.5,0.75) node [left] {$\scriptstyle \CosIrrMod{0}$};
\draw (-2.5,0.25) node [left] {$\scriptstyle \tfunc{\varsigma}{\CosIrrMod{0}}$};
\end{tikzpicture}
\hspace{2cm}
\begin{tikzpicture}[auto]
\draw [<->,gray] (-2.5,0) -- (2.5,0) node [above,black] {$\scriptstyle e$};
\draw [<->,gray] (0,-1.5) -- (0,1.5) node [right,black] {$\scriptstyle n$};
\foreach \x in {-2,-1,0,1,2}
 \foreach \y in {-1,-0.5,0,0.5,1}
  \draw [gray] (\x,\y) circle (1.5pt);
\foreach \x in {-1.5,-0.5,0.5,1.5}
 \foreach \y in {-0.75,-0.25,0.25,0.75}
  \filldraw [black] (\x,\y) circle (1.5pt);
\draw [black,thick] (-1.5,1.25) -- (2.5,-0.75);
\draw [black,thick] (-2.5,1.25) -- (2.5,-1.25);
\draw [black,thick] (-2.5,0.75) -- (1.5,-1.25);
\draw (-1.5,1.25) node [above] {$\scriptstyle \tfunc{\varsigma^{-1}}{\CosOthMod{0}}$};
\draw (-2.5,1.25) node [left] {$\scriptstyle \CosOthMod{0}$};
\draw (-2.5,0.75) node [left] {$\scriptstyle \tfunc{\varsigma}{\CosOthMod{0}}$};
\end{tikzpicture}
\caption{Schematic illustrations of the possible spectra of the proposed logarithmic and non-logarithmic $\AKMSA{gl}{1}{1}$-theories.  At left is the possible spectrum of the atypicals $\AffAtypMod{n,e}$ in both (and for the logarithmic theory, the indecomposables $\AffProjMod{n,e}$ too).  At right is the possible spectrum of typicals $\AffTypMod{n,e}$ in the logarithmic theory (the atypical points are marked in white for comparison).  Modules corresponding to points on the same line yield the same coset module as indicated.} \label{figSpec}
\end{figure}

\section{Extended Algebras} \label{secExtAlg}

We show in \appref{sec:real} that there are real forms of $\AKMSA{gl}{1}{1}$ which are compact, at least in the bosonic directions.  The spectrum of the corresponding theory should therefore be discrete.  This is corroborated by  the coset considerations of the previous section --- both the $\beta \gamma$ ghosts and symplectic fermions are believed to admit orbifolds/compactifications resulting in discrete spectra \cite{KauSym00,RidFus10}.  On the other hand, the characters of the symmetry algebra are supposed to form a representation of the modular group and we saw in \secref{sec:modular} that this is only possible with $\AKMSA{gl}{1}{1}$ for a continuous spectrum.  Proposing a consistent discrete spectrum therefore requires changing the symmetry algebra.  In this section, we construct several extended chiral algebras for $\AKMSA{gl}{1}{1}$.  We then show that such proposals can be valid by using one of these extended algebras to deduce a discrete spectrum that is closed under fusion, yet admits a well-defined action of the modular group upon combining the characters appropriately.

\subsection{Chiral Algebra Extensions}


Our search for extended algebras is guided by the following considerations:  First, note that if we choose to extend by a zero-grade field associated to any irreducible $\AKMSA{gl}{1}{1}$-module, then we must include the rest of its zero-grade fields in the extension.  Second, the fields we extend by should be closed under conjugation.  Third, extending by fields from typical irreducibles will lead to logarithmic behaviour in the extended chiral algebra because fusing typicals with their conjugates yields staggered indecomposables of the form $\AffProjMod{n,\ell k}$.

It seems then that the most tractable extensions will involve zero-grade fields from atypical modules $\AffAtypMod{n,\ell k}$ and their conjugates $\AffAtypMod{-n,-\ell k}$.  The simplest extension we could hope for would involve a single atypical and its conjugate and have the further property that these extension fields generate no new fields at the level of the commutation relations.  This may be achieved for extension fields of integer or half-integer conformal dimension by requiring that the \opes{} of the zero-grade fields of $\AffAtypMod{n,\ell k}$ are regular.  From the fusion rule \eqref{eqnFusL0xL0SF}, we obtain
\begin{equation}
\AffAtypMod{n,\ell k} \fuse \AffAtypMod{n,\ell k} = \AffAtypMod{2n - \func{\eps}{\ell},2\ell k},
\end{equation}
from which it follows that the zero-grade fields of $\AffAtypMod{n,\ell k}$ will have regular \opes{} with one another if $2 \Delta_{n,\ell k} \leqslant \Delta_{2n - \func{\eps}{\ell},2\ell k}$, that is, if
\begin{equation}
\abs{\ell} \leqslant 2 \Delta_{n,\ell k}.
\end{equation}
Extending by fields of dimension $\Delta_{n,\ell k} = \tfrac{1}{2}$ therefore gives $\ell = \pm 1$ ($\ell = 0$ would give extension fields of dimension $0$) and extending by fields of dimension $1$ requires $\ell = \pm 1, \pm 2$ (and so on).  In each case, the $\pm$ merely accounts for our ability to exchange $\AffAtypMod{n,\ell k}$ with its conjugate.

Extending by $\AffAtypMod{n,\ell k}$ and its conjugate amounts to augmenting the affine currents of $\AKMSA{gl}{1}{1}$ by four zero-grade fields.  Two will be bosonic and two will be fermionic.  As the parity of a field should match that of its conjugate, there are just two ways of assigning these parities.  Computing the extended algebra defined by a given choice of parity is straight-forward as the atypical irreducibles define \emph{simple currents} in the fusion ring of $\AKMSA{gl}{1}{1}$.  The formalism for simple current extensions developed in \cite{RidSU206,RidMin07} may therefore be applied directly.  However, the computations for higher-dimensional extensions become rather intricate, so we shall instead use the well-known free field realisation of $\AKMSA{gl}{1}{1}$ as a subalgebra of the direct sum of the symplectic fermion algebra with those of the bosons $\varphi$ and $\phi$ defined in \eqnref{eqnDefBosons}.  This also has the advantage of choosing a parity for us (which turns out to be important as we discuss in \secref{secDim1/2Ext}).  For completeness, and to fix notation, we briefly review this free field realisation in \appref{appFreeFields}.

We remark that it is also permissible to extend by \emph{twisted} atypicals $\tfunc{\sigma^{-1/2}}{\AffAtypMod{n,e}}$ and their conjugates.  These modules have one-dimensional zero-grade subspaces, so this would lead to only two extension fields.  Such extensions are, however, less satisfying than those discussed below as the extension fields are only able to generate a proper subset of the $\AKMSA{gl}{1}{1}$ currents.  We will therefore not pursue this possibility here.

\subsection{Dimension $\tfrac{1}{2}$ Extensions} \label{secDim1/2Ext}

It follows from the above analysis that when the extension fields have conformal dimension $\tfrac{1}{2}$, there is a unique choice for the atypical module (up to conjugation) by whose fields we may extend.  The four extension fields must in fact belong to the atypicals $\AffAtypMod{0,k}$ and $\AffAtypMod{0,-k}$, hence will have weights $\brac{\pm \tfrac{1}{2} , \pm k}$.

In the free field realisation, the zero-grade fields of the atypical module $\AffAtypMod{0,k}$ are given by the vertex operators $\bc = \vertop{Y/2 + Z}$ and $\bg = -\vertop{-Y/2 + Z} \chi^-$ constructed from the symplectic fermions $\chi^{\pm}$ and the bosons $\varphi = Z + \tfrac{1}{2} Y$ and $\phi = Z - \tfrac{1}{2} Y$ (see \appref{appFreeFields}).  Those of $\AffAtypMod{0,-k}$ are $\bar{\bc} = \vertop{-Y/2 - Z}$ and \mbox{$\bb = +\vertop{Y/2 - Z} \chi^+$}.  The \opes{} which have singular terms are then
\begin{subequations} \label{eqnDim=1/2ExtAlg}
\begin{align}
\func{\bc}{z} \func{\bar{\bc}}{w} &= \frac{1}{z-w} + \func{\partial \varphi}{w} + \ldots = -\func{\bar{\bc}}{w} \func{\bc}{z} \\
\text{and} \qquad \func{\bb}{z} \func{\bg}{w} &= \frac{-1}{z-w} + \func{\partial \phi}{w} + \ldots = \func{\bg}{w} \func{\bb}{z}. \label{BetaGammaOPE}
\end{align}
\end{subequations}
These fields therefore define a free complex fermion $\brac{\bc,\bar{\bc}}$ and a $\beta \gamma$ ghost system.  Because the mixed \opes{} are regular, this extended algebra decomposes into the direct sum (as Lie superalgebras) of the chiral algebras of these theories.

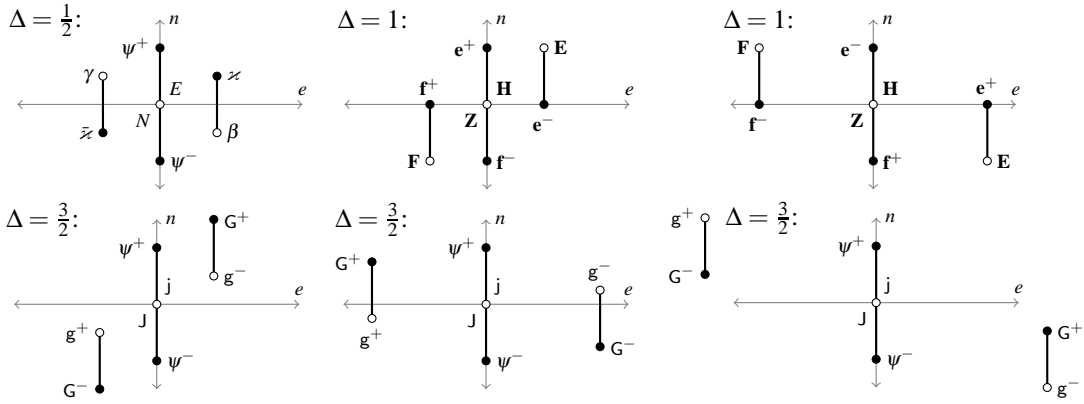
\begin{figure}
\centering
\begin{tikzpicture}[auto,scale=0.75]
\draw [<->,gray] (-2.5,0) -- (2.5,0) node [above,black] {$\scriptstyle e$};
\draw [<->,gray] (0,-1.5) -- (0,1.5) node [right,black] {$\scriptstyle n$};
\draw [black,thick] (-1,0.5) node [left] {$\scriptstyle \bg$} -- (-1,-0.5) node [left] {$\scriptstyle \bar{\bc}$};
\draw [black,thick] (1,0.5) node [right] {$\scriptstyle \bc$} -- (1,-0.5) node [right] {$\scriptstyle \bb$};
\draw [black,thick] (0,1) node [left] {$\scriptstyle \psi^+$} -- (0,-1) node [right] {$\scriptstyle \psi^-$};
\draw (0,0) node [below left] {$\scriptstyle N$};
\draw (0,0) node [above right] {$\scriptstyle E$};
\foreach \x in {(-1,-0.5),(0,1),(0,-1),(1,0.5)}
 \filldraw [black] \x circle (2pt);
\foreach \x in {(-1,0.5),(0,0),(1,-0.5)}
 \draw [black,fill=white] \x circle (2pt);
\draw (-2,1.5) node {$\Delta = \tfrac{1}{2}$:};
\end{tikzpicture}
\hspace{0mm}
\begin{tikzpicture}[auto,scale=0.75]
\draw [<->,gray] (-2.5,0) -- (2.5,0) node [above,black] {$\scriptstyle e$};
\draw [<->,gray] (0,-1.5) -- (0,1.5) node [right,black] {$\scriptstyle n$};
\draw [black,thick] (-1,0) node [above] {$\scriptstyle \Bf^+$} -- (-1,-1) node [left] {$\scriptstyle \BF$};
\draw [black,thick] (1,1) node [right] {$\scriptstyle \BE$} -- (1,0) node [below] {$\scriptstyle \Be^-$};
\draw [black,thick] (0,1) node [left] {$\scriptstyle \Be^+$} -- (0,-1) node [right] {$\scriptstyle \Bf^-$};
\draw (0,0) node [below left] {$\scriptstyle \BZ$};
\draw (0,0) node [above right] {$\scriptstyle \BH$};
\foreach \x in {(-1,0),(0,1),(0,-1),(1,0)}
 \filldraw [black] \x circle (2pt);
\foreach \x in {(-1,-1),(0,0),(1,1)}
 \draw [black,fill=white] \x circle (2pt);
\draw (-2,1.5) node {$\Delta = 1$:};
\end{tikzpicture}
\hspace{0mm}
\begin{tikzpicture}[auto,scale=0.75]
\draw [<->,gray] (-2.5,0) -- (2.5,0) node [above,black] {$\scriptstyle e$};
\draw [<->,gray] (0,-1.5) -- (0,1.5) node [right,black] {$\scriptstyle n$};
\draw [black,thick] (-2,1) node [left] {$\scriptstyle \BF$} -- (-2,0) node [below] {$\scriptstyle \Bf^-$};
\draw [black,thick] (2,0) node [above] {$\scriptstyle \Be^+$} -- (2,-1) node [right] {$\scriptstyle \BE$};
\draw [black,thick] (0,1) node [left] {$\scriptstyle \Be^-$} -- (0,-1) node [right] {$\scriptstyle \Bf^+$};
\draw (0,0) node [below left] {$\scriptstyle \BZ$};
\draw (0,0) node [above right] {$\scriptstyle \BH$};
\foreach \x in {(-2,0),(0,1),(0,-1),(2,0)}
 \filldraw [black] \x circle (2pt);
\foreach \x in {(-2,1),(0,0),(2,-1)}
 \draw [black,fill=white] \x circle (2pt);
\draw (-2,1.5) node {$\Delta = 1$:};
\draw [white,thick] (-3,0.5) node [left] {$\scriptstyle \mathsf{g}^+$} -- (-3,-0.5) node [left] {$\scriptstyle \mathsf{G}^-$}; 
\draw [white,thick] (3,-0.5) node [right] {$\scriptstyle \mathsf{G}^+$} -- (3,0.5) node [right] {$\scriptstyle \mathsf{g}^-$}; 
\end{tikzpicture}
\\
\begin{tikzpicture}[auto,scale=0.75]
\draw [<->,gray] (-2.5,0) -- (2.5,0) node [above,black] {$\scriptstyle e$};
\draw [<->,gray] (0,-1.5) -- (0,1.5) node [right,black] {$\scriptstyle n$};
\draw [black,thick] (-1,-0.5) node [left] {$\scriptstyle \mathsf{g}^+$} -- (-1,-1.5) node [left] {$\scriptstyle \mathsf{G}^-$};
\draw [black,thick] (1,1.5) node [right] {$\scriptstyle \mathsf{G}^+$} -- (1,0.5) node [right] {$\scriptstyle \mathsf{g}^-$};
\draw [black,thick] (0,1) node [left] {$\scriptstyle \psi^+$} -- (0,-1) node [right] {$\scriptstyle \psi^-$};
\draw (0,0) node [below left] {$\scriptstyle \mathsf{J}$};
\draw (0,0) node [above right] {$\scriptstyle \mathsf{j}$};
\foreach \x in {(-1,-1.5),(0,1),(0,-1),(1,1.5)}
 \filldraw [black] \x circle (2pt);
\foreach \x in {(-1,-0.5),(0,0),(1,0.5)}
 \draw [black,fill=white] \x circle (2pt);
\draw (-2,1.5) node {$\Delta = \tfrac{3}{2}$:};
\end{tikzpicture}
\hspace{0mm}
\begin{tikzpicture}[auto,scale=0.75]
\draw [<->,gray] (-2.5,0) -- (2.5,0) node [above,black] {$\scriptstyle e$};
\draw [<->,gray] (0,-1.5) -- (0,1.5) node [right,black] {$\scriptstyle n$};
\draw [black,thick] (-2,-0.25) node [below] {$\scriptstyle \mathsf{g}^+$} -- (-2,0.75) node [left] {$\scriptstyle \mathsf{G}^+$};
\draw [black,thick] (2,-0.75) node [right] {$\scriptstyle \mathsf{G}^-$} -- (2,0.25) node [above] {$\scriptstyle \mathsf{g}^-$};
\draw [black,thick] (0,1) node [left] {$\scriptstyle \psi^+$} -- (0,-1) node [right] {$\scriptstyle \psi^-$};
\draw (0,0) node [below left] {$\scriptstyle \mathsf{J}$};
\draw (0,0) node [above right] {$\scriptstyle \mathsf{j}$};
\foreach \x in {(-2,0.75),(0,1),(0,-1),(2,-0.75)}
 \filldraw [black] \x circle (2pt);
\foreach \x in {(-2,-0.25),(0,0),(2,0.25)}
 \draw [black,fill=white] \x circle (2pt);
\draw (-2,1.5) node {$\Delta = \tfrac{3}{2}$:};
\draw [white,thick] (-1,-0.5) node [left] {$\scriptstyle \mathsf{g}^+$} -- (-1,-1.5) node [left] {$\scriptstyle \mathsf{G}^-$}; 
\end{tikzpicture}
\hspace{0mm}
\begin{tikzpicture}[auto,scale=0.75]
\draw [<->,gray] (-2.5,0) -- (2.5,0) node [above,black] {$\scriptstyle e$};
\draw [<->,gray] (0,-1.5) -- (0,1.5) node [right,black] {$\scriptstyle n$};
\draw [black,thick] (-3,1.5) node [left] {$\scriptstyle \mathsf{g}^+$} -- (-3,0.5) node [left] {$\scriptstyle \mathsf{G}^-$};
\draw [black,thick] (3,-0.5) node [right] {$\scriptstyle \mathsf{G}^+$} -- (3,-1.5) node [right] {$\scriptstyle \mathsf{g}^-$};
\draw [black,thick] (0,1) node [left] {$\scriptstyle \psi^+$} -- (0,-1) node [right] {$\scriptstyle \psi^-$};
\draw (0,0) node [below left] {$\scriptstyle \mathsf{J}$};
\draw (0,0) node [above right] {$\scriptstyle \mathsf{j}$};
\foreach \x in {(-3,0.5),(0,1),(0,-1),(3,-0.5)}
 \filldraw [black] \x circle (2pt);
\foreach \x in {(-3,1.5),(0,0),(3,-1.5)}
 \draw [black,fill=white] \x circle (2pt);
\draw (-2,1.5) node {$\Delta = \tfrac{3}{2}$:};
\end{tikzpicture}
\caption{The weights $\brac{n,e}$ of the fields generating the extended chiral algebras along with the weights of $\SLSA{gl}{1}{1}$ itself (we label the weights by the corresponding fields).  White weights are bosonic and black weights are fermionic.  The first weight diagram corresponds to the direct sum of a $\beta \gamma$ ghost system with a complex fermion, the second to $\AKMSA{sl}{2}{1}_1$, the third to $\AKMSA{sl}{2}{1}_{-1/2}$ and the last three to the $W$-algebras of \secref{secDim3/2Ext}.} \label{figExtAlgWts}
\end{figure}

It follows from \eqnDref{eqnDefBosons}{eqnDim=1/2ExtAlg} that the bosonic $\AKMSA{gl}{1}{1}$ fields $N$ and $E$ are generated as linear combinations of normally-ordered products of ghosts and fermions.  It is likewise easy to check that the fermionic fields $\psi^{\pm}$ are also generated in this way:
\begin{equation} \label{eqnGhostFermionOPEs}
\func{\bb}{z} \func{\bc}{w} = +\frac{\func{\psi^+}{w}}{\sqrt{k}} + \ldots \qquad \text{and} \qquad \func{\bg}{z} \func{\bar{\bc}}{w} = -\frac{\func{\psi^-}{w}}{\sqrt{k}} + \ldots
\end{equation}
This then completes the realisation of $\AKMSA{gl}{1}{1}$ as a subalgebra of the extended algebra.\footnote{We remark that the resulting free field realisation of $\AKMSA{gl}{1}{1}$ in terms of $\beta \gamma$ ghosts and a complex fermion was actually the starting point of the analysis of \cite{KacSup87}.  The authors refer to this observation as the \emph{super boson-fermion correspondence}.}  This realisation may be used, for instance, to verify that the energy-momentum tensors of the ghosts and the fermions sum precisely to that of $\AKMSA{gl}{1}{1}$ given in \eqnref{eqnDefT}.  For convenience, we illustrate the weights of these extension fields in the first diagram of \figref{figExtAlgWts}.

We remark that the above identification allows us to specify which adjoint on $\AKMSA{gl}{1}{1}$ is consistent with the algebra extension.  Specifically, the ghost adjoint and the complex fermion adjoint are given by
\begin{equation} \label{eqnGhostFermionAdjoints}
\bb_r^{\dag} = \bg_{-r}, \quad \bg_r^{\dag} = \bb_{-r} \qquad \text{and} \qquad \bc_r^{\dag} = \bar{\bc}_{-r}, \quad \bar{\bc}_r^{\dag} = \bc_{-r}.
\end{equation}
The \opes{} \eqref{eqnDim=1/2ExtAlg} and \eqref{eqnGhostFermionOPEs} then imply the \emph{generalised} commutation relations
\begin{subequations}
\begin{gather}
\sum_{\ell = 0}^{\infty} \Bigl[ \bc_{r-\ell} \bar{\bc}_{s+\ell} - \bar{\bc}_{s-\ell-1} \bc_{r+\ell+1} \Bigr] = +\Bigl( r+\frac{1}{2} \Bigr) \delta_{r+s,0} + N_{r+s} + \frac{1}{2k} E_{r+s}, \\
\sum_{\ell = 0}^{\infty} \Bigl[ \bb_{r-\ell} \bg_{s+\ell} + \bg_{s-\ell-1} \bb_{r+\ell+1} \Bigr] = -\Bigl( r+\frac{1}{2} \Bigr) \delta_{r+s,0} + N_{r+s} - \frac{1}{2k} E_{r+s}, \\
\sum_{\ell = 0}^{\infty} \Bigl[ \bb_{r-\ell} \bc_{s+\ell} + \bc_{s-\ell-1} \bb_{r+\ell+1} \Bigr] = +\frac{\psi^+_{r+s}}{\sqrt{k}},
\qquad
\sum_{\ell = 0}^{\infty} \Bigl[ \bg_{r-\ell} \bar{\bc}_{s+\ell} + \bar{\bc}_{s-\ell-1} \bg_{r+\ell+1} \Bigr] = -\frac{\psi^-_{r+s}}{\sqrt{k}},
\end{gather}
\end{subequations}
from which \eqref{eqnGhostFermionAdjoints} gives
\begin{equation}
N_r^{\dag} = N_{-r}, \qquad E_r^{\dag} = E_{-r}, \qquad \brac{\psi^+_r}^{\dag} = -\psi^-_{-r} \qquad \text{and} \qquad \brac{\psi^-_r}^{\dag} = -\psi^+_{-r}.
\end{equation}
Up to the signs for the fermions (we assume $k>0$; the signs would be positive for $k<0$), this is the adjoint \eqref{eqnDefAdjoint} of the unitary superalgebra $\AKMSA{u}{1}{1}$.

Finally, we wish to mention a subtlety that occurs when constructing this extended algebra abstractly.  Recall that we had a parity choice for the affine primary field of weight $\brac{\tfrac{1}{2},k}$.  The free field realisation takes this field to be a fermion (and this fixes the parities of the other affine primaries).  If we had instead declared that this field is a boson, denoting it by $\bb'$, then its conjugate $\bg'$ of weight $\brac{-\tfrac{1}{2},-k}$ would also be bosonic.  One could then normalise these fields so that their \ope{} takes the form \eqref{BetaGammaOPE} and so we would again conclude that this extended algebra includes a copy of the ghost system.  However, one can check that the boson $\partial \phi' = \normord{\bb' \bg'}$ is now \emph{euclidean}, rather than lorentzian.  Indeed, further study uncovers contradictions (including failure of associativity) in the structure of this extended algebra.  We must therefore conclude that this parity choice is in fact \emph{inconsistent}!\footnote{Such failures of associativity are, however, not uncommon in abstract constructions of extended algebras.  In certain cases \cite{JacQua06,RidMin07}, associativity can be restored by introducing auxiliary operators on an \emph{ad hoc} basis.  We expect that such treatments will also work in this case.}

\subsection{Dimension $1$ Extensions} \label{secDim1Ext}

If we choose to extend by dimension $1$ fields, then there are two distinct choices:  Either we use the zero-grade fields of $\AffAtypMod{1/2,k}$ and $\AffAtypMod{-1/2,-k}$, or those of $\AffAtypMod{1/2,-2k}$ and $\AffAtypMod{-1/2,2k}$ (using both generates additional fields of dimension $0$).  Choosing the parity so that the field of weight $\brac{1,\ell k}$  ($\ell = 1,-2$) is bosonic,\footnote{We remark that if we had instead chosen the field of weight $\brac{1,\ell k}$ to be \emph{fermionic}, then the linear combination giving $\BH$ and $\BZ$ does not lead to \ope{} coefficients which match (a rescaled version of) the Killing form of $\SLSA{sl}{2}{1}$.  We have not pursued the implications of this choice further, but suspect that it leads to an inconsistent extended algebra as in \secref{secDim1/2Ext}.} we obtain the second and third weight diagrams pictured in \figref{figExtAlgWts}.  As these fields all have dimension $1$, we expect that the extended algebras will be of affine type and we recognise the weight diagrams as those of the simple Lie superalgebra $\SLSA{sl}{2}{1}$ (up to a linear transformation).  Indeed, if we define
\begin{equation}
\BH = N + \frac{E}{\ell k} \qquad \text{and} \qquad \BZ = N - \frac{E}{\ell k},
\end{equation}
then we precisely recover the weight diagram given in \figref{figSL21Wts} (we report our conventions for $\SLSA{sl}{2}{1}$ and its affinisation in \appref{appSL21}).  Moreover, this definition leads to
\begin{equation}
\func{\BH}{z} \func{\BH}{w} = \frac{2 / \ell}{\brac{z-w}^2} + \ldots, \qquad \func{\BZ}{z} \func{\BZ}{w} = -\frac{2 / \ell}{\brac{z-w}^2} + \ldots
\end{equation}
and $\func{\BH}{z} \func{\BZ}{w}$ regular, which suggests, upon comparing with the $\SLSA{sl}{2}{1}$ Killing form in \eqnref{eqnSL21Killing}, that the extended algebra will be $\AKMSA{sl}{2}{1}$ at level $1 / \ell$.

We remark immediately that the case $\ell = -2$, giving $\AKMSA{sl}{2}{1}_{-1/2}$ as the expected extended algebra, is a subalgebra of the extended algebra considered in \secref{secDim1/2Ext}.  This follows from the fusion rules
\begin{equation}
\AffAtypMod{0,k} \fuse \AffAtypMod{0,k} = \AffAtypMod{-1/2,2k} \qquad \text{and} \qquad \AffAtypMod{0,-k} \fuse \AffAtypMod{0,-k} = \AffAtypMod{1/2,-2k}.
\end{equation}
We will therefore not pursue this extended algebra, but content ourselves with noting that the $\beta \gamma$ ghosts of \secref{secDim1/2Ext} give rise to the bosonic subalgebra $\AKMA{sl}{2}_{-1/2} \subset \AKMSA{sl}{2}{1}_{-1/2}$, the complex fermion gives rise to the $\AKMA{u}{1}$-subalgebra and combining the two yields the remaining fermionic fields.

We therefore turn to verifying that the extended algebra when $\ell = 1$ is indeed $\AKMSA{sl}{2}{1}_1$.  Again, we shall accomplish this by using the free field realisation of \appref{appFreeFields}.  However, there is some subtlety to this computation.  To ensure that the $\AKMSA{sl}{2}{1}_1$ currents have the correct parities, we shall introduce an operator-valued function $\mu$ which is required to satisfy
\begin{equation}
\mu_{a,b} \mu_{c,d} = (-1)^{ad} \mu_{a+b,c+d}, \qquad \text{($a,b,c,d \in \ZZ$).}
\end{equation}
Note that the algebra generated by these operators has unit $\mu_{0,0}$.  The currents are then given by
\begin{subequations}
\begin{align}
\BE &= +\mu_{1,1} \vertop{Y+Z}, & \BH &= \partial Z + \partial Y, & \BZ &= \partial Z - \partial Y, & \BF &= -\mu_{-1,-1} \vertop{-Y-Z}, \\
\Be^+ &= -\mu_{1,0} \vertop{Y} \chi^+, & \Bf^+ &= \mu_{0,-1} \vertop{-Z} \chi^+, & \Be^- &= \mu_{0,1} \vertop{Z} \chi^-, & \Bf^- &= +\mu_{-1,0} \vertop{-Y} \chi^-,
\end{align}
\end{subequations}
and routine computation now verifies that the resulting extended algebra is indeed $\AKMSA{sl}{2}{1}_1$.

We remark that the $\mu_{a,b}$ are all that is required in order to extend the free field realisation of $\AKMSA{gl}{1}{1}$ to $\AKMSA{sl}{2}{1}_1$.  The \emph{ad hoc} addition of such a family of operators is a sign that our choice of adjoint, the super-unitary one \eqref{eqnDefAdjoint}, is not adapted to this extended algebra.  Indeed, the spectrum of $\AKMSA{sl}{2}{1}_1$ predicted in \cite{Saleur:2006tf} has both a discrete and continuous nature:  The subalgebra $\AKMA{sl}{2}_1$ seems to be compact (its quantum numbers take discrete labels), whereas the subalgebra $\AKMA{u}{1}$ generated by $\BZ$ is non-compact (continuous labels).  Of course, there are real forms of $\SLSA{gl}{1}{1}$ that have compact and non-compact directions (see \appref{sec:real}).

\subsection{Dimension $\tfrac{3}{2}$ Extensions} \label{secDim3/2Ext}

There are three distinct choices for extensions of conformal dimension $\tfrac{3}{2}$, corresponding to the zero-grade fields of the atypical irreducibles $\AffAtypMod{1,k}$, $\AffAtypMod{-1/4,2k}$ and $\AffAtypMod{-1,3k}$ (as well as their respective conjugates).  The latter choice again results in an extended algebra which is a subalgebra of that considered in \secref{secDim1/2Ext} because
\begin{equation}
\AffAtypMod{0,k} \fuse \AffAtypMod{0,k} \fuse \AffAtypMod{0,k} = \AffAtypMod{-1,3k}.
\end{equation}
We remark that the free field realisation of \appref{appFreeFields} suggests a parity choice in each case.  Specifically, we should take the extension field of weight $\brac{\tfrac{3}{2},k}$ to be \emph{fermionic}, that of weight $\brac{\tfrac{1}{4},2k}$ to be \emph{bosonic}, and that of weight $\brac{-\tfrac{1}{2},3k}$ to be \emph{fermionic}.  This results in the fourth, fifth and sixth weight diagrams pictured in \figref{figExtAlgWts} (respectively).

We start with the extended algebra for $\AffAtypMod{1,k}$.  Consider first the fermionic dimension $\tfrac{3}{2}$ fields $\mathsf{G}^{\pm} = \sqrt{\tfrac{2}{3}} \vertop{\pm \brac{3Y/2 + Z}}$.  This normalisation yields \opes{} which are regular except for
\begin{equation}
\func{\mathsf{G}^+}{z} \func{\mathsf{G}^-}{w} = \frac{2/3}{\brac{z-w}^3} + \frac{2 \func{\mathsf{J}}{w}}{\brac{z-w}^2} + \frac{3 \normord{\func{\mathsf{J}}{w} \func{\mathsf{J}}{w}} + \func{\partial \mathsf{J}}{w}}{z-w} + \ldots \ ,
\end{equation}
as follows from \eqnref{eqnVertexOPE}.  Here, $\mathsf{J} = \tfrac{1}{2} \partial Y + \tfrac{1}{3} \partial Z = \tfrac{1}{3} N + \tfrac{1}{2k} E$ is a euclidean boson.  With $\mathsf{T} = \tfrac{3}{2} \normord{\mathsf{J} \mathsf{J}}$, it is now routine to check that the $\mathsf{G}^{\pm}$, $\mathsf{J}$ and $\mathsf{T}$ generate the $\mathcal{N} = 2$ superconformal algebra $\alg{sVir}^{\brac{2}}$ of central charge $1$.  The bosonic dimension $\tfrac{3}{2}$ fields $\mathsf{g}^{\pm} = \sqrt{3} \vertop{\mp \brac{Y/2 + Z}} \chi^{\pm}$ likewise have regular \opes{} except for
\begin{equation}
\func{\mathsf{g}^+}{z} \func{\mathsf{g}^-}{w} = \frac{3}{\brac{z-w}^3} + \frac{3 \func{\mathsf{j}}{w}}{\brac{z-w}^2} + \frac{\tfrac{3}{2} \normord{\func{\mathsf{j}}{w} \func{\mathsf{j}}{w}} + \tfrac{3}{2} \func{\partial \mathsf{j}}{w} + 3 \normord{\func{\chi^+}{w} \func{\chi^-}{w}}}{z-w} + \ldots \ ,
\end{equation}
where $\mathsf{j}$ is the euclidean boson $-\tfrac{1}{2} \partial Y - \partial Z = -N - \tfrac{1}{2k} E$.  Taking $\mathsf{t} = \tfrac{1}{2} \normord{\mathsf{j} \mathsf{j}} - \normord{\chi^+ \chi^-}$, one finds that the $\mathsf{g}^{\pm}$, $\mathsf{j}$ and $\mathsf{t}$ generate the level $0$ Bershadsky-Polyakov algebra $W_3^{\brac{2}}$ with central charge $-1$.  The full extended algebra is not the direct sum $\alg{sVir}^{\brac{2}} \oplus W_3^{\brac{2}}$ --- the non-regular mixed \opes{} generate the fermionic $\AKMSA{gl}{1}{1}$ currents:
\begin{equation}
\func{\mathsf{g}^{\pm}}{z} \func{\mathsf{G}^{\pm}}{w} = \sqrt{\frac{2}{k}} \frac{\func{\psi^{\pm}}{w}}{\brac{z-w}^2} + \ldots
\end{equation}

The second choice $\AffAtypMod{-1/4,2k}$ yields a slightly different extended algebra.  
This time, the (appropriately normalised) fermionic dimension $\tfrac{3}{2}$ fields 
$\mathsf{G}^{\pm} =\pm \tfrac{1}{\sqrt{3}} \vertop{\pm \brac{3Y/4 -2 Z}} \chi^\pm \partial\chi^\pm$, the lorentzian boson $\mathsf{J} = -\tfrac{1}{4} \partial Y + \tfrac{2}{3} \partial Z$ and the energy-momentum tensor $\mathsf{T} = -\tfrac{3}{2} \normord{\mathsf{J} \mathsf{J}} - \normord{\chi^+ \chi^-}$ yield the $\mathcal{N} = 2$ superconformal algebra of central charge $-1$.  In particular, we have
\begin{equation}
\func{\mathsf{G}^+}{z} \func{\mathsf{G}^-}{w} = \frac{-2/3}{\brac{z-w}^3} + \frac{2 \func{\mathsf{J}}{w}}{\brac{z-w}^2} + \frac{-3 \normord{\func{\mathsf{J}}{w} \func{\mathsf{J}}{w}} + \func{\partial \mathsf{J}}{w} - 2 \normord{\func{\chi^+}{w} \func{\chi^-}{w}}}{z-w} + \ldots
\end{equation}
The bosonic subalgebra, now comprising the dimension $\tfrac{3}{2}$ fields $\mathsf{g}^{\pm} = \sqrt{3} \vertop{\mp \brac{Y/4 +2 Z}} \chi^{\pm}$, the euclidean boson $\mathsf{j} = -\tfrac{1}{4} \partial Y - 2 \partial Z$ and the dimension $2$ field $\mathsf{t} = \tfrac{1}{2} \normord{\mathsf{j} \mathsf{j}} - \normord{\chi^+ \chi^-}$, is again the Bershadsky-Polyakov algebra of level $0$ and central charge $-1$.  As the central charges of these subalgebras do not sum to zero, it is obvious that the full extended algebra is not a direct sum.  The fermionic $\AKMSA{gl}{1}{1}$ currents are again generated by the mixed \opes{}:
\begin{equation}
\func{\mathsf{g}^{\mp}}{z} \func{\mathsf{G}^{\pm}}{w} = \frac{2}{\sqrt{k}} \frac{\func{\psi^{\pm}}{w}}{\brac{z-w}^2}  + \ldots
\end{equation}

With both these choices, the result is an extended algebra which contains subalgebras isomorphic to an $\mathcal{N} = 2$ superconformal algebra and a Bershadsky-Polyakov algebra.  For completeness, we mention that the third choice $\AffAtypMod{-1,3k}$, which derives from the extended algebra of \secref{secDim1/2Ext}, also results in an extended algebra with similar subalgebras.  Specifically, we have a subalgebra isomorphic to $\alg{sVir}^{\brac{2}}$ with central charge $-1$ and a subalgebra isomorphic to $W_3^{\brac{2}}$ with level $-\tfrac{5}{3}$ and central charge $-1$.

It is telling that the same structure is found with other $W$-algebras including, in particular, that obtained from the Drin'feld-Sokolov reduction of the affine Kac-Moody superalgebra $\AKMSA{sl}{3}{1}$ that is induced by the embedding $\SLA{sl}{2} \hookrightarrow \SLSA{sl}{3}{1}$ given by $A \mapsto \brac{\begin{smallmatrix} A & 0 \\ 0 & 0 \end{smallmatrix}}$.  This $W$-algebra does not appear to coincide with any of the extended algebras constructed here, though it appears to be related to that obtained by changing the parity of our second extension (assuming that this leads to a consistent extended algebra).

\subsection{Extended Characters and the Spectrum}

Let us consider the dimension $\tfrac{1}{2}$ extension of \secref{secDim1/2Ext}. We saw that the simple current modules $\AffAtypMod{0,k}$ and $\AffAtypMod{0,-k}$ extend $\AKMSA{gl}{1}{1}$ to the direct sum of the $\beta\gamma$ ghost algebra and that of a complex fermion. The vacuum module of this extended algebra may then be identified with the orbit of the $\AKMSA{gl}{1}{1}$ vacuum module under fusion with these simple currents.  \eqnref{eqnFusL0xL0SF} gives this orbit as the set $\tset{\AffAtypMod{-\ell / 2 + \func{\eps}{\ell} , \ell k} \st \ell \in \ZZ}$, so we are led to consider the sum
\begin{align} \label{eqnGhostFermionVacChar}
\sum_{\ell \in \ZZ} \nch{\AffAtypMod{-\ell / 2 + \func{\eps}{\ell} , \ell k}}{x;y;z;q} &= x^k \sum_{\ell \in \ZZ} \frac{1}{\qnum{q}{\infty}^2} \sum_{m_1, m_2 = 0}^{\infty} \frac{q^{m_1 m_2 + \brac{m_1 + m_2} / 2 + \brac{m_1 - m_2 + \ell}^2/2}}{\qnum{q}{m_1} \qnum{q}{m_2}} z^{m_1 - m_2 + \ell/2} y^{\ell k} \notag \\
&= \frac{x^k}{\qnum{q}{\infty}^2} \sum_{m_1, m_2 = 0}^{\infty} \frac{q^{m_1 m_2 + \brac{m_1 + m_2} / 2}}{\qnum{q}{m_1} \qnum{q}{m_2}} z^{\brac{m_1 - m_2} / 2} y^{\brac{m_2 - m_1} k} \sum_{\ell \in \ZZ} y^{\ell k} z^{\ell / 2} q^{\ell^2 / 2} \notag \\
&= x^k \sum_{m_1, m_2 = 0}^{\infty} \frac{q^{\brac{m_1 + m_2} / 2}}{\qnum{q}{m_1} \qnum{q}{m_2}} \bigl( y^{-k} z^{1/2} \bigr)^{m_1 - m_2} \cdot \frac{1}{\qnum{q}{\infty}} \sum_{\ell \in \ZZ} \bigl( y^k z^{1/2} \bigr)^{\ell} q^{\ell^2 / 2} \notag \\
&= x^k \frac{\func{\eta}{q}}{\Jth{4}{y^{-k} z^{1/2} ; q}} \cdot \frac{\Jth{3}{y^k z^{1/2} ; q}}{\func{\eta}{q}}.
\end{align}
Here, we have used \eqnTref{eqnAtypCharDecomp}{eqnToBeProven}{eqnGhostVacChar}, in that order, to arrive at the product of the ghost vacuum character and the complex fermion (Neveu-Schwarz) vacuum character.  The factors $y^{-k} z^{1/2}$ and $y^k z^{1/2}$ appearing above account for the zero-modes being traced over in the ghost and fermion characters being $N_0 - E_0 / 2k$ and $N_0 + E_0 / 2k$, respectively.

The modular properties of the character of this fusion orbit are particularly pleasing.  Applying the generators $\mathsf{S}$ and $\mathsf{T}$ as in \secref{sec:modular} gives the six ratios
\begin{equation} \label{ModularHexagon}
\parbox[c]{0.8\textwidth}{
\begin{center}
\begin{tikzpicture}[auto,thick,scale=2]
\node (f34) at (-0.9,1) [] {$x^k \frac{\Jth{3}{y^{-k} z^{1/2} ; q}}{\Jth{4}{y^k z^{1/2} ; q}}$};
\node (f32) at (-1.9,0) [] {$x^k \frac{\Jth{3}{y^{-k} z^{1/2} ; q}}{\Jth{2}{y^k z^{1/2} ; q}}$};
\node (f43) at (0.9,1) [] {$x^k \frac{\Jth{4}{y^{-k} z^{1/2} ; q}}{\Jth{3}{y^k z^{1/2} ; q}}$};
\node (f42) at (-0.9,-1) [] {$x^k \frac{\Jth{4}{y^{-k} z^{1/2} ; q}}{\Jth{2}{y^k z^{1/2} ; q}}$};
\node (f23) at (1.9,0) [] {$x^k \frac{\Jth{2}{y^{-k} z^{1/2} ; q}}{\Jth{3}{y^k z^{1/2} ; q}}$};
\node (f24) at (0.9,-1) [] {$x^k \frac{\Jth{2}{y^{-k} z^{1/2} ; q}}{\Jth{4}{y^k z^{1/2} ; q}}$};
\draw [<->] (f34) to node {$\mathsf{T}$} (f43);
\draw [<->] (f43) to node {$\mathsf{S}$} (f23);
\draw [<->] (f23) to node {$\ee^{-\ii \pi / 4} \mathsf{T}$} (f24);
\draw [<->] (f24) to node {$\mathsf{S}$} (f42);
\draw [<->] (f42) to node {$\ee^{\ii \pi / 4} \mathsf{T}$} (f32);
\draw [<->] (f32) to node {$\mathsf{S}$} (f34);
\end{tikzpicture}
\end{center}
}
\end{equation}
on which one verifies that $\mathsf{C} = \mathsf{S}^2 = \brac{\mathsf{S} \mathsf{T}}^3$ acts as the identity (the extended vacuum module is self-conjugate).  Identifying these ratios as characters of the extended algebra is not difficult, but the interpretation in terms of $\AKMSA{gl}{1}{1}$ is somewhat less straight-forward.  For example, $\jth{4} / \jth{3}$ corresponds to inserting $\brac{-1}^F$ and $\brac{-1}^G$ into the complex fermion and ghost vacuum characters, where $F$ and $G$ are the fermion and ghost number operators, respectively.  However, this does not amount to replacing the $\AKMSA{gl}{1}{1}$-characters in \eqnref{eqnGhostFermionVacChar} by the corresponding supercharacters.  Rather, we should modify the usual characters by inserting an extra factor of $\brac{-1}^{E_0 / k}$.

To identify the $\AKMSA{gl}{1}{1}$ content of the other ratios, it is convenient to note that $\AffAtypMod{-\ell / 2 + \func{\eps}{\ell} , \ell k} = \tfunc{\sigma^{\ell}}{\AffAtypMod{\ell / 2 , 0}}$.  One can now obtain the ratio $\jth{2} / \jth{4}$ by replacing the sum over $\ell \in \ZZ$ in \eqref{eqnGhostFermionVacChar} by a sum over $\ell \in \ZZ + \tfrac{1}{2}$.  The $\AKMSA{gl}{1}{1}$-characters appearing in this sum are therefore those of certain twisted atypicals.  The ratio $\jth{2} / \jth{3}$ may be obtained by replacing these twisted $\AKMSA{gl}{1}{1}$-characters by the corresponding supercharacters.  $\jth{3} / \jth{2}$ also decomposes as a sum over twisted atypical $\AKMSA{gl}{1}{1}$-supercharacters, but this time the supercharacters are those of $\tfunc{\sigma^{\ell}}{\AffAtypMod{\brac{\ell - 1} / 2 , 0}}$ (with $\ell \in \ZZ + \tfrac{1}{2}$).  Finally, $\jth{4} / \jth{2}$ decomposes into the sum of the characters of these twisted atypicals, modified by an insertion of $\brac{-1}^{E_0 / k - 1/2}$.

The extended algebra characters of \eqref{ModularHexagon} therefore correspond to linear combinations of the characters of the $\AKMSA{gl}{1}{1}$ atypicals $\tfunc{\sigma^{\ell}}{\AffAtypMod{\ell / 2 , 0}}$, with $\ell \in \tfrac{1}{2} \ZZ$, and $\tfunc{\sigma^{\ell}}{\AffAtypMod{\brac{\ell - 1} / 2 , 0}}$, with $\ell \in \ZZ + \tfrac{1}{2}$.  The first set is closed under fusion, but the second is not.  Indeed, it is easy to check that the fusion ring generated by these atypicals consists of the genuine atypicals $\tfunc{\sigma^{\ell}}{\AffAtypMod{n , 0}}$, with $\ell \in \ZZ$ and $n \in \tfrac{1}{2} \ZZ$, and their twisted counterparts with $\ell \in \ZZ + \tfrac{1}{2}$ and $n \in \tfrac{1}{2} \ZZ + \tfrac{1}{4}$.

At the level of the extended algebra, fusion therefore generates an infinite number of additional extended modules.  The number of linearly independent characters, however, turns out to be finite.  Beyond those given in \eqref{ModularHexagon}, one finds only three more.  Their modular transformations are summarised by
\begin{equation} \label{ModularChain}
\parbox[c]{0.83\textwidth}{
\begin{center}
\begin{tikzpicture}[auto,thick,scale=2]
\node (f21) at (-1.8,0) [] {$\ii x^k \frac{\Jth{2}{y^{-k} z^{1/2} ; q}}{\Jth{1}{y^k z^{1/2} ; q}}$};
\node (f41) at (0,0) [] {$\ii x^k \frac{\Jth{4}{y^{-k} z^{1/2} ; q}}{\Jth{1}{y^k z^{1/2} ; q}}$};
\node (f31) at (1.8,0) [] {$\ii x^k \frac{\Jth{3}{y^{-k} z^{1/2} ; q}}{\Jth{1}{y^k z^{1/2} ; q}}$};
\draw [<->] (f21) to node {$\ii \mathsf{S}$} (f41);
\draw [<->] (f41) to node {$\ee^{\ii \pi / 4} \mathsf{T}$} (f31);
\draw (f21) to [in=160,out=200,distance=6mm,loop] node {$\mathsf{T}$} ();
\draw (f31) to [in=-20,out=20,distance=6mm,loop] node {$-\ii \mathsf{S}$} ();
\end{tikzpicture}
\end{center}
}
\ ,
\end{equation}
which shows that $\mathsf{C}$ acts as minus the identity on these characters.  These nine ratios of theta functions then constitute the (linearly independent) characters of the \emph{minimal} set of extended algebra modules which are closed under fusion and modular transformations.  We emphasise that this minimal set is nevertheless infinite.  Its elements decompose as $\AKMSA{gl}{1}{1}$-modules into a \emph{discrete} set of atypicals and twisted atypicals.  Of course, it remains to see if this may be extended by a discrete set of typicals, for example, while maintaining modular invariance.  We will not consider this here.  The point is merely to show how the extended algebra formalism allows one to restrict to a discrete spectrum of $\AKMSA{gl}{1}{1}$-modules and still exhibit closure under fusion and a well-defined action of the modular group on the characters.  A similar analysis should be possible for the other extended algebras.  We hope to return to this in the future.

\section{Summary and Outlook}\label{secoutlook}

In this article, we have analysed a collection of logarithmic conformal field theories stemming from $\AKMSA{gl}{1}{1}$.  We have used a combination of representation theoretic and free field methods, each being employed as appropriate.  We began with an analysis of the Lie superalgebra $\AKMSA{gl}{1}{1}$, its representations, fusion ring and modular properties.  With this in hand, we determined that the $\beta\gamma$ ghosts may be constructed as a $\AKMA{u}{1}$-coset of $\AKMSA{gl}{1}{1}$. This was achieved by identifying the coset algebra at the level of states with that of the ghost system.  Free field considerations also predict this result, though we noted that they must be supplemented by the analysis of the vacuum character.  We completed the description of the coset theory by identifying how each $\AKMSA{gl}{1}{1}$ character decomposes into a ghost character.  This was followed by a summary of our results identifying other cosets of $\AKMSA{gl}{1}{1}$ by different $\AKMA{u}{1}$-subalgebras.  In particular, we found the $\mathcal{N} = 2$ super-Virasoro algebra at central charge $c=-1$ and the Bershadsky-Polyakov-algebra at level $k=0$.  These are, however, just some of the infinitely many coset algebras that one can construct.
 
The remainder of the article was devoted to finding chiral algebras extending $\AKMSA{gl}{1}{1}$.  Avoiding logarithmic singularities constrains the extensions to adjoining certain atypical modules and we have computed these algebras explicitly for all possible extensions of conformal dimension $\tfrac{1}{2}$, $1$ and $\tfrac{3}{2}$.  The results include a complex free fermion plus the $\beta\gamma$ ghosts, the Lie superalgebra $\AKMSA{sl}{2}{1}$ at levels $-\tfrac{1}{2}$ and $1$, and certain new $W$-superalgebras which combine the $\mathcal{N} = 2$ super-Virasoro algebra and the Bershadsky-Polyakov algebra.  Our search for extended algebras was motivated by the question of real forms of $\SLSA{gl}{1}{1}$ and discrete versus continuous spectra.  While the modular properties of the characters of $\AKMSA{gl}{1}{1}$ require a continuous spectrum, we have demonstrated that by combining characters of $\AKMSA{gl}{1}{1}$ into extended algebra characters, it is possible to find a discrete spectrum that is closed under fusion and modular transformations.

It is worthwhile emphasising that although the theories we consider are often non-rational with a \emph{continuum} of irreducible representations, this is not problematic in the slightest because they are naturally parametrised in such a way that all standard features (characters, fusion rules, modular transformations) vary continuously.  Indeed, it is only at a non-generic set of parameter values that one has to worry about indecomposability and logarithmic behaviour.  From this point of view, the logarithmic complexity encountered in these non-rational theories is no worse than those of their rational cousins and, in fact, taking simple current extensions of non-rational theories can result in rational (logarithmic) models.

As stated in the introduction, the most important conclusion that we draw from the results presented here is that \emph{almost all} of the logarithmic CFTs that researchers regard as archetypal are in fact all closely related to one another.  This then raises the important question of whether these paradigms are indeed prototypical examples or if other models are going to reveal qualitatively new features.  Certainly for applications, it is absolutely necessary to have a good feel for the typical behaviour of this class of CFTs.  It is thus crucial to attempt to understand other examples of logarithmic CFT.  We remark that there is some recent progress in this direction with the $c=0$ triplet model \cite{FeiLog06,RasWEx08,GabFus09,AdaWAl09}.  However, the algebraic complexities of the $\func{W}{2,3}$ algebra have meant that much of this progress is indirect.  More tractable candidates, in our opinion, include the $\AKMSA{sl}{2}{1}$ WZW model (at level $k \neq -\tfrac{1}{2}, 1$) and fractional level $\AKMA{sl}{2}$ theories (with $k \neq -\tfrac{1}{2}$).  We expect that  the fruitful combination of representation theoretic and free field methods will lead to progress and insights in these examples and beyond, and hope to report on this in the future.

Another area which deserves further investigation is the theory of the $W$-algebras and $W$-superalgebras which arise in the study of affine Lie superalgebras.  In \cite{Feigin:2004wb}, certain algebras denoted by $W^{\brac{n}}_2$ were constructed as commutants of $\AKMA{sl}{n}_k$ in $\AKMSA{sl}{n}{1}_k$. Moreover, when the level is critical, these algebras arise in the study of sigma models whose target space is a fermionic coset of $\SLSG{PSL}{n}{n}$ \cite{CGL}.  The $W$-superalgebras which arise as extensions of $\AKMSA{gl}{1}{1}$ appear to share several key features with these commutant algebras.  However, they are much more accessible algebraically and one may likewise expect that they will be of some use in string theoretic applications.  We also mention that these extended algebras share many features in common with certain algebras \cite{Odake:1988bh} related to manifolds with $\SLG{SU}{n}$ holonomy. These are extensions of the $\mathcal{N} = 2$ super-Virasoro algebra by two bosonic and two fermionic fields.\footnote{We thank Andrew Linshaw for this comment.}  As we have seen, the $\mathcal{N} = 2$ super-Virasoro algebra may be related to $\AKMSA{gl}{1}{1}$ in several ways (see \cite{Creutzig:2010ne} for another), so it is natural to ask whether there are extensions of both these algebras which are related in a similar manner.

\section*{Acknowledgements}

TC is grateful to Peter R\o{}nne, Louise Dolan and Andrew Linshaw for discussions.
DR likewise thanks Peter Bouwknegt and Pierre Mathieu for the same.  
The work of TC was partially supported by U.S.\ Department of Energy,
Grant No.\ DE-FG02-06ER-4141801, Task A.
The research of DR is supported by an Australian Research Council Discovery Project DP0193910.

\appendix

\section{Real Forms} \label{sec:real}

In this appendix, we discuss the real forms of the Grassmann envelope of the Lie superalgebra $\SLSA{gl}{1}{1}$ (for real forms of Lie superalgebras, see \cite{Parker:1980af}).
This envelope is the complexification of the Lie \emph{algebra} of the (real) Lie supergroup $\SLSG{GL}{1}{1}$ and consists of the elements $aN + bE + c_+ \psi^+ + c_- \psi^-$, where $a$ and $b$ are Grassmann-even and the $c_{\pm}$ are Grassmann-odd \cite{Berezin:1987wh}.  
Define a semimorphism of a complex Lie algebra to be a conjugate-linear transformation which preserves the Lie bracket. 
Given an involutive semimorphism $\omega$, a real form is then obtained as the set of Lie algebra elements satisfying $\omega(X) = X$.
Note that for Lie algebras, involutive semimorphisms and \emph{adjoints} are in $1$-$1$ correspondence via $\omega \leftrightarrow -\omega$.
 

One may extend complex conjugation to the Grassmann algebra as an involutive antiautomorphism $\mathcal T$, that is, a parity-preserving map of the Grassmann algebra satisfying \cite{Berezin:1987wh}:
\begin{enumerate}
\item $\func{\mathcal{T}}{\lambda f + \mu g} = \bar{\lambda} \func{\mathcal{T}}{f} + \bar{\mu} \func{\mathcal{T}}{g}$,
\item $\func{\mathcal{T}}{fg} = \func{\mathcal{T}}{g} \func{\mathcal{T}}{f}$,
\item $\mathcal{T}^2 = \id$.
\end{enumerate} 
Here, $\lambda, \mu \in \CC$, $f$ and $g$ are Grassmann numbers, and $\bar{\lambda}$ denotes the ordinary complex conjugate of $\lambda$.  An involutive semimorphism of the Grassmann envelope of $\SLSA{gl}{1}{1}$ is then obtained by composing $\mathcal{T}$ with the Grassmann-linear map $\alpha$ defined by
\begin{equation}
\func{\alpha}{N} = N, \qquad \func{\alpha}{E} = E, \qquad \func{\alpha}{\psi^{\pm}} = \pm \psi^{\pm}.
\end{equation}


Twisting the above development by an automorphism of $\SLSA{gl}{1}{1}$ allows us to construct other involutive semimorphisms.  Such automorphisms include
\begin{subequations} \label{eqnautomorphisms}
\begin{align}
\func{\Pi}{N} &= -N, & \func{\Pi}{E} &= +E, & \func{\Pi}{\psi^{\pm}} &= \psi^{\mp}, \\
\func{\Omega}{N} &= -N, & \func{\Omega}{E} &= -E, & \func{\Omega}{\psi^{\pm}} &= \pm \psi^{\mp}, \\
\func{\omega_{\lambda}}{N} &= +N, & \func{\omega_{\lambda}}{E} &= +E, & \func{\omega_{\lambda}}{\psi^{\pm}} &= \lambda^{\pm1} \psi^{\pm}.
\end{align}
\end{subequations}
Note that $\Pi$ does not preserve the metric; rather, it negates it.  Its lift to $\AKMSA{gl}{1}{1}$ therefore negates the level.  As the subalgebras fixed by $\Pi$ and $\Pi\circ\Omega$, and their analogues for $\SLSA{gl}{n}{n}$, are called strange Lie superalgebras, we will refer to $\Pi$ as the strange automorphism.  It is of order two. The automorphism $\Omega$, on the other hand, preserves the metric and is of order four, squaring to $\omega_{-1}$.  It is the conjugation automorphism of $\SLSA{gl}{1}{1}$ (compare \eqnref{eqnDefConj}). The last family $\omega_{\lambda}$, $\lambda \in \CC \setminus \set{0}$, consists of inner automorphisms $\omega_{\lambda}$, so it is of little direct importance to real form considerations.  However, $\omega_{\ii}$ will be used below to ensure that the order of the semimorphism constructed from $\Pi$ is indeed $2$.

The resulting involutive semimorphisms of the Grassmann envelope of $\SLSA{gl}{1}{1}$ are listed in \tabref{tabRealForms}, along with their real forms and the topology of the bosonic submanifold of the Lie supergroup.  
The semimorphisms of the two strange forms are probably unphysical, at least for our considerations, because a corresponding adjoint for the affine superalgebra is only possible for imaginary level.
{
\begin{table}[h]
\setlength{\extrarowheight}{4pt}
\begin{center}
\begin{tabular}{|c|c|c|c|}
\hline
Semimorphism & Bosonic Submanifold & Real Form \\[1mm]
\hline\hline
$\mathcal{T} \circ \alpha \circ \Omega$ & $S^1_E \times S^1_N$ & unitary superalgebra $\SLSA{u}{1}{1}$ \\[1mm] \hline
$\mathcal{T} \circ \alpha$ & $\R_E \times \R_N$ & real superalgebra $\SLSA{gl}{1}{1}_\R$ \\[1mm] \hline
$\mathcal{T} \circ \alpha \circ \Pi \circ \omega_{\ii}$ & $\R_E \times S^1_N$ & strange form $\SLSA{gl}{1}{1}_A$ \\[1mm] \hline
$\mathcal{T} \circ \alpha \circ \Pi \circ \Omega$ & $S^1_E \times \R_N$ & strange form $\SLSA{gl}{1}{1}_B$ \\[1mm] \hline
\end{tabular}
\vspace{3mm}
\caption{The real forms of the Grassmann envelope of $\SLSA{gl}{1}{1}$, given along with their involutive semimorphisms and the topologies of the bosonic submanifold of the corresponding Lie supergroups.} \label{tabRealForms}
\end{center}
\end{table}
}

\section{A Combinatorial Identity} \label{appIdentity}

In this appendix, we prove the identity \eqref{eqnToBeProven} for $m \geqslant 0$.  The case $m < 0$ then follows from symmetry under $m \leftrightarrow -m$.  This identity follows readily\footnote{We thank Pierre Mathieu for pointing this out in the case $m=0$.} from a hypergeometric identity of Heine \cite[Eq.~(1.4.3)]{GasBas04}:
\begin{equation} \label{eqnHeine}
\hypergeom{a}{b}{c}{q,z} = \frac{\qnum{abz/c}{\infty}}{\qnum{z}{\infty}} \hypergeom{c/a}{c/b}{c}{q,\frac{abz}{c}}.
\end{equation}
Recall that Heine's basic hypergeometric series is defined by
\begin{equation}
\hypergeom{a}{b}{c}{q,z} = \sum_{n=0}^{\infty} \frac{\qnum{a}{n} \qnum{b}{n}}{\qnum{c}{n} \qnum{q}{n}} z^n,
\end{equation}
where $\qnum{a}{n}$ denotes the standard $q$-factorial $\prod_{i=0}^{n-1} \brac{1-aq^i}$.

We start by noting that
\begin{equation}
\lim_{a,b \rightarrow \infty} \hypergeom{a}{b}{q^{2m+1}}{q,\frac{q^{2m+2}}{ab}} = \lim_{a,b \rightarrow \infty} \sum_{j=0}^{\infty} \frac{\qnum{a}{j} \qnum{b}{j}}{a^j b^j} \frac{q^{\brac{2m+2} j}}{\qnum{q}{j} \qnum{q^{2m+1}}{j}} = \sum_{j=0}^{\infty} \frac{q^{j^2 + \brac{2m+1} j}}{\qnum{q}{j} \qnum{q^{2m+1}}{j}},
\end{equation}
since $\qnum{a}{j} / a^j = \prod_{i=0}^{j-1} \brac{a^{-1} - q^i} \rightarrow \brac{-1}^j q^{j \brac{j-1} / 2}$ as $a \rightarrow \infty$.  Applying \eqref{eqnHeine} therefore gives
\begin{align}
\sum_{j=0}^{\infty} \frac{q^{j^2 + \brac{2m+1} j}}{\qnum{q}{j} \qnum{q^{2m+1}}{j}} &= \lim_{a,b \rightarrow \infty} \frac{\qnum{q}{\infty}}{\qnum{q^{2m+2}/ab}{\infty}} \hypergeom{q^{2m+1}/a}{q^{2m+1}/b}{q^{2m+1}}{q,q} \nonumber \\
&= \lim_{a,b \rightarrow \infty} \frac{\qnum{q}{\infty}}{\qnum{q^{2m+2}/ab}{\infty}} \sum_{j=0}^{\infty} \frac{\qnum{q^{2m+1}/a}{j} \qnum{q^{2m+1}/b}{j}}{\qnum{q}{j} \qnum{q^{2m+1}}{j}} q^j \nonumber \\
&= \qnum{q}{\infty} \sum_{j=0}^{\infty} \frac{q^j}{\qnum{q}{j} \qnum{q^{2m+1}}{j}}.
\end{align}
Multiplying both sides by $q^m$ and dividing by $\qnum{q}{2m}$ and $\qnum{q}{\infty}$ then gives, up to a shift in the summation index $j$, \eqnref{eqnToBeProven} for $m \geqslant 0$, as required.

\section{Free Field Realisations for $\AKMSA{gl}{1}{1}$} \label{appFreeFields}

The affine Kac-Moody superalgebra $\AKMSA{gl}{1}{1}$ has two well-known free field realizations, the standard Wakimoto realization \cite{SalGL106} and one constructed from symplectic fermions and two bosons, one euclidean and one lorentzian \cite{Guruswamy:1999hi}.  An explicit equivalence between the two realisations was established in \cite{CR09}.  Here, we review (and motivate) the symplectic fermion realization for computations in \secref{secExtAlg}.

Recall that in \eqnref{eqnDefBosons}, we identified two commuting bosons
\begin{equation}
\func{\partial \varphi}{z} = \func{N}{z} + \frac{1}{2k} \func{E}{z} \qquad \text{and} \qquad \func{\partial \phi}{z} = \func{N}{z} - \frac{1}{2k} \func{E}{z},
\end{equation}
the first being euclidean and the second lorentzian.  Taking the coset of $\AKMSA{gl}{1}{1}$ by the $\AKMA{u}{1}^2$-subalgebra generated by these bosons gives the theory of symplectic fermions, suggesting that a free field realisation may be built from these fields.  Indeed, let us take the fermions $\chi^{\pm}$ to satisfy
\begin{equation}
\func{\chi^+}{z} \func{\chi^-}{w} = \frac{1}{\brac{z-w}^2} + \text{ regular terms},
\end{equation}
with $\func{\chi^{\pm}}{z} \func{\chi^{\pm}}{w}$ regular, and construct bosonic fields $Z = \tfrac{1}{2} \brac{\varphi + \phi}$ and $Y = \varphi - \phi$ so that
\begin{equation}
\func{\partial Y}{z} \func{\partial Z}{w} = \frac{1}{\brac{z-w}^2} + \text{ regular terms}
\end{equation}
and $\func{\partial Y}{z} \func{\partial Y}{w}$ and $\func{\partial Z}{z} \func{\partial Z}{w}$ are regular.  Then, the $\AKMSA{gl}{1}{1}$ current fields are expressed as
\begin{equation} \label{eqnGL11FFR}
\func{E}{z} = k \func{\partial Y}{z}, \qquad \func{N}{z} = \func{\partial Z}{z} \qquad \text{and} \qquad \func{\psi^{\pm}}{z} = \sqrt{k} \normord{\ee^{\pm \func{Y}{z}}} \func{\chi^{\pm}}{z},
\end{equation}
and a moderately tedious computation shows that the $\AKMSA{gl}{1}{1}$ energy momentum tensor \eqref{eqnDefT} indeed corresponds to the sum of those for the boson and symplectic fermion systems.

It remains to explicitly construct the $\AKMSA{gl}{1}{1}$ primary fields.  When these fall into typical representations, this requires the \emph{twist fields} that define the symplectic fermion primaries \cite{KauSym00}.  However, these fields are surplus to our needs (details may be found in \cite{CR09}).  \secref{secExtAlg} in fact only requires the zero-grade fields of certain atypical modules with $\abs{e/k} = 1$ and $2$.  These are particularly easy to construct.

First, note that the vertex operator
\begin{equation}
\vertop{n \func{Y}{w} + e \func{Z}{w} / k} = \vertop{\brac{n+e/2k} \func{\varphi}{w}} \vertop{-\brac{n-e/2k} \func{\phi}{w}}
\end{equation}
has affine weight $\brac{n,e}$, using \eqref{eqnGL11FFR}, but conformal dimension $ne/k$.  Restricting to $e=k$, the atypical irreducibles $\AffAtypMod{n,k}$ correspond to zero-grade fields of weights $\brac{n \pm \tfrac{1}{2} , k}$ and conformal dimensions $n + \tfrac{1}{2}$.  It follows that $\vertop{\brac{n+1/2} \func{Y}{w} + \func{Z}{w}}$ realises one of these fields in this case.  The other is obtained by acting with $\psi^-$, again using \eqref{eqnGL11FFR}:
\begin{equation}
\func{\psi^-}{z} \vertop{\brac{n+1/2} \func{Y}{w} + \func{Z}{w}} = \frac{\sqrt{k} \vertop{\brac{n-1/2} \func{Y}{w} + \func{Z}{w}} \func{\chi^-}{w}}{z-w} + \ldots
\end{equation}
With the conjugate fields ($e=-k$), this completes the determination when $\abs{e/k} = 1$:
\begin{equation} \label{eqnAffPrimFFR}
\begin{matrix}
\vertop{\brac{n+1/2} Y + Z}, & \hspace{20mm} & \vertop{\brac{n-1/2} Y + Z} \func{\chi^-}{w}, & \hspace{20mm} & \text{($e=k$),} \\
\vertop{-\brac{n-1/2} Y - Z} \func{\chi^+}{w}, & \hspace{20mm} & \vertop{-\brac{n+1/2} Y - Z} & \hspace{20mm} & \text{($e=-k$).}
\end{matrix}
\end{equation}
The corresponding quadruplet for $\abs{e/k} = 2$ may be taken to be
\begin{equation} \label{eqnAffPrimFFR2}
\vphantom{f} 
\begin{matrix}
\vertop{\brac{n+1/2} Y + 2 Z} \func{\chi^-}{w}, & \hspace{20mm} & \vertop{\brac{n-1/2} Y + 2 Z} \func{\chi^-}{w} \func{\partial \chi^-}{w}, & \hspace{20mm} & \text{($e=2k$),} \\
\vertop{-\brac{n-1/2} Y - 2 Z} \func{\chi^+}{w} \func{\partial \chi^+}{w}, & \hspace{20mm} & \vertop{-\brac{n+1/2} Y - 2 Z} \func{\chi^+}{w} & \hspace{20mm} & \text{($e=-2k$).}
\end{matrix}
\end{equation}
We also record here the \ope{} for the vertex operators.  Let $X_{n,e} = n Y + e Z / k$.  Then,
\begin{multline} \label{eqnVertexOPE}
\vertop{\func{X_{n,e}}{z}} \vertop{\func{X_{n',e'}}{w}} = \brac{z-w}^{\brac{ne'+n'e}/k} \biggl[ \vertop{\func{X_{n+n',e+e'}}{w}} + \normord{\func{\partial X_{n,e}}{w} \Vertop{\func{X_{n+n',e+e'}}{w}}} \brac{z-w} \Biggr. \\
\Biggl. + \frac{1}{2} \normord{\Bigl( \func{\partial X_{n,e}}{w} \func{\partial X_{n,e}}{w} + \func{\partial^2 X_{n,e}}{w} \Bigr) \Vertop{\func{X_{n+n',e+e'}}{w}}} \brac{z-w}^2 + \ldots \biggr].
\end{multline}
It follows from this that $\vertop{n \func{Y}{w} + e \func{Z}{w} / k}$ and $\vertop{n' \func{Y}{w} + e' \func{Z}{w} / k}$ will be mutually bosonic when $ne' + n'e$ is an even multiple of $k$ and mutually fermionic when $ne' + n'e$ is an odd multiple of $k$.

\section{Spin-$\tfrac{3}{2}$ Algebras} \label{appSpin3/2Alg}

In this appendix, we give for convenience the defining \opes{} for certain algebras generated by two fields of conformal dimension $\tfrac{3}{2}$ and $\AKMA{u}{1}$-charge $\pm 1$.  The first family is that of the $\mathcal{N} = 2$ superconformal algebras $\alg{sVir}^{\brac{2}}$ in which the generating fields $\mathsf{G}^{\pm}$ are fermionic.  Our conventions are captured by $\func{\mathsf{G}^{\pm}}{z} \func{\mathsf{G}^{\pm}}{w}$ being regular and
\begin{equation}
\func{\mathsf{G}^+}{z} \func{\mathsf{G}^-}{w} = \frac{2c/3}{\brac{z-w}^3} + \frac{2 \func{\mathsf{J}}{w}}{\brac{z-w}^2} + \frac{2 \func{\mathsf{T}}{w} + \func{\partial \mathsf{J}}{w}}{z-w} + \ldots
\end{equation}
Here, $\mathsf{J}$ is the (dimension $1$) $\AKMA{u}{1}$-current under which the $\mathsf{G}^{\pm}$ are charged, normalised so that
\begin{equation}
\func{\mathsf{J}}{z} \func{\mathsf{J}}{w} = \frac{c/3}{\brac{z-w}^2} + \text{ regular terms,}
\end{equation}
and $\mathsf{T}$ is the energy-momentum tensor.  The central charge $c$ is a free parameter.

The second family is obtained when the dimension $\tfrac{3}{2}$ fields $\mathsf{g}^{\pm}$ are bosonic.  These are the (level $k$) Bershadsky-Polyakov algebras $W_3^{\brac{2}}$ which are obtained as the Drin'feld-Sokolov reductions of $\AKMA{sl}{3}_k$ induced by the embedding $\SLA{sl}{2} \hookrightarrow \SLA{sl}{3}$ given by $A \mapsto \brac{\begin{smallmatrix} A & 0 \\ 0 & 0 \end{smallmatrix}}$ \cite{PolGau90,BerCon91}.  The product $\func{\mathsf{g}^{\pm}}{z} \func{\mathsf{g}^{\pm}}{w}$ is again regular, but we have
\begin{equation}
\func{\mathsf{g}^+}{z} \func{\mathsf{g}^-}{w} = \frac{\brac{k+1} \brac{2k+3}}{\brac{z-w}^3} + \frac{3 \brac{k+1} \func{\mathsf{j}}{w}}{\brac{z-w}^2} + \frac{3 \normord{\func{\mathsf{j}}{w} \func{\mathsf{j}}{w}} + \tfrac{3}{2} \brac{k+1} \func{\partial \mathsf{j}}{w} - \brac{k+3} \func{\mathsf{t}}{w}}{z-w} + \ldots \ ,
\end{equation}
where now $\mathsf{j}$ is the $\AKMA{u}{1}$-current charging the $\mathsf{g}^{\pm}$ and $\mathsf{t}$ is the energy-momentum tensor.  The current turns out to satisfy
\begin{equation}
\func{\mathsf{j}}{z} \func{\mathsf{j}}{w} = \frac{\brac{2k+3}/3}{\brac{z-w}^2} + \text{ regular terms,}
\end{equation}
and the central charge is given in terms of the $\AKMA{sl}{3}$ level $k$ by
\begin{equation}
c = -\frac{\brac{2k+3} \brac{3k+1}}{k+3}.
\end{equation}
The critical level $k=-3$ is excluded from this family for obvious reasons.  

\section{The Lie Superalgebras $\SLSA{sl}{2}{1}$ and $\AKMSA{sl}{2}{1}$} \label{appSL21}

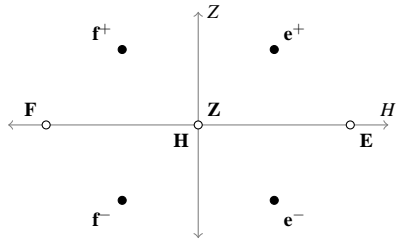
\begin{figure}
\centering
\begin{tikzpicture}[auto]
\draw [<->,gray] (-2.5,0) -- (2.5,0) node [above,black] {$\scriptstyle H$};
\draw [<->,gray] (0,-1.5) -- (0,1.5) node [right,black] {$\scriptstyle Z$};
\draw (-1,1) node [above left] {$\scriptstyle \Bf^+$}
      (1,1) node [above right] {$\scriptstyle \Be^+$}
      (1,-1) node [below right] {$\scriptstyle \Be^-$}
      (-1,-1) node [below left] {$\scriptstyle \Bf^-$}
      (-2,0) node [above left] {$\scriptstyle \BF$}
      (2,0) node [below right] {$\scriptstyle \BE$}
      (0,0) node [above right] {$\scriptstyle \BZ$}
            node [below left] {$\scriptstyle \BH$};
\foreach \x in {(-1,1),(1,1),(-1,-1),(1,-1)}
 \filldraw [black] \x circle (1.5pt);
\foreach \x in {(-2,0),(0,0),(2,0)}
 \draw [black,fill=white] \x circle (1.5pt);
\end{tikzpicture}
\caption{The weights ($\BH$- and $\BZ$-eigenvalues) $\brac{H,Z}$ of the simple Lie superalgebra $\SLSA{sl}{2}{1}$, labelled by the corresponding weight vectors.  White weights are bosonic and black weights are fermionic.} \label{figSL21Wts}
\end{figure}

Here, we briefly outline our conventions for the simple Lie superalgebra $\SLSA{sl}{2}{1}$ and its affinisation.  Recall that $\SLSA{sl}{2}{1}$ is defined to be the vector space of supertraceless endomorphisms of $\CC^{2|1}$, equipped with the standard graded commutator.  Letting $E_{ij}$ denote the matrix with all zero entries but for a $1$ in row $i$ and column $j$, we choose a basis as follows:
\begin{equation}
\begin{matrix}
\BE = E_{12}, \quad \BH = E_{11} - E_{22}, \quad \BF = E_{21}, \quad \BZ = E_{11} + E_{22} + 2 E_{33} & \qquad & \text{(bosons),} \\
\Be^+ = E_{32}, \quad \Bf^+ = -E_{31}, \quad \Be^- = E_{13}, \quad \Bf^- = E_{23} & \qquad & \text{(fermions).}
\end{matrix}
\end{equation}
The bosons form a subalgebra isomorphic to $\SLA{sl}{2} \oplus \SLA{u}{1}$, with $\BZ$ spanning the second summand.  The fermions split into two irreducible representations of the bosonic subalgebra which may be identified with the tensor product of the fundamental representation of $\SLA{sl}{2}$ and the $\SLA{u}{1}$-irreducible of charge ($\BZ$-eigenvalue) $1$ or $-1$.  We normalise the fermions so that $\comm{\BF}{\Be^{\pm}} = \Bf^{\pm}$ and use the superscript $\pm$ to indicate the $\SLA{u}{1}$-charge.  The weight diagram of $\SLSA{sl}{2}{1}$ therefore looks as in \figref{figSL21Wts}.

As usual, we take the invariant bilinear form to be given by the supertrace in the defining representation.  This yields
\begin{equation} \label{eqnSL21Killing}
\begin{matrix}
\killing{\BH}{\BH} = 2, \qquad 
\killing{\BZ}{\BZ} = -2, \qquad 
\killing{\BE}{\BF} = \killing{\BF}{\BE} = 1, \\
-\killing{\Be^+}{\Bf^-} = \killing{\Bf^-}{\Be^+} = 1, \qquad
-\killing{\Be^-}{\Bf^+} = \killing{\Bf^+}{\Be^-} = 1,
\end{matrix}
\end{equation}
with all other combinations of basis elements vanishing.  We can now construct the affinisation $\AKMSA{sl}{2}{1}$ in the usual way and use the Sugawara construction to realise the energy-momentum tensor.  In this way, one checks that the dual Coxeter number of $\SLSA{sl}{2}{1}$ is $1$ and that the central charge always vanishes.

\raggedright


\begin{thebibliography}{10}

\bibitem{Maldacena:1997re}
J~Maldacena.
\newblock {The Large $N$ Limit of Superconformal Field Theories and
  Supergravity}.
\newblock {\em Adv. Theo. Math. Phys.}, 2:231--252, 1998.
\newblock \textsf{arXiv:\mbox{hep-th}/9711200}.

\bibitem{Zirnbauer:1999ua}
M~Zirnbauer.
\newblock {Conformal Field Theory of the Integer Quantum Hall Plateau
  Transition}.
\newblock \textsf{arXiv:\mbox{hep-th}/9905054}.

\bibitem{Guruswamy:1999hi}
S~Guruswamy, A~LeClair, and A~Ludwig.
\newblock {$gl \left(N \middle\vert N \right)$ Super Current Algebras for
  Disordered Dirac Fermions in Two-Dimensions}.
\newblock {\em Nucl. Phys.}, B583:475--512, 2000.
\newblock \textsf{arXiv:cond-mat/9909143}.

\bibitem{Quella:2007sg}
T~Quella, V~Schomerus, and T~Creutzig.
\newblock {Boundary Spectra in Superspace Sigma-Models}.
\newblock {\em JHEP}, 0810:024, 2008.
\newblock \textsf{arXiv:0712.3549 [\mbox{hep-th}]}.

\bibitem{Ashok:2009xx}
S~Ashok, R~Benichou, and J~Troost.
\newblock {Conformal Current Algebra in Two Dimensions}.
\newblock {\em JHEP}, 0906:017, 2009.
\newblock \textsf{arXiv:0903.4277 [\mbox{hep-th}]}.

\bibitem{Creutzig:2010hr}
T~Creutzig.
\newblock {Yangian Superalgebras in Conformal Field Theory}.
\newblock {\em Nucl. Phys.}, B849:636--653, 2011.
\newblock \textsf{arXiv:1011.6424 [\mbox{hep-th}]}.

\bibitem{Candu:2011hu}
C~Candu and V~Schomerus.
\newblock {Exactly Marginal Parafermions}.
\newblock{\em Phys. Rev.}, D84:051704, 2011.
\newblock \textsf{arXiv:1104.5028 [\mbox{hep-th}]}.

\bibitem{Quella:2007hr}
T~Quella and V~Schomerus.
\newblock {Free Fermion Resolution of Supergroup WZNW Models}.
\newblock {\em JHEP}, 0709:085, 2007.
\newblock \textsf{arXiv:0706.0744 [\mbox{hep-th}]}.

\bibitem{Hikida:2007sz}
Y~Hikida and V~Schomerus.
\newblock {Structure Constants of the $OSP \left( 1 \middle\vert 2 \right)$
  WZNW Model}.
\newblock {\em JHEP}, 0712:100, 2007.
\newblock \textsf{arXiv:0711.0338 [\mbox{hep-th}]}.

\bibitem{Creutzig:2011qm}
T~Creutzig, Y~Hikida, and P~R\o{}nne.
\newblock {Supergroup-Extended Super Liouville Correspondence}.
\newblock {\em JHEP}, 1106:063, 2011.
\newblock \textsf{arXiv:1103.5753 [\mbox{hep-th}]}.

\bibitem{Creutzig:2010ne}
T~Creutzig and P~R\o{}nne.
\newblock {From World-Sheet Supersymmetry to Super Target Spaces}.
\newblock {\em JHEP}, 1011:021, 2010.
\newblock \textsf{arXiv:1006.5874 [\mbox{hep-th}]}.

\bibitem{Creutzig:2008ag}
T~Creutzig.
\newblock {Geometry of Branes on Supergroups}.
\newblock {\em Nucl. Phys.}, B812:301--321, 2009.
\newblock \textsf{arXiv:0809.0468 [\mbox{hep-th}]}.

\bibitem{Gotz:2006qp}
G~Gotz, T~Quella, and V~Schomerus.
\newblock {The WZNW Model on $PSU \left( 1,1 \middle\vert 2 \right)$}.
\newblock {\em JHEP}, 0703:003, 2007.
\newblock \textsf{arXiv:\mbox{hep-th}/0610070}.

\bibitem{Saleur:2006tf}
H~Saleur and V~Schomerus.
\newblock {On the $SU \left(2 \middle\vert 1 \right)$ WZW Model and its
  Statistical Mechanics Applications}.
\newblock {\em Nucl. Phys.}, B775:312--340, 2007.
\newblock \textsf{arXiv:\mbox{hep-th}/0611147}.

\bibitem{Creutzig:2010zp}
T~Creutzig and Y~Hikida.
\newblock {Branes in the $OSP \left( 1 \middle\vert 2 \right)$ WZNW Model}.
\newblock {\em Nucl. Phys.}, B842:172--224, 2011.
\newblock \textsf{arXiv:1004.1977 [\mbox{hep-th}]}.

\bibitem{Giribet:2009eb}
G~Giribet, Y~Hikida, and T~Takayanagi.
\newblock {Topological String on $OSP \left(1 \middle\vert 2 \right) / U \left(
  1 \right)$}.
\newblock {\em JHEP}, 0909:001, 2009.
\newblock \textsf{arXiv:0907.3832 [\mbox{hep-th}]}.

\bibitem{Creutzig:2009fh}
T~Creutzig, P~R\o{}nne, and V~Schomerus.
\newblock {$N=2$ Superconformal Symmetry in Super Coset Models}.
\newblock {\em Phys. Rev.}, D80:066010, 2009.
\newblock \textsf{arXiv:0907.3902 [\mbox{hep-th}]}.

\bibitem{Rozansky:1992td}
L~Rozansky and H~Saleur.
\newblock {S and T Matrices for the Super $U \left( 1,1 \right)$ WZW model:
  Application to Surgery and Three Manifolds Invariants Based on the
  Alexander-Conway Polynomial}.
\newblock {\em Nucl. Phys.}, B389:365--423, 1993.
\newblock \textsf{arXiv:\mbox{hep-th}/9203069}.

\bibitem{RozQua92}
L~Rozansky and H~Saleur.
\newblock {Quantum Field Theory for the Multivariable Alexander-Conway
  Polynomial}.
\newblock {\em Nucl. Phys.}, B376:461--509, 1992.

\bibitem{SalGL106}
H~Saleur and V~Schomerus.
\newblock {The $GL \left( 1 \mid 1 \right)$ WZW Model: From Supergeometry to
  Logarithmic CFT}.
\newblock {\em Nucl. Phys.}, B734:221--245, 2006.
\newblock \textsf{arXiv:\mbox{hep-th}/0510032}.

\bibitem{CS09}
T~Creutzig and V~Schomerus.
\newblock {Boundary {C}orrelators in {S}upergroup {WZNW} {M}odels}.
\newblock {\em Nucl. Phys.}, B807:471--494, 2009.
\newblock \textsf{arXiv:0804.3469 [\mbox{hep-th}]}.

\bibitem{Creutzig:2007jy}
T~Creutzig, T~Quella, and V~Schomerus.
\newblock {Branes in the $GL \left( 1 \middle\vert 1 \right)$ WZNW-Model}.
\newblock {\em Nucl. Phys.}, B792:257--283, 2008.
\newblock \textsf{arXiv:0708.0583 [\mbox{hep-th}]}.

\bibitem{CR09}
T~Creutzig and P~R\o{}nne.
\newblock {The $GL \left( 1 \middle| 1 \right)$-Symplectic Fermion
  Correspondence}.
\newblock {\em Nucl. Phys.}, B815:95--124, 2009.
\newblock \textsf{arXiv:0812.2835 [\mbox{hep-th}]}.

\bibitem{GurLog93}
V~Gurarie.
\newblock {Logarithmic Operators in Conformal Field Theory}.
\newblock {\em Nucl. Phys.}, B410:535--549, 1993.
\newblock \texttt{arXiv:\mbox{hep-th}/9303160}.

\bibitem{CarLog99}
J~Cardy.
\newblock {Logarithmic Correlations in Quenched Random Magnets and Polymers}.
\newblock \texttt{arXiv:cond-mat/9911024}.

\bibitem{PirPre04}
G~Piroux and P~Ruelle.
\newblock {Pre-Logarithmic and Logarithmic Fields in a Sandpile Model}.
\newblock {\em J. Stat. Mech.}, 0410:P005, 2004.
\newblock \texttt{arXiv:\mbox{hep-th}/0407143}.

\bibitem{PeaLog06}
P~Pearce, J~Rasmussen, and J-B Zuber.
\newblock {Logarithmic Minimal Models}.
\newblock {\em J. Stat. Mech.}, 0611:017, 2006.
\newblock \texttt{arXiv:\mbox{hep-th}/0607232}.

\bibitem{ReaAss07}
N~Read and H~Saleur.
\newblock {Associative-Algebraic Approach to Logarithmic Conformal Field
  Theories}.
\newblock {\em Nucl. Phys.}, B777:316--351, 2007.
\newblock \texttt{arXiv:\mbox{hep-th}/0701117}.

\bibitem{RidPer07}
P~Mathieu and D~Ridout.
\newblock {From Percolation to Logarithmic Conformal Field Theory}.
\newblock {\em Phys. Lett.}, B657:120--129, 2007.
\newblock \texttt{arXiv:0708.0802 [\mbox{hep-th}]}.

\bibitem{RidPer08}
D~Ridout.
\newblock {On the Percolation BCFT and the Crossing Probability of Watts}.
\newblock {\em Nucl. Phys.}, B810:503--526, 2009.
\newblock \texttt{arXiv:0808.3530 [\mbox{hep-th}]}.

\bibitem{VasInd11}
R~Vasseur, J~Jacobsen, and H~Saleur.
\newblock {Indecomposability Parameters in Chiral Logarithmic Conformal Field
  Theory}.
\newblock {\em Nucl. Phys.}, B851:314--345, 2011.
\newblock \textsf{arXiv:1103.3134 [\mbox{hep-th}]}.

\bibitem{KytFro08}
K~Kyt\"{o}l\"{a}.
\newblock {SLE Local Martingales in Logarithmic Representations}.
\newblock {\em J. Stat. Mech.}, 0908:P08005, 2009.
\newblock \texttt{arXiv:0804.2612 [math-ph]}.

\bibitem{GruIns08}
D~Grumiller and N~Johansson.
\newblock {Instability in Cosmological Topologically Massive Gravity at the
  Chiral Point}.
\newblock {\em JHEP}, 0807:134, 2008.
\newblock \texttt{arXiv:0805.2610 [\mbox{hep-th}]}.

\bibitem{FloBit03}
M~Flohr.
\newblock {Bits and Pieces in Logarithmic Conformal Field Theory}.
\newblock {\em Int. J. Mod. Phys.}, A18:4497--4592, 2003.
\newblock \textsf{arXiv:\mbox{hep-th}/0111228}.

\bibitem{GabInd96}
M~Gaberdiel and H~Kausch.
\newblock {Indecomposable Fusion Products}.
\newblock {\em Nucl. Phys.}, B477:293--318, 1996.
\newblock \textsf{arXiv:\mbox{hep-th}/9604026}.

\bibitem{GabRat96}
M~Gaberdiel and H~Kausch.
\newblock {A Rational Logarithmic Conformal Field Theory}.
\newblock {\em Phys. Lett.}, B386:131--137, 1996.
\newblock \textsf{arXiv:\mbox{hep-th}/9606050}.

\bibitem{FeiLog06}
B~Feigin, A~Gainutdinov, A~Semikhatov, and I~Yu Tipunin.
\newblock {Logarithmic Extensions of Minimal Models: Characters and Modular
  Transformations}.
\newblock {\em Nucl. Phys.}, B757:303--343, 2006.
\newblock \textsf{arXiv:\mbox{hep-th}/0606196}.

\bibitem{EbeVir06}
H~Eberle and M~Flohr.
\newblock {Virasoro Representations and Fusion for General Augmented Minimal
  Models}.
\newblock {\em J. Phys.}, A39:15245--15286, 2006.
\newblock \textsf{arXiv:\mbox{hep-th}/0604097}.

\bibitem{GabLog06}
M~Gaberdiel and I~Runkel.
\newblock {The Logarithmic Triplet Theory with Boundary}.
\newblock {\em J. Phys.}, A39:14745--14780, 2006.
\newblock \textsf{arXiv:\mbox{hep-th}/0608184}.

\bibitem{AdaTri08}
D~Adamovi\'{c} and A~Milas.
\newblock {On the Triplet Vertex Algebra $\mathcal{W} \left(p\right)$}.
\newblock {\em Adv. Math.}, 217:2664--2699, 2008.
\newblock \texttt{arXiv:0707.1857 [math.QA]}.

\bibitem{RidLog07}
P~Mathieu and D~Ridout.
\newblock {Logarithmic $M \left( 2,p \right)$ Minimal Models, their Logarithmic
  Couplings, and Duality}.
\newblock {\em Nucl. Phys.}, B801:268--295, 2008.
\newblock \texttt{arXiv:0711.3541 [\mbox{hep-th}]}.

\bibitem{PeaInt08}
P~Pearce, J~Rasmussen, and P~Ruelle.
\newblock {Integrable Boundary Conditions and $W$-Extended Fusion in the
  Logarithmic Minimal Models $LM \left( 1,p \right)$}.
\newblock {\em J. Phys.}, A41:295201, 2008.
\newblock \textsf{arXiv:0803.0785 [\mbox{hep-th}]}.

\bibitem{GabFus01}
M~Gaberdiel.
\newblock {Fusion Rules and Logarithmic Representations of a WZW Model at
  Fractional Level}.
\newblock {\em Nucl. Phys.}, B618:407--436, 2001.
\newblock \textsf{arXiv:\mbox{hep-th}/0105046}.

\bibitem{LesLog04}
F~Lesage, P~Mathieu, J~Rasmussen, and H~Saleur.
\newblock {Logarithmic Lift of the $\widehat{su} \left( 2 \right)_{-1/2}$
  Model}.
\newblock {\em Nucl. Phys.}, B686:313--346, 2004.
\newblock \textsf{arXiv:\mbox{hep-th}/0311039}.

\bibitem{AdaCon05}
D~Adamovi\'{c}.
\newblock {A Construction of Admissible $A_1^{\left(1\right)}$-Modules of level
  $-\frac{4}{3}$}.
\newblock {\em J. Pure Appl. Alg.}, 196:119--134, 2005.
\newblock \textsf{arXiv:math.QA/0401023}.

\bibitem{RidFus10}
D~Ridout.
\newblock {Fusion in Fractional Level $\widehat{\mathfrak{sl}} \left( 2
  \right)$-Theories with $k=-\tfrac{1}{2}$}.
\newblock {\em Nucl. Phys.}, B848:216--250, 2011.
\newblock \textsf{arXiv:1012.2905 [\mbox{hep-th}]}.

\bibitem{LesSU202}
F~Lesage, P~Mathieu, J~Rasmussen, and H~Saleur.
\newblock {The $\widehat{su} \left( 2 \right)_{-1/2}$ WZW Model and the $\beta
  \gamma$ System}.
\newblock {\em Nucl. Phys.}, B647:363--403, 2002.
\newblock \textsf{arXiv:\mbox{hep-th}/0207201}.

\bibitem{FucNon04}
J~Fuchs, S~Hwang, A~Semikhatov and I~Yu~Tipunin.
\newblock {Nonsemisimple Fusion Algebras and the Verlinde Formula}.
\newblock {\em Comm. Math. Phys.}, 247:713--742, 2004.
\newblock \textsf{arXiv:\mbox{hep-th}/0306274}.

\bibitem{RidSL208}
D~Ridout.
\newblock {$\widehat{\mathfrak{sl}} \left( 2 \right)_{-1/2}$: A Case Study}.
\newblock {\em Nucl. Phys.}, B814:485--521, 2009.
\newblock \textsf{arXiv:0810.3532 [\mbox{hep-th}]}.

\bibitem{FjeDua11}
J~Fjelstad.
\newblock {On Duality and Extended Chiral Symmetry in the $SL \left( 2 ,
  \mathbb{R} \right)$ WZW Model}.
\newblock {J. Phys.}, A44:235404, 2011.
\newblock \texttt{arXiv:1102.4196 [\mbox{hep-th}]}.

\bibitem{RohRed96}
F~Rohsiepe.
\newblock {On Reducible but Indecomposable Representations of the Virasoro
  Algebra}.
\newblock \texttt{arXiv:\mbox{hep-th}/9611160}.

\bibitem{FjeLog02}
J~Fjelstad, J~Fuchs, S~Hwang, A~Semikhatov, and I~Yu Tipunin.
\newblock {Logarithmic Conformal Field Theories via Logarithmic Deformations}.
\newblock {\em Nucl. Phys.}, B633:379--413, 2002.
\newblock \texttt{arXiv:\mbox{hep-th}/0201091}.

\bibitem{FeiMod06}
B~Feigin, A~Gainutdinov, A~Semikhatov, and I~Yu Tipunin.
\newblock {Modular Group Representations and Fusion in Logarithmic Conformal
  Field Theories and in the Quantum Group Center}.
\newblock {\em Comm. Math. Phys.}, 065:47--93, 2006.
\newblock \texttt{arXiv:\mbox{hep-th}/0504093}.

\bibitem{FeiKaz06}
B~Feigin, A~Gainutdinov, A~Semikhatov, and I~Yu Tipunin.
\newblock {Kazhdan-Lusztig Correspondence for the Representation Category of
  the Triplet $W$-Algebra in Logarithmic CFT}.
\newblock {\em Theo. Math. Phys.}, 2006:1210--1235, 2006.
\newblock \textsf{arXiv:math/0512621}.

\bibitem{HuaLog07}
Y-Z Huang, J~Lepowsky, and L~Zhang.
\newblock {Logarithmic Tensor Product Theory for Generalized Modules for a
  Conformal Vertex Algebra}.
\newblock \textsf{arXiv:0710.2687 [math.QA]}.

\bibitem{RidSta09}
K~Kyt\"{o}l\"{a} and D~Ridout.
\newblock {On Staggered Indecomposable Virasoro Modules}.
\newblock {\em J. Math. Phys.}, 50:123503, 2009.
\newblock \textsf{arXiv:0905.0108 [math-ph]}.

\bibitem{AdaStr10}
D~Adamovic and A~Milas.
\newblock {The Structure of Zhu's Algebras for Certain $W$-Algebras}.
\newblock {\em Adv. Math.}, 227:2425--2456, 2011.
\newblock \textsf{arXiv:1006.5134 [math.QA]}.

\bibitem{Kac:1977em}
V~Kac.
\newblock {Lie Superalgebras}.
\newblock {\em Adv. Math.}, 26:8--96, 1977.

\bibitem{Kac:1977qb}
V~Kac.
\newblock {A Sketch of Lie Superalgebra Theory}.
\newblock {\em Comm. Math. Phys.}, 53:31--64, 1977.

\bibitem{KacSup87}
V~Kac and J~van~de~Leur.
\newblock {Super Boson-Fermion Correspondence}.
\newblock {\em Ann. Inst. Fourier}, 37:99--137, 1987.

\bibitem{RidSL210}
D~Ridout.
\newblock {$\widehat{\mathfrak{sl}} \left( 2 \right)_{-1/2}$ and the Triplet
  Model}.
\newblock {\em Nucl. Phys.}, B835:314--342, 2010.
\newblock \textsf{arXiv:1001.3960 [\mbox{hep-th}]}.

\bibitem{KauSym00}
H~Kausch.
\newblock {Symplectic Fermions}.
\newblock {\em Nucl. Phys.}, B583:513--541, 2000.
\newblock \textsf{arXiv:\mbox{hep-th}/0003029}.

\bibitem{Creutzig:2009zz}
T~Creutzig.
\newblock {\em {Branes in Supergroups}}.
\newblock PhD thesis, DESY Theory Group, 2009.
\newblock \textsf{arXiv:0908.1816 [\mbox{hep-th}]}.

\bibitem{HumRep08}
J~Humphreys.
\newblock {\em {Representations of Semisimple Lie Algebras in the BGG Category
  $\mathcal{O}$}}, volume~94 of {\em Graduate Studies in Mathematics}.
\newblock American Mathematical Society, Providence, 2008.

\bibitem{GotRep07}
G~Gotz, T~Quella, and V~Schomerus.
\newblock {Representation Theory of $\mathfrak{sl} \left( 2 \middle\vert 1
  \right)$}.
\newblock {\em J. Alg.}, 312:829--848, 2007.
\newblock \textsf{arXiv:\mbox{hep-th}/0504234}.

\bibitem{NahQua94}
W~Nahm.
\newblock {Quasirational Fusion Products}.
\newblock {\em Int. J. Mod. Phys.}, B8:3693--3702, 1994.
\newblock \textsf{arXiv:\mbox{hep-th}/9402039}.

\bibitem{DiFCon97}
P~Di Francesco, P~Mathieu, and D~S\'{e}n\'{e}chal.
\newblock {\em {Conformal Field Theory}}.
\newblock Graduate Texts in Contemporary Physics. Springer-Verlag, New York,
  1997.

\bibitem{SemEmb97}
A~Semikhatov and V~Sirota.
\newblock {Embedding Diagrams of $N=2$ Verma Modules and Relaxed $\widehat{sl}
  \left( 2 \right)$ Verma Modules}.
\newblock \textsf{arXiv:\mbox{hep-th}/9712102}.

\bibitem{GasBas04}
G~Gasper and M~Rahman.
\newblock {\em {Basic Hypergeometric Series}}, volume~96 of {\em Encyclopedia
  of Mathematics and its Applications}.
\newblock Cambridge University Press, Cambridge, 2004.

\bibitem{RidSU206}
P~Mathieu and D~Ridout.
\newblock {The Extended Algebra of the $SU \left( 2 \right)$ Wess-Zumino-Witten
  Models}.
\newblock {\em Nucl. Phys.}, B765:201--239, 2007.
\newblock \textsf{arXiv:\mbox{hep-th}/0609226}.

\bibitem{RidMin07}
P~Mathieu and D~Ridout.
\newblock {The Extended Algebra of the Minimal Models}.
\newblock {\em Nucl. Phys.}, B776:365--404, 2007.
\newblock \textsf{arXiv:\mbox{hep-th}/0701250}.

\bibitem{JacQua06}
P~Jacob and P~Mathieu.
\newblock {A Quasi-Particle Description of the $M \left( 3 , p \right)$
  Models}.
\newblock {\em Nucl. Phys.}, B733:205--232, 2006.
\newblock \textsf{arXiv:\mbox{hep-th}/0506074}.

\bibitem{RasWEx08}
J~Rasmussen and P~Pearce.
\newblock {W-Extended Fusion Algebra of Critical Percolation}.
\newblock {\em J. Phys.}, A41:295208, 2008.
\newblock \textsf{arXiv:0804.4335 [\mbox{hep-th}]}.

\bibitem{GabFus09}
M~Gaberdiel, I~Runkel, and S~Wood.
\newblock {Fusion Rules and Boundary Conditions in the $c=0$ Triplet Model}.
\newblock {\em J. Phys.}, A42:325403, 2009.
\newblock \textsf{arXiv:0905.0916 [\mbox{hep-th}]}.

\bibitem{AdaWAl09}
D~Adamovic and A~Milas.
\newblock {On $W$-Algebras Associated to $\left( 2,p \right)$ Minimal Models
  and their Representations}.
\newblock {Int. Math. Res. Not. IMRN}, 2010:3896--3934.
\newblock \textsf{arXiv:0908.4053 [math.QA]}.

\bibitem{Feigin:2004wb}
B~Feigin and A~Semikhatov.
\newblock {$W^{\left( 2 \right)}_n$ Algebras}.
\newblock {\em Nucl. Phys.}, B698:409--449, 2004.
\newblock \textsf{arXiv:math/0401164}.

\bibitem{CGL}
T~Creutzig, P~Gao, and A~Linshaw.
\newblock {A commutant realization of $W^{(2)}_n$ at critical level}.
\newblock To appear in {Int. Math. Res. Not. IMRN}.
\newblock \textsf{arXiv:1109.4065}.

\bibitem{Odake:1988bh}
S~Odake.
\newblock {Extension of $N=2$ Superconformal Algebra and Calabi-Yau
  Compactification}.
\newblock {\em Mod. Phys. Lett.}, A4:557, 1989.

\bibitem{Parker:1980af}
M.~Parker.
\newblock {Classification of Real Simple Lie Superalgebras of Classical Type}.
\newblock {\em J. Math. Phys.}, 21:689--697, 1980.

\bibitem{Berezin:1987wh}
F~Berezin, A~Kirillov (Ed), and D~Leites (Ed).
\newblock {\em {Introduction to Superanalysis}}, volume~9 of {\em Mathematical
  Physics and Applied Mathematics}.
\newblock D Reidel, Dordrecht, 1987.

\bibitem{PolGau90}
A~Polyakov.
\newblock {Gauge Transformations and Diffeomorphisms}.
\newblock {\em Int. J. Mod. Phys.}, A5:833--842, 1990.

\bibitem{BerCon91}
M~Bershadsky.
\newblock {Conformal Field Theories via Hamiltonian Reduction}.
\newblock {\em Comm. Math. Phys.}, 139:71--82, 1991.

\end{thebibliography}

\end{document}